\pdfoutput=1

\documentclass{tcibook}
\usepackage{everypage}
\usepackage{fancyhea}
\usepackage{multirow}
\usepackage{amsmath}
\usepackage{subfigure}
\usepackage{work}
\usepackage{bm}       
\usepackage{graphicx}
\usepackage{lineno}
\usepackage{mathrsfs}
\usepackage{multirow}
\usepackage{longtable}
\usepackage{tabularx}
\usepackage{float}
\usepackage{slashed}
\usepackage{xspace}
\usepackage{lscape}
\usepackage{rotating}
\usepackage{array}
\usepackage{booktabs}
\usepackage{dcolumn}
\usepackage{enumerate}
\usepackage[section]{placeins}
\usepackage{appendix}

\newcommand{\nc}{\newcommand}  

\def\beq{\begin{equation}}
\def\eeq#1{\label{#1}\end{equation}}
\def\eeqn{\end{equation}}

\newenvironment{Eqnarray}%
   {\arraycolsep 0.14em\begin{eqnarray}}{\end{eqnarray}}
\def\beqa{\begin{Eqnarray}}
\def\eeqa#1{\label{#1}\end{Eqnarray}}
\def\eeqan{\end{Eqnarray}}

\nc{\ra}{\rightarrow}  
\nc{\slsh}{\slash\hspace*{-0.22cm}}
\def\Re{{\cal R \mskip-4mu \lower.1ex \hbox{\it e}\,}}
\def\Im{{\cal I \mskip-5mu \lower.1ex \hbox{\it m}\,}}

\nc{\vev}[1]{ \left\langle {#1} \right\rangle }
\nc{\bra}[1]{ \langle {#1} | }
\nc{\ket}[1]{ | {#1} \rangle }
\nc{\fb}{\,{\rm fb}^{-1}}
\nc{\ev}{{\rm eV}}
\nc{\kev}{{\rm keV}}
\nc{\Mev}{{\rm MeV}}
\nc{\gev}{{\rm GeV}}
\nc{\tev}{{\rm TeV}}
\nc{\mev}{{\rm MeV}}

\def\del{\partial}
\def\Dslash{\not{\hbox{\kern-4pt $D$}}}
\def\dslash{\not{\hbox{\kern-2pt $\del$}}}
\def\pslash{\not{\hbox{\kern-2pt $p$}}}
\def\ETmiss{ \not{\hbox{\kern-4pt $E$}}_T }

\def\msb{{\bar{\ssstyle M \kern -1pt S}}}

\begin{document}

\def\bibname{References}
\bibliographystyle{plain}

\raggedbottom

\pagenumbering{roman}

\parindent=0pt
\parskip=8pt
\setlength{\evensidemargin}{0pt}
\setlength{\oddsidemargin}{0pt}
\setlength{\marginparsep}{0.0in}
\setlength{\marginparwidth}{0.0in}
\marginparpush=0pt


\pagenumbering{arabic}

\renewcommand{\chapname}{chap:intro_}
\renewcommand{\chapterdir}{.}
\renewcommand{\arraystretch}{1.25}
\addtolength{\arraycolsep}{-3pt}
\def\missET {{\not\!\! E_T}}



 
\chapter{New Particles Working Group Report}
\label{chap:newParticles}

\begin{center}\begin{boldmath}



\begin{center}

\begin{large} {\bf Conveners: Yuri Gershtein, Markus Luty, Meenakshi Narain,\\ Lian-Tao Wang, Daniel Whiteson} \end{large}

Authors: K. Agashe, L. Apanasevich, G. Artoni, A. Avetisyan, H.
Baer, C. Bartels, M. Bauer, D. Berge, M. Berggren,  S. Bhattacharya,  K. Black, T. Bose, J. Brau,
R. Brock, E. Brownson, M. Cahill-Rowley, A. Cakir, A. Chaus,
T. Cohen, B. Coleppa, R. Cotta,
N. Craig, K. Dienes, B. Dobrescu, D. Duggan, R. Essig, J. Evans, 
A. Drlica-Wagner, S. Funk, Y. Gershtein, J. George, F. Goertz, T. Golling, T. Han, A. Haas, M. Hance, D. Hayden,
U. Heintz, A. Henrichs, J. Hewett, J. Hirschauer, K. Howe,
A. Ismail, K. Kaadze, Y. Kats, F. Kling, D. Kolchmeyer, 
D. Kr{\"u}cker, K.C. Kong, A. Kumar, G. Kribs, P.
Langacker, A. Lath, S. J. Lee, J. List, T. Lin, L. Linssen, T. Liu, Z.
Liu, A. Lobanov, J. Loyal, M. Luty, A. Martin, I. Melzer-Pellmann, 
M. Narain, M.M. Nojiri, S. Padhi, N. Parashar, B. Penning, M. Perelstein, M. Peskin, 
A. Pierce, W. Porod, C. Potter, T. Rizzo, G. Sciolla, J. Stupak III, S. Su, T.M.P. Tait,  T.
Tanabe, B. Thomas, S. Thomas, S. Upadhyay, N. Varelas, E. Varnes, L. Vecchi,
A. Venturini, B. Vormwald,
J. Wacker, M. Walker,  L.-T. Wang, D. Whiteson, M. Wood, F. Yu, N. Zhou
\end{center}


\end{boldmath}\end{center}


\begin{center}
{\bf Abstract}
\end{center}
This report summarizes the work of the Energy Frontier New Physics working group of the 2013 Community Summer Study (Snowmass).

\section{Executive Summary}
\begin{itemize}

\item With the discovery of the Higgs, we have experimentally
established the standard model with a scalar particle that
appears to be elementary. This gives us a model that can be
extrapolated to very high energy scales and forces the
question of the naturalness of elementary scalars. 
Additional motivation for further exploration of the
TeV scale comes from  supersymmetry, Higgs compositeness, 
and dark matter, as well as connections to the other frontiers
through flavor and neutrino physics.

\item The LHC run 1 new physics program is extremely broad, and
has out-performed expectations due to innovative search
techniques and advances in theory. It has provided strong
constraints on a wide variety of new physics models.

\item 14 TeV LHC with 300~fb$^{-1}$ will provide an 
enormous gain in sensitivity to a wide range of new physics models 
due to increase of both energy and luminosity. 
Roughly this corresponds to an order of
magnitude in tuning in supersymmetry and composite models.

\item At the high-luminosity LHC (14~TeV with 3000~fb$^{-1}$, 
any preceding 
LHC run 2 discovery can be extensively studied.
The high-luminosity LHC also extends the reach for new physics.
For most models the improvements are in the 
electroweak sector and improvement in tuning can be achieved by 
a factor of 2 to 4 from the supersymmertic sector.
A 33 TeV upgrade would allow any hint of new physics in the 14 TeV program to be extensively studied, as well as providing significant additional reach for new discoveries.

\item The ILC new physics program has been studied in great
detail, and has excellent capabilities to discover and
measure the properties of new physics, including dark matter, 
with almost no loopholes. 
A necessary requirement is that the new physics
must be accessible. Essentially this means particles at
sufficiently low mass missed by LHC due to blind spots, or
heavy physics indirectly accessible through precision
measurement. Discovery of physics beyond the standard model
at LHC that is accessible at ILC would make the case even
more compelling.

\item 
A 100 TeV $pp$ collider has unprecedented and robust reach for new physics 
that is evident even with the preliminary level of studies performed so far. 
It can probe an additional two orders of magnitude in fine-tuning in 
supersymmetry compared to LHC14, and can discover WIMP dark matter up to the 
TeV mass scale. 
Any discovery at the LHC would be accessible at this machine and could be 
better studied there, making the case for these options even more compelling. 


\item High energy $e^+ e^-$ colliders such as CLIC
and muon colliders offer a long-term
program that can extend precision and reach of a wide range
of physics.
\end{itemize}

\begin{figure}[h!]
\hspace*{-0.6in}
\subfigure{
\includegraphics[width=0.7\textwidth]{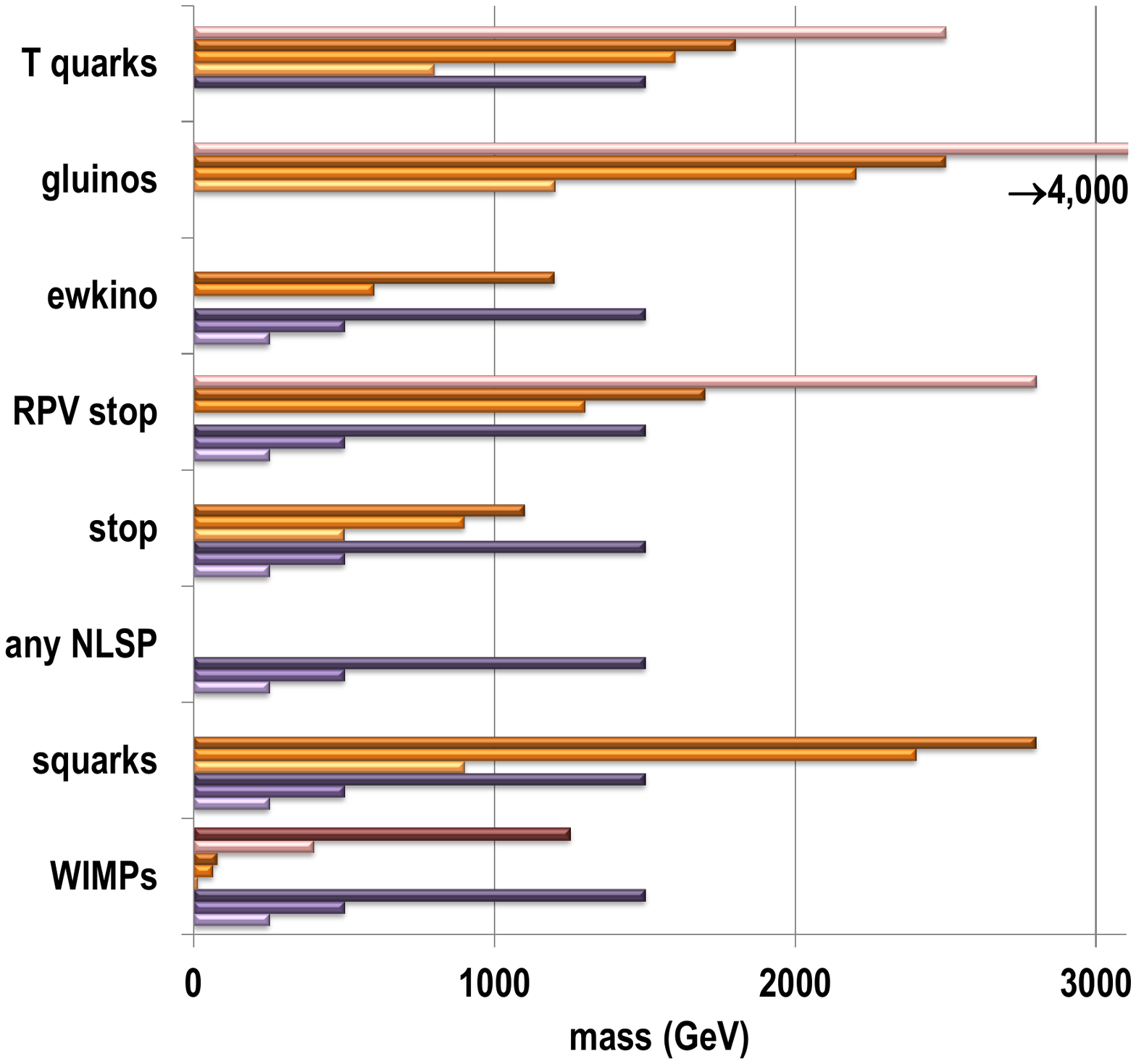} }
\hspace*{-1.5in}
\subfigure{
\includegraphics[width=0.7\textwidth]{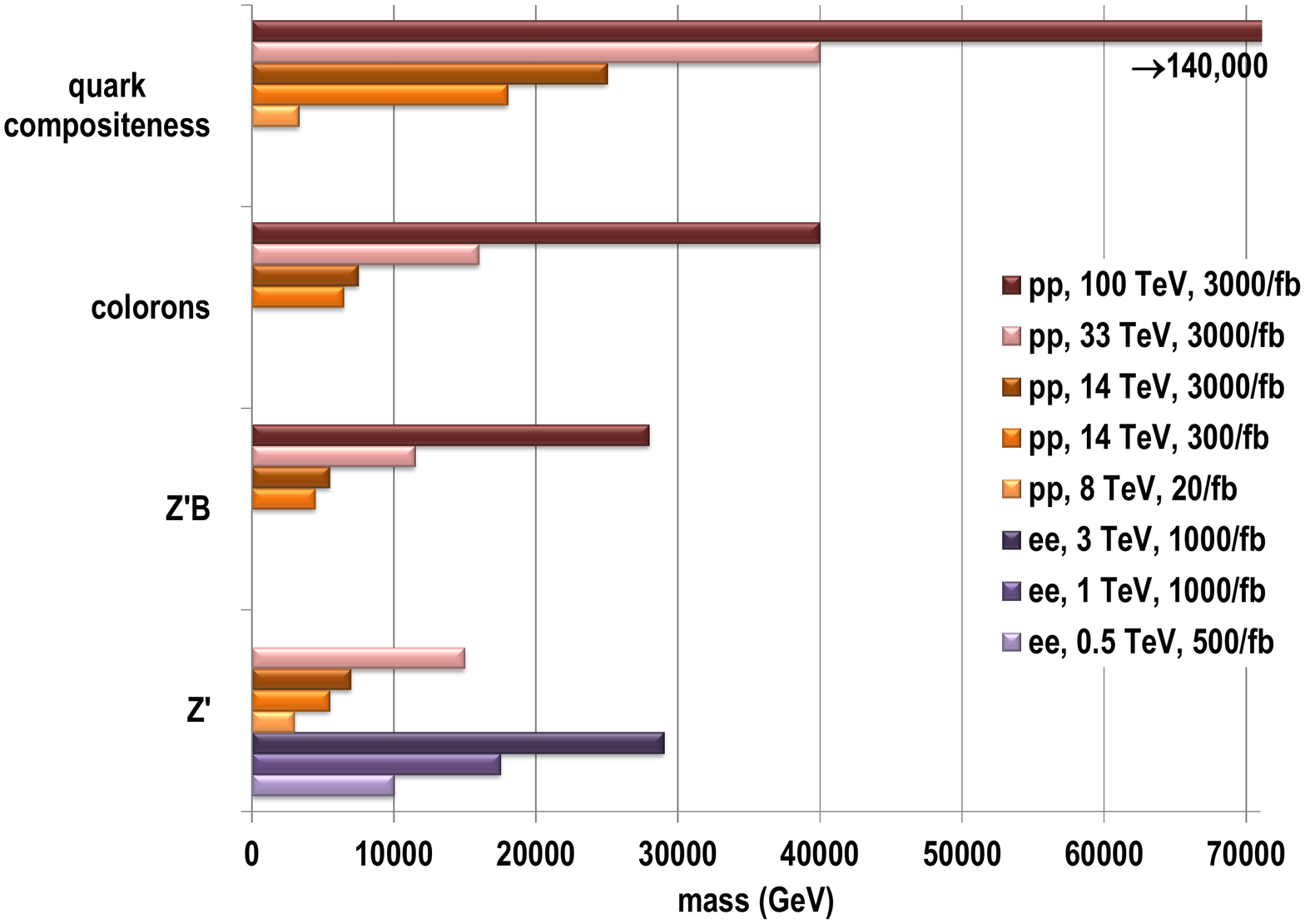}}
\caption[]{95\% confidence level upper limits for masses of new particles
beyond the standard model expected from $pp$ and $e^+e^-$ colliders at 
different energies.
Although upper mass reach is generally higher at $pp$ colliders, these
searches often have low-mass loopholes, while $e^+ e^-$ collider searches
are remarkably free of such loopholes.}
\label{fig:summary}
\end{figure}

A summary of the energy reach for a range of
physics beyond the SM at various proposed facilities is shown in 
Fig.~\ref{fig:summary}.
This is a highly simplified plot.
In particular, although the mass reach of hadron colliders is generally
very impressive, hadron colliders searches often have blind spots,
for example due to compressed spectra or suppressed couplings.
Searches at $e^+ e^-$ colliders are much more model
independent, but generally have more limited mass reach.
Many examples of this complementarity 
are discussed in the body of this report.

\section{Introduction}


Searches for new particles at high-energy particle
accelerators have historically been one of the most 
fruitful paths to discovery of new fundamental particles and interactions, and
the establishment of new laws of nature.
Particles with any possible quantum numbers can be produced in 
particle-antiparticle pairs, provided only that they are kinematically accessible 
and couple with sufficient strength to the colliding particles.
General-purpose particle detectors measure the kinematics of all
particles with strong or electromagnetic couplings, 
as well as the `missing' momentum due to weakly interacting particles.
This gives them the ability to discover almost any possible decay
of new particles, as well as new stable particles.
An immense body of theoretical and phenomenological work has given
a detailed understanding of the effects of the standard model
particles and interactions.
This allows new particles to be discovered above standard model
backgrounds, and their detailed properties to be measured.
This is exemplified by the discovery of the Higgs boson.
Within a year of the initial discovery, the Higgs program has progressed to 
detailed measurements of Higgs properties, and the standard model with
a Higgs has been experimentally established, at least as a 
leading approximation.

This is a beginning, not an end.
There are many reasons to think that the standard model is incomplete,
and that there is new physics to be discovered in searches at the
energy frontier.
Many of these are discussed in the while papers submitted to the new
particles group.
The number, diversity, and quality of these white papers attests to the 
intellectual vigor of this area of research.
Rather than attempt a summary of this work, 
we have decided to illustrate the many
exciting possibilities with some examples.
This report is therefore organized around a number of well-motivated
signals where signs of new physics may be found.
To illustrate the impact of such a discovery and the possibilities
for further study, we consider in each case a particular model
where a discovery can be made at LHC Run 2 (14~TeV with a luminosity of
300/fb).
In each case, such a discovery suggests one or more natural candidate
models that can be studied in more detail at future experimental facilities.
These `discovery stories' rely heavily on the white papers, which give
more comprehensive treatments of the subject.
We also consider the case for continuing the search for new physics
at the energy frontier if there is no discovery at LHC run 2.



\subsection{Physics Motivation}
With the discovery of the Higgs boson, particle physics is entering a new era: 
we now have a theory that can be consistently extrapolated to scales many orders
of magnitude beyond those that we can directly probe experimentally. 
At the same time, there has been no observation of physics beyond the standard
model at high-energy colliders.
This raises the question of whether there are in fact
discoveries to be made at the TeV energy scale that are accessible
at the energy frontier.

Our answer is that there is strong motivation to continue the search
for new physics at the TeV scale and beyond.
The impetus for this 
comes from both {\it big questions} and {\it big ideas.}
For some of the big questions, the
answers must lie at the TeV scale, while for others this is only suggested by the
principle of minimality of scales in nature.
The big ideas arose from the necessity of reconciling the
highly constrained theoretical framework of quantum field theory
with the phenomena observed in nature,
as well as from theoretical investigations, especially string theory.

Some of the big questions:
\begin{itemize}

\item {\it What is the dark matter?} 
The cosmological and astrophysical evidence for dark matter is incontrovertible,
but its particle origin is completely unknown.
The most compelling candidate is a weakly interacting massive particle
(a WIMP), a thermal relic with mass at the electroweak
scale, whose interactions with ordinary matter determine its cosmological density
today.
In this scenario, dark matter can be directly produced and studied at energy
frontier colliders.


\item {\it Is the Higgs boson solely responsible for electroweak symmetry breaking
and the origin of mass?}
%
%
%
The 125~GeV Higgs boson appears to be the first scalar elementary particle
observed in nature.
Its measured couplings make it
clear that it plays a central role in breaking electroweak symmetry and giving
mass to the other elementary particles.
But the Higgs boson may be wholly or partially composite, 
and/or there may be additional Higgs bosons
as part of a larger Higgs sector.
These possibilities can be explored by detailed study of the 125~GeV
Higgs boson and direct searches for extended Higgs sectors.

\item {\it Are fundamental parameters finely tuned?} 
The mass of an elementary Higgs boson is sensitive to physics at high energy
scales.
If there is no physics beyond the standard model, the fundamental
Higgs mass parameter must be adjusted
to an accuracy order 1 part in  $10^{32}$ in order to explain the separation between 
the TeV scale and the Planck scale.
Avoiding this fine-tuning is one of the main motivations for physics beyond
the standard model.
Models that eliminate this tuning predict new particles at the TeV scale
that couple to the Higgs and can be explored at collider experiments.

\item {\it What is the origin of the matter-antimatter asymmetry?} 
Antimatter has almost mirror-image properties with matter, and yet the
universe is made almost entirely of matter.
Explaining this requires new physics to provide baryon number violation,
$CP$ violation, and out-of-equilibrium physics in the early universe.
If this takes place at the electroweak scale, this implies new sources
of $CP$ violation and new scalar particles at the TeV scale.

%

\item {\it What is the origin of quark, lepton, and neutrino 
mass hierarchies and mixing angles?} 
These `flavor' parameters account for most of the fundamental parameters of particle
physics, and their pattern remains mysterious.
New particles at the TeV scale with 
flavor-dependent couplings are present in many models, and observation of 
such particles would provide additional important clues to this puzzle.

\item {\it Are there new fundamental forces in nature?} 
Candidates for fundamental theories such as string theory generally predict
additional gauge forces and other interactions that can arise at the TeV
scale.
Discovering these will yield invaluable clues to the structure of the
fundamental interactions of nature.


\item {\it Are `elementary' particles really composite?}
This possibility is motivated by the fact that many of the particles we
observe are composite states of underlying dynamics, and by attempts
to address the other big questions listed above.
Uncovering evidence of compositeness at the TeV scale would be another
window on new forces in nature.
\end{itemize}

\begin{figure}
\begin{center}
\includegraphics[scale=0.4]{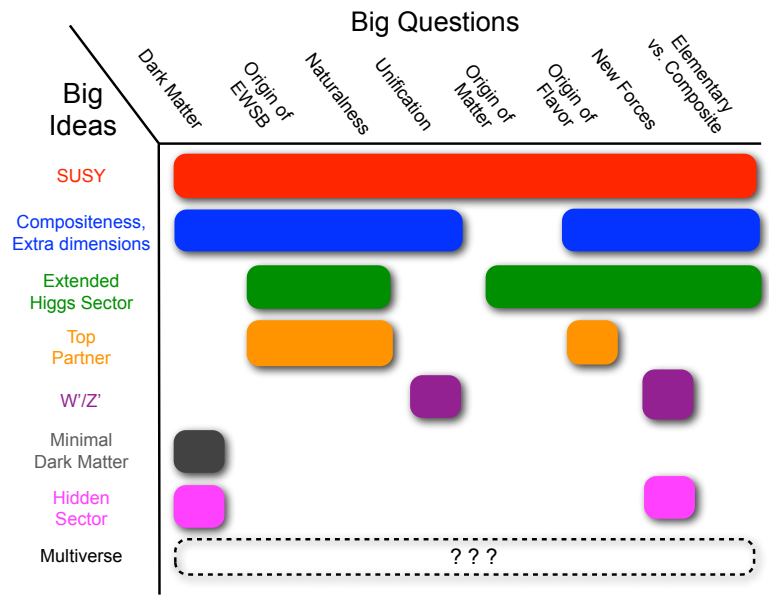}
\caption{\label{fig:QA}
Overlap between the questions and ideas discussed in the text.}
\end{center}
\end{figure}

Over the last few decades, advances in theoretical physics have also led to
big ideas about fundamental physics that can be probed at the energy
frontier.
Some of them are specifically designed to address one of the big questions
given above, others have even broader implications. 

\begin{itemize}

\item{\it Supersymmetry.} 
This is an extension of Einstein's symmetry of space and time
that relates particles with different spins.
It is required for consistency of string theory, and holds out the promise of
unifying all the fundamental forces of nature.
Supersymmetry the unique framework that allows an elementary Higgs boson without 
fine-tuning, provided that supersymmetry is broken at the TeV scale.
Minimal versions of supersymmetry automatically predict
gauge coupling unification, and provides a candidate for dark matter.

\item{\it Extra dimensions and compositeness.} 
Additional dimensions of space that are too small to be seen directly are a
ubquitous feature of string theory.
Excitations of these extra dimensions can manifest themselves in new particles
and interactions.
Remarkably, some theories with extra dimensions are equivalent (or `dual')
to composite theories.
This has led to a deeper understanding of both extra dimensions and compositeness,
and led to many interesting and detailed proposals for new phyics based
on these ideas.

\item{\it Unification of forces.}
The idea that all elementary interactions have a unified origin
goes back to Einstein, and has its modern form in grand unification and
string theory.
There is experimental evidence for the unification of gauge couplings
at short distances, and string theory generally predicts additional
interactions that may exist at the TeV scale.

\item{\it Hidden Sectors.}
Additional particle sectors that interact very weakly with standard model
particles are a generic feature of string theory, and may play an
important role in cosmology, for example dark matter.

\item{\it `Smoking Gun' Particles.}
Some kinds of new particles give especially important clues about
the big questions and ideas discussed here.
Top partners are required in most solutions to the naturalness problem;
additional Higgs bosons are present in many models of electroweak symmetry
breaking; contact interactions of dark matter with standard model particles
are the minimal realization of WIMP dark matter; 
and unified theories often predict new gauge bosons ($W'/Z'$) that mix with the
electroweak gauge bosons.

\item{\it The Multiverse.}
String theory apparently predicts a `landscape' of vacua, and eternal inflation
gives a plausible mechanism for populating them in the universe.
The implications of this for particle physics and cosmology are far from
clear, but it has the potential to account for apparently unnatural
phenomena, such as fine-tuning.

\end{itemize}

These questions and ideas are summarized in Fig.~\ref{fig:QA},
along with the connections between them.

\section{Discovery Stories}

\subsection{Higgs Beyond the Standard Model}
%
%

The discovery of the 125~GeV Higgs boson has made a fundamental change in our view of
particle physics.
The properties of the Higgs already measured at the LHC are in agreement
with those of an elementary standard model Higgs boson at the $10\%$ level,
forcing the question of the naturalness of elementary scalars.
The naturalness of the Higgs boson requires new physics at the TeV
scale.
The observed Higgs particle may itself be the harbinger of new physics:
it may mix with other Higgs bosons in an extended Higgs sector,
have modified couplings due to standard model states,
or couple to new physics such as dark matter.
In this section we discuss the opportunities for discovery
in some of these scenarios.

\paragraph{Composite Higgs models:}
Besides supersymmetry, the other major idea for solving the naturalness
problem is compositeness of the Higgs sector.
This category includes models with extra dimensions at the TeV scale,
such as Randall-Sundrum models.
The most basic indication that the physics of extra dimensions is closely related 
to compositeness is that both predict towers of
resonances at the TeV scale.
This connection is a major theme of modern particle theory, extending
all the way from phenomenology and model-building to string theory
via the AdS/CFT correspondence.


After the discovery of the 125~GeV Higgs boson, the only natural version of
Higgs sector compositeness is that the observed Higgs boson is a
pseudo Nambu-Goldstone boson (PNGB) arising from
spontaneous breaking of a global symmetry at a scale above the TeV scale.
There are corrections to the couplings of the 125~GeV Higgs boson that are
proportional to $\xi = (v/f)^2$, where
$v = 246$~GeV is the vacuum expectation value of the Higgs boson and
$f$ is a compositeness scale.
The ratio $\xi$ is also a direct measure of the fine-tuning required in the 
model, and is presently  less than approximately $0.1$ due to
Higgs coupling measurements.
Improvements on Higgs coupling measurements will improve bounds on this 
parameter.

These PNGB Higgs models can be probed at hadron colliders in several ways.
First, additional naturalness considerations motivate the presence
of top partners in these models.
These can be searched for at hadron colliders, as discussed in 
\S\ref{sec:toppartners} below.
Second, the heavy resonances expected from the symmetry breaking sector
at the scale $f$ can be probed.
Limits from LHC run 1 for these particles require them to have masses above
2~TeV, and the LHC run 2 and HL-LHC will have additional reach for these
models.

At lepton colliders such as ILC and CLIC, the most sensitive probe
of the parameter $\xi$ is the Higgs coupling fit.
For example, these can reach down to $\xi \sim 0.002$
at 3~TeV CLIC with a luminosity of 1$/$ab \cite{clicwp,cliccdr}.
The impact of other future colliders for the Higgs coupling fit
is discussed in the Higgs working group report.
In addition, CLIC has a high sensitivity to enhanced double Higgs production,
which is a direct consequence of the compositeness of the Higgs
boson and therefore a `smoking gun' for composite modes.
This also directly measures the parameter $\xi$ and can be probed
down to $\xi \simeq 0.03$ \cite{clicwp}.

%
%

\paragraph{Exotic Higgs decays:}
An essential component of the research program of the LHC, and any other 
future high-energy collider is the search for non-standard (`exotic') decays 
of the 125-GeV Higgs boson, {\it i.e.}~decays of a SM-like Higgs boson into 
one or more particles beyond the SM.
Non-standard Higgs decays have always been a well-motivated possibility as evidenced 
by an extensive existing, and growing, literature 
(see~\cite{exotics} for a comprehensive survey of the possibilities and 
an extensive list of references).
They remain a well-motivated possibility even with the discovery of a Higgs particle 
that is consistent with the simplest SM expectations, and they must be searched
for explicitly as they are often
unconstrained by other analyses.

The Higgs is the only SM particle that can have renormalizable couplings to 
SM gauge singlet fields.
This, together with the strikingly narrow width of the SM Higgs boson, 
implies that a small coupling of the Higgs to a new, 
light state can easily give rise to sizable branching ratios to non-standard decay modes.  For example, the addition of a singlet scalar field $S$ to the SM, which couples to 
the Higgs through a quartic interaction of the form
$\lambda S^2 H^2$ is a common building block in models of extended Higgs sectors.
Even a coupling as small as $\lambda = 5\times 10^{-3}$ yields a $10\%$ 
branching ratio of $h\to SS$ (for $m_S < m_h/2$).  
Exotic Higgs decays are a generic feature of extensions of the SM that contain light
states, and naturally arise in many models of electroweak symmetry breaking, 
such as the NMSSM and Little Higgs models. 
A detailed experimental characterization of the Higgs boson at 
125~GeV includes searches for
{\it e.g.}~$h \to {\rm invisible}$, $(1,2)\gamma+{\rm MET}$, 
$\gamma+Z'$, $Z'+{\rm MET}$, lepton- or photon-jets, isolated leptons+MET, 
$4x, 2x2y$ (with $x, y = e, \mu, \tau$-leptons, MET, jets, or $b$-jets), 
with or without displaced vertices.

One of the best-motivated examples
of exotic Higgs decays is the decay of the 125~GeV Higgs to invisible
particles.
This is strongly motivated by the coupling of the Higgs to dark matter.
Indeed, such a coupling gives a direct detection cross section of the
size that is currently being probed in dark matter direct detection
experiments.
LHC limits on the branching ratio $h \to \mbox{invisible}$
are $0.65$ (ATLAS) \cite{ATLAS-CONF-2013-011} 
and $0.75$ (CMS) \cite{CMS-PAS-HIG-13-018}.
This illustrates that there is a great deal of room for new physics
in this channel.
The Snowmass Higgs working group concludes that the
LHC14 with 300/fb will probe branching fractions in the range
$0.17$--$0.28$, while the HL-LHC will probe branching
fractions in the range $0.06$--$0.17$.
The quoted ranges depend on the control of systematics.
Overall, the HL-LHC provides an order of magnitude improvement
in the reach for this mode.
The invisible decay of the Higgs can be cleanly studied at
$e^+ e^-$ colliders via $Zh$ production.
For example, the ILC can detect invisible decays with a branching ration as 
low as $0.69$ \cite{HiggsWG}.
In fact, the reconstruction of Higgs boson recoiling from the $Z$
allows the measurement of any possible exotic decay in a model-independent
way, while also giving accurate measurements of the Higgs
mass and couplings.
%
%
%
%
There are no studies of invisible decaying Higgs for VLHC, but the LHC results
provide a proof of concept that this is possible.

In an extended Higgs sector with singlets, it is very common to have
lighter Higgs bosons with suppressed couplings to the $Z$, but which
can be seen at an $e^+ e^-$ collider either through direct production
or by decays of the 125~GeV Higgs boson.
A specific example of this kind 
in the next-to-minimal SUSY standard model
(NMSSM)
that has been studied for this report
is the cascade decay $h_1 \to aa \to (\tau\tau)(\tau\tau)$
at the ILC \cite{Liu:2013gea}.
This is an interesting example of an early discovery mode at a 250~GeV ILC.
In addition to discovery,
the masses can be measured to better than $1\%$.

The SM higgs boson may well be part of an extended Higgs sector that
facilitates dark matter annihilations.  The Higgs
portal~\cite{Patt:2006fw} features new Higgs bosons that undergo a
hidden sector Higgs mechanism.  These new Higgs bosons mix with the SM
Higgs boson and modify it's coupling strength by the cosine of the
mixing angle.  If the dark matter annihilates through the Higgs
portal, \cite{Walker:2013hka} shows that Xenon1T covers almost all of
the Higgs portal parameter space.  Given a direct detection signal,
accelerator searches would be required to understand if the Higgs
portal is responsible for the dark matter annihilation.  The smoking
gun for this mechanism is the observation of a new Higgs-like boson
that decays invisibly.  This is possible at the LHC14, VLHC and/or the
ILC.  Because the SM Higgs coupling strength is modified by the Higgs
mixing angle, deviations from the SM value can be probed by the 1 TeV
ILC for 1000 fb$^{-1}$ to cover most of the Higgs portal parameter
space.

\paragraph{Connection with neutrino masses:}
A more speculative but very interesting possibility is that the Higgs
discovered at the LHC is part of an extended Higgs sector at the TeV
scale that also generates naturally small neutrino masses.
A very interesting example of this possibility is the type-II see-saw
model consisting of a Higgs triplet coupling to left-handed leptons
and the Higgs doublet.
This model contains a doubly-charged Higgs scalar with lepton number
violating decays to same-sign $ee$, $\mu\mu$, and $\tau\tau$.
Searches at LHC have placed limits of $m_{H^{++}} > 400$~GeV
using $4.7/$fb at 7 TeV \cite{ATLAS:2012hi}.
LHC run 2 can extend the reach for the charged Higgs to approximately
1~TeV in favorable cases.
The branching ratios to various charged lepton flavors are correlated
with the neutrino masses, so that the neutrino mass hieararchy can be
determined from collider data.
This provides an exciting connection with the neutrino physics program.
A well-motivated extension of this model is left-right symmetric model,
which contains a right-handed $W$ boson that can be observed at the
HL-LHC with masses up to 4~TeV.

\begin{figure}[h!]
\includegraphics[scale=0.40]{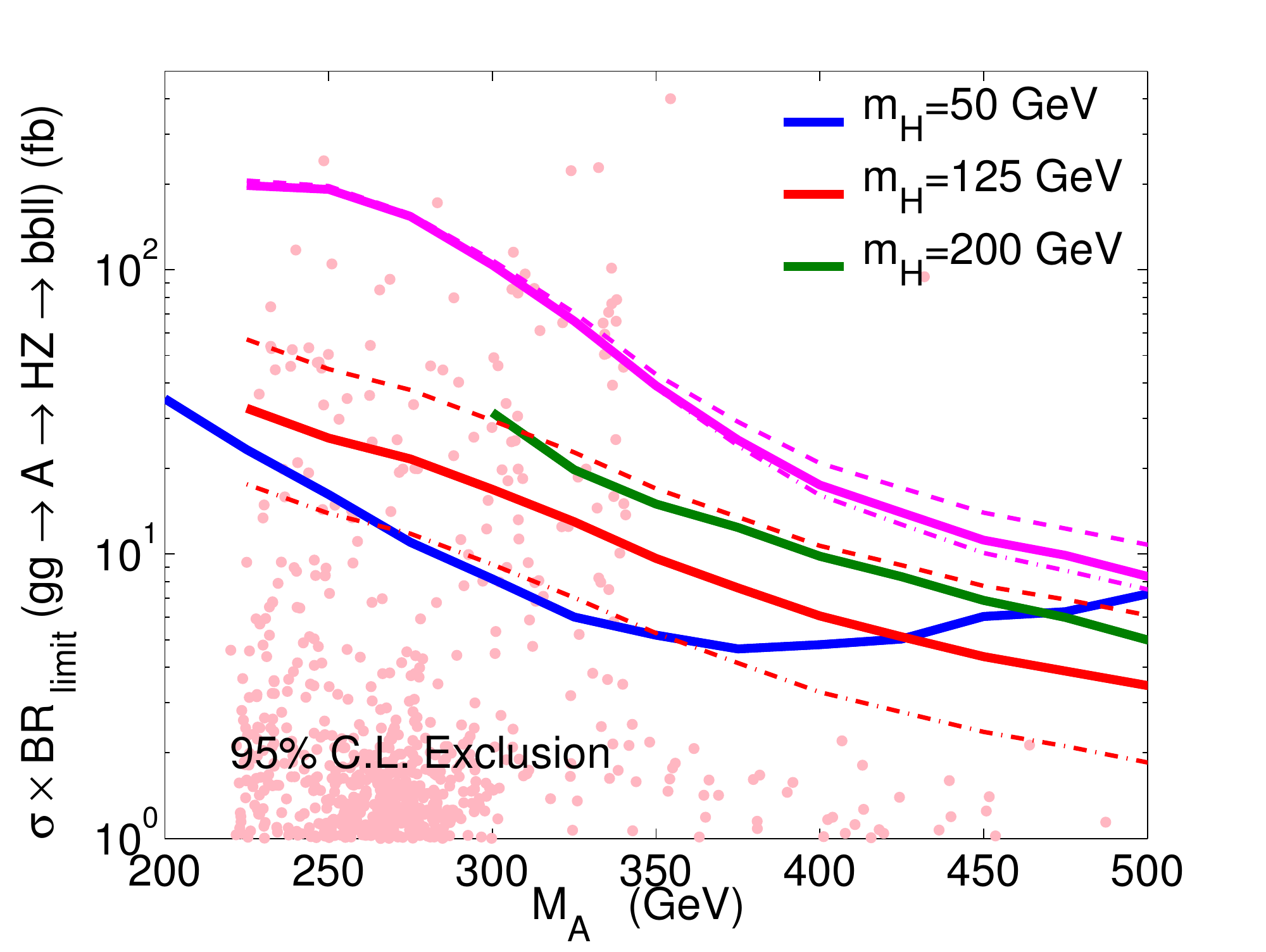}
\includegraphics[scale=0.40]{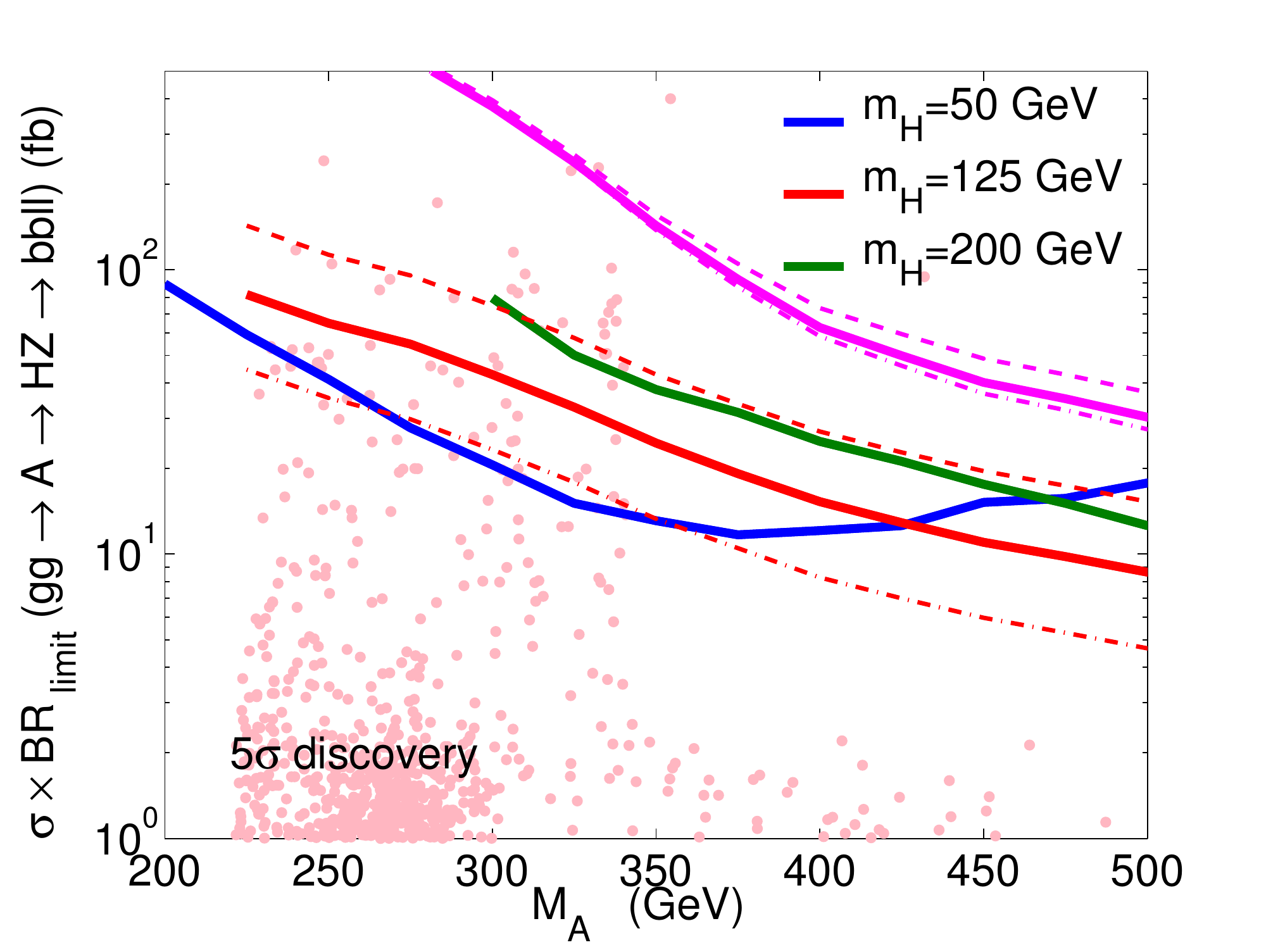}
 \caption{ 95\% C.L. exclusion (left panel) and 5$\sigma$ reach (right panel)  for $A \rightarrow hZ \rightarrow bb \ell \ell)$ for LHC14 with 300 ${\rm fb}^{-1}$ and  $m_h=$ 50, 125 and 200 GeV.  Also shown are the reach 
of  100 ${\rm fb}^{-1}$ (dashed lines) and 1000 ${\rm fb}^{-1}$ 
(dash-dotted lines) for the case of $m_h=125$ GeV, and the reach  
assuming 10\% systematic error for the backgrounds (purple lines).  
Scattering dots are the possible $\sigma \times BR$ range  in the 
NMSSM for the daughter Higgs mass being $125\pm2$ GeV~\cite{Coleppa:2013dya}.}
\label{fig:AZhcslimitNMSSM}
\end{figure}

\paragraph{Extended Higgs sectors:}
Additional Higgs bosons are predicted by many well-motivated extensions of the 
Standard Model, including supersymmetry and some composite Higgs models. 
Searches for these states are also motivated simply from the need to
fully explore the Higgs sector.
An example of a model-independent search is the search for $A \to ZH$,
where $A$ is a neutral pseudoscalar and $H$ is a scalar that may
or may not be the 125~GeV Higgs.
These states are present in SUSY models, and the
possibility of $H$ below 125~GeV is natural for example in the NMSSM.
This has been studied in \cite{Coleppa:2013dya}, and the reach for LHC14
results are shown in  Fig.~\ref{fig:AZhcslimitNMSSM}.
This shows that there is extensive reach for discovery in this mode.

Another aspect of extended Higgs sectors is that 
mixing between the 125~GeV Higgs and additional
Higgs bosons modifies the couplings of the observed
125~GeV Higgs compared to standard model values.
Therefore, searches for new Higgs states and precision Higgs coupling measurements
are complementary approaches to the important problem of studying the newly-found
Higgs particle.
We will discuss this in the context of a 2 Higgs doublet model (2HDM),
in which the observed 125~GeV state is the lightest CP-even neutral scalar $h$.
In addition, this model contains another neutral CP-even scalar $H$,
a CP-odd neutral scalar $A$, and a charged Higgs $H^\pm$.
In addition to the masses of the various Higgs bosons, the model
depends on the CP-even mixing angle $\alpha$ and the ratio of 
vacuum expectation values of the 2 Higgs doublets
$\tan\beta = v_1/v_2$.

The Higgs coupling constant fits for the 2HDM restrict the model
to small values of $\cos(\beta - \alpha)$.
This is naturally realized in the decoupling limit 
where one linear combination of the Higgs doublets is heavy and has a
small vacuum expectation value.
The couplings of the lightest Higgs $h$ approach their standard model
values as $\cos(\beta - \alpha) \to 0$, so these searches also probe
deviations from the decoupling limit.
Direct searches for new Higgs bosons can probe to smaller values
of $\cos(\beta - \alpha)$ provided the masses of the new states are
sufficiently light.
Therefore, direct Higgs searches and precision Higgs coupling measurements
are complementary probes of new physics in this scenario.

Precision Higgs couplings are extensively discussed in the Higgs
working group report in these proceedings, so we focus on direct 
searches here.
Higgs bosons generally couple most strongly to particles with the
largest mass,  motivating  searches for new particles that decay to
vector bosons,  the $h$, or heavy fermions (top/bottom/tau).  These
couplings also lead to appreciable production rates at both hadron and
lepton  colliders. 

At hadron colliders, the $H$ and $A$ can be copiously singly produced
through gluon fusion or $b \bar b$ and $t \bar t$  associated
production.  Charged Higgs bosons can be produced singly via $t b$
associated production  at $pp$ colliders  or pair-produced at $e^+
e^-$ colliders.  Decays of these additional states may be observed in
standard Higgs search channels or novel resonant channels involving
the 125~GeV Higgs such as $Wh, Zh,$ or $hh$.

\begin{figure}
\centering
\includegraphics[width=0.4\linewidth]{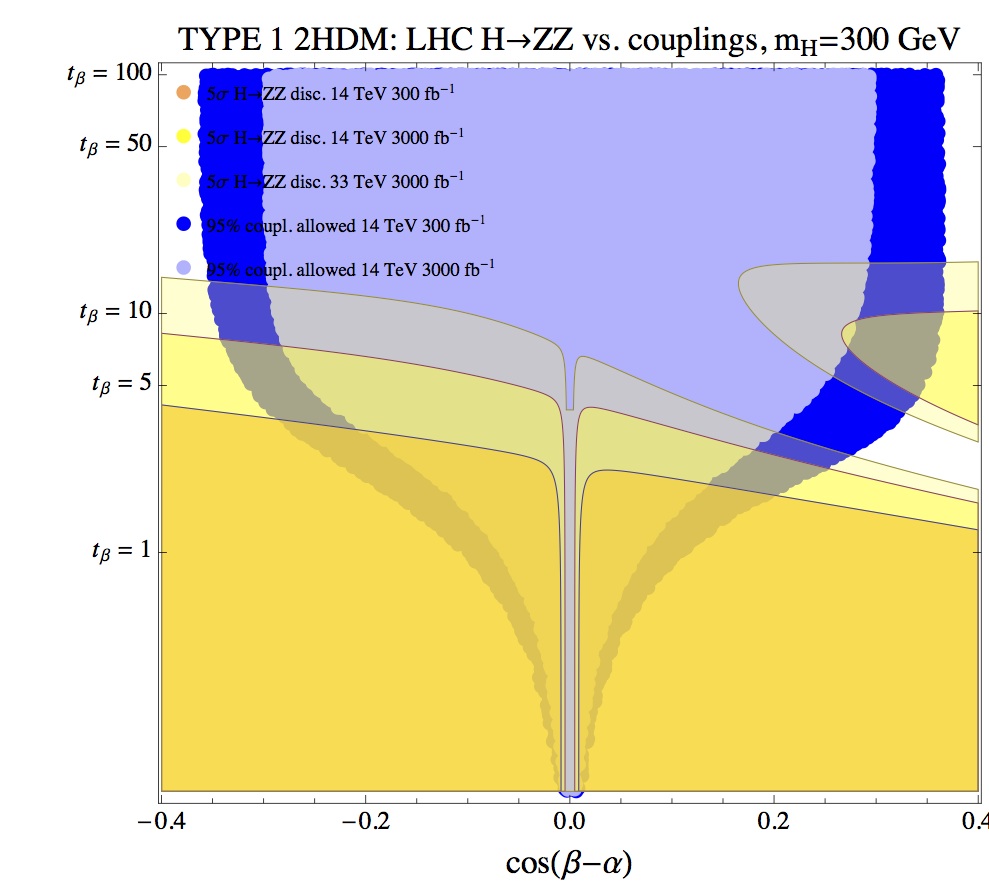}
\includegraphics[width=0.4\linewidth]{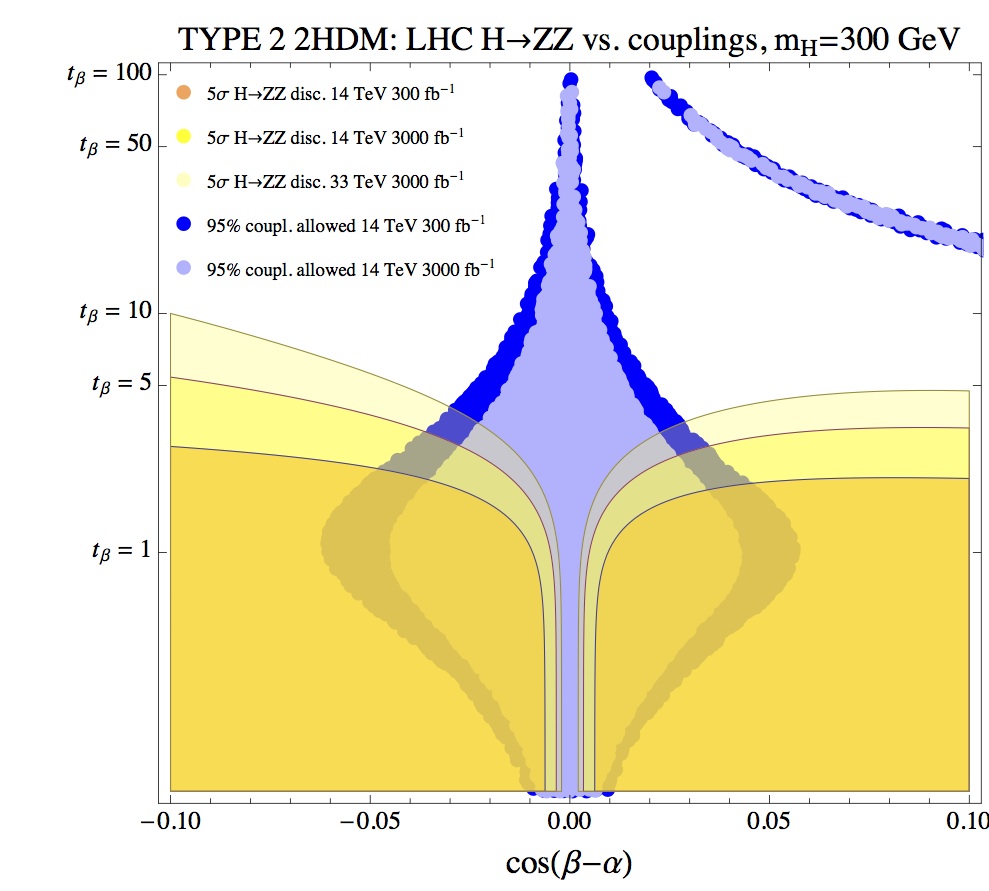}
\caption{$5\sigma$ discovery reach for 300~GeV $H$ 
decaying via $H \rightarrow ZZ \rightarrow 4\ell$.
The Type I and Type II 2HDM models as shown in the left  
and right panels respectively. Dark and the two light  yellow regions
are the $5\sigma$  reach by direct searches at LHC14 
with 300~fb$^{-1}$, 3000~fb$^{-1}$) and HE-LHC with 3000~fb$^{-1}$ respectively~\cite{2HDMwp}.
The  region expected to be allowed at a 
95\% CL by complementary precision Higgs coupling measurements, is shown 
as dark (light) blue for 300~${\rm fb}^{-1}$~(3000~${\rm fb}^{-1}$)~\cite{Barger:2013ofa}.
}
\label{fig:HZZreach}
\end{figure}

\begin{figure}
\centering
\includegraphics[width=0.4\linewidth]{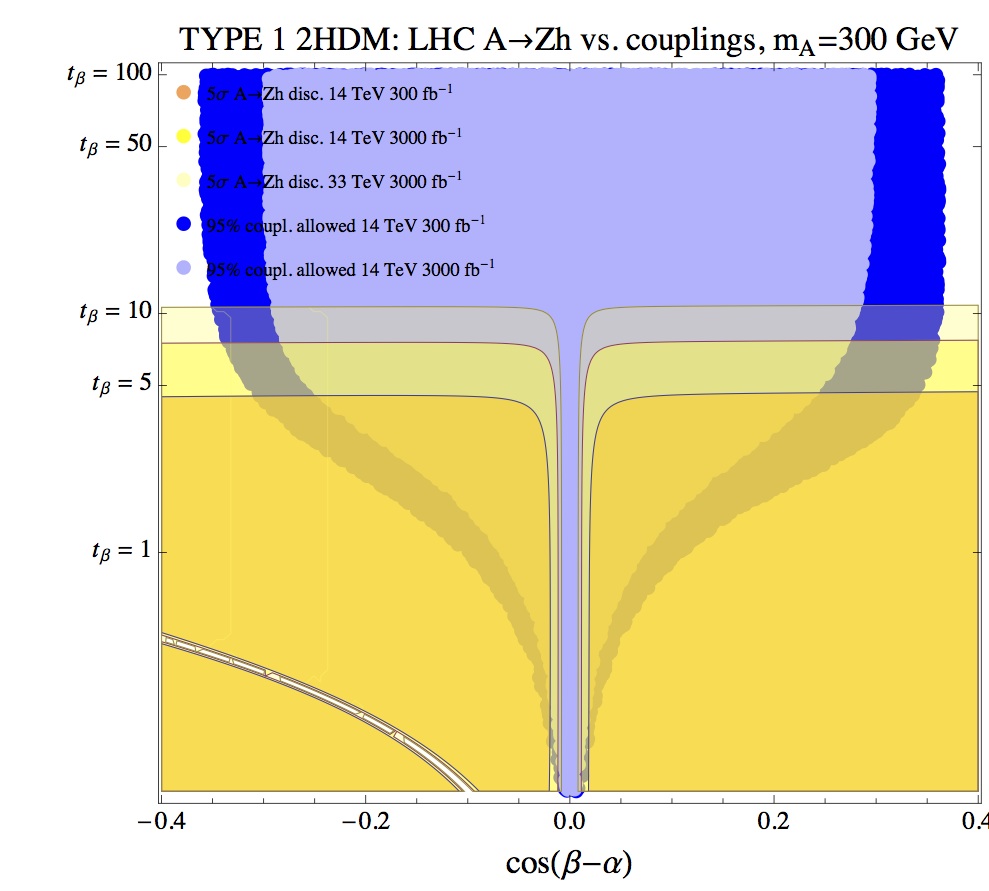}
\includegraphics[width=0.4\linewidth]{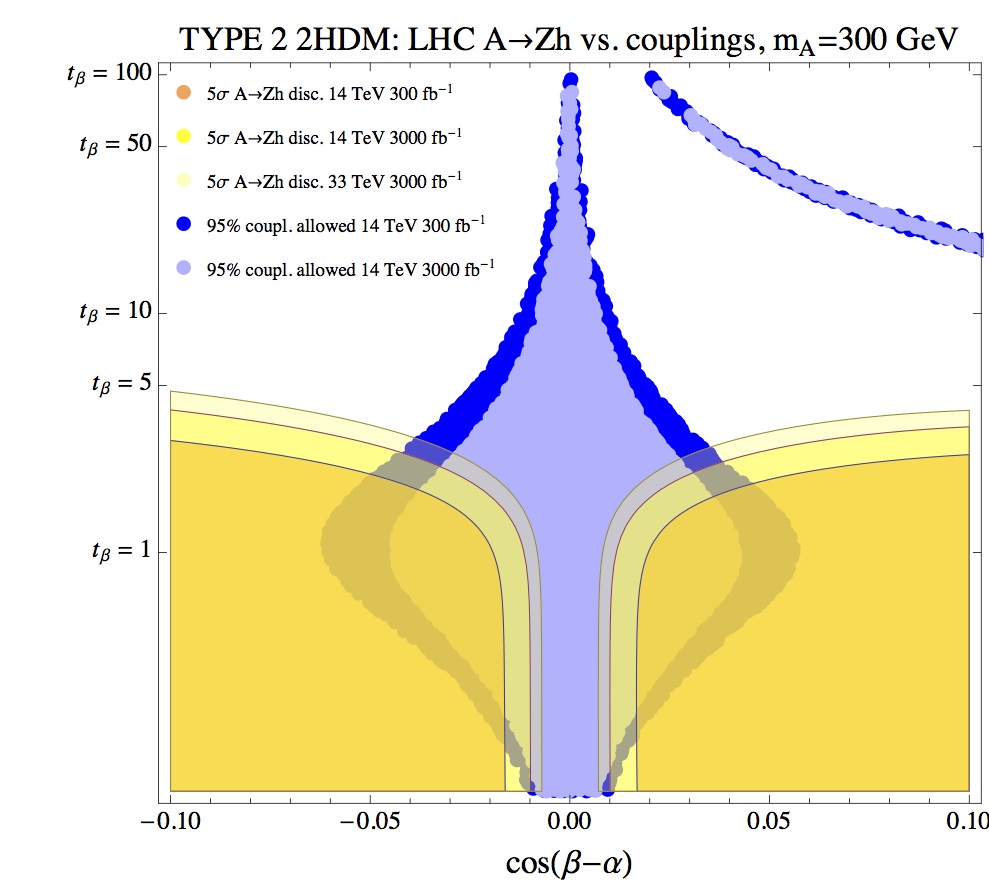}
\caption{$5\sigma$ discovery reach for 300~GeV 
 $A \rightarrow Z h \rightarrow (\ell\ell)(bb)\mbox{ or } (\tau\tau$) 
The Type I and Type II 2HDM models as shown in the left  
and right panels respectively. 
 Dark and the two light  yellow regions
are the $5\sigma$  reach by direct searches at LHC14 
with 300~fb$^{-1}$, 3000~fb$^{-1}$) and HE-LHC with 3000~fb$^{-1}$ respectively~\cite{2HDMwp}.
The  region expected to be allowed at a 
95\% CL by complementary precision Higgs coupling measurements, is shown 
as dark (light) blue for 300~${\rm fb}^{-1}$~(3000~${\rm fb}^{-1}$)~\cite{Barger:2013ofa}.
}
\label{fig:AZhreach}
\end{figure}

At hadron colliders, two senstive searches are
$H \to ZZ \to 4\ell$ and $A \to Zh$ followed by $Z \to \ell\ell$
and $h \to bb$ or $\tau\tau$~\cite{}. 
These studies have 
been carried out using the Snowmass LHC detector 
simulation~\cite{SnowmassDetSim} and backgrounds~\cite{SnowmassBackgrounds,SnowmassOSG}. The reach for the first signal in the  
$\tan\beta$ vs. $\cos(\beta-\alpha)$ plane of 
Type I and Type II 2HDM  is given for the 14 TeV LHC in
Fig.~\ref{fig:HZZreach}, and the second in 
Fig.~\ref{fig:AZhreach}~\cite{2HDMwp}. 
For HE-LHC at $\sqrt{s}=33$ TeV and 3000${\rm fb}^{-1}$, the reach is 
extended considerably~\cite{2HDMwp}. 
In both the figures, the  expected 95\% CL allowed region 
for LHC14 with 300${\rm fb}^{-1}$ and 3000~${\rm fb}^{-1}$, determined from the 
complemetary approach of precision Higgs coupling measurements is also shown.
There is a considerable region near the alignment limit which can 
only be excluded through direct search.

At $e^+ e^-$ colliders,
for a wide range of models
$H^+ H^-$ and $H A$ can be observed as long as the sum of the masses
is below the center of mass energy of the collider.
Production of heavy Higgs bosons in association with a $Z$ is suppressed
in the decoupling limit.
Furthermore, the masses of the observed states
can be measured to better than $1\%$ \cite{clicwp}.
Consequently, 3~TeV CLIC is ideally suited for the study of Higgs bosons
with masses up to $1.5$~TeV.


\paragraph{Extended Higgs sector discovery story:}
We now discuss one discovery story in detail.
We consider a signal at LHC14/300 for $A \to Zh \to (\ell\ell)(bb)$.
There is significant reach for discovery in this channel, as can be seen in
Fig.~\ref{fig:AZhreach}.
For this scenario, we assume a type-II 2HDM with $m_A = 300$~GeV,
and $\alpha = -0.475$, $\tan \beta = 2$.

By the end of the 14 TeV run with 300 fb$^{-1}$ of integrated luminosity, an excess is 
observed in searches for anomalous $Zh$ production in the $\ell \ell bb$ final state consistent with a production cross section times branching ratio of $\sim 10$~fb. 
The full $m_{\ell \ell b b}$ invariant mass distribution peaks around 300 GeV. 
The lepton pair production is consistent with the leptonic decay of a $Z$ boson, 
while the invariant mass of the bottom quark pair is consistent with the decays of the observed SM-like Higgs boson at 125 GeV. 
The signal significance is $\sim 2.5 \sigma$. 
There is also a mild $\sim 1 \sigma$ excess in the $\ell\ell\tau^+ \tau^-$ channel 
where the lepton pairs are again consistent with leptonic decay of a $Z$ boson, 
but without sufficient mass resolution to conclusively relate to the excess in the 
$\ell \ell bb$ final state. 
The final states and approximate mass reconstruction in the $\ell \ell bb$  
final state  are consistent with the production of a pseudoscalar Higgs boson with 
decay to $Zh$. 

At the same time, there are no meaningful excesses in searches for resonant $WW$ and 
$ZZ$ di-boson production in this mass range, nor are there any meaningful signals 
in the ongoing searches for additional MSSM Higgs bosons in the $bb$ and $\tau \tau$ 
final states at large $\tan \beta$.  
In the context of a two-Higgs-doublet model, the natural interpretation is the 
CP-odd pseudoscalar $A$ at low $\tan \beta$, where the branching ratio for $A \to Zh$ 
may be appreciable but the rates for gluon fusion or $bbA$ associated production with 
$A \to bb, \tau \tau$ are not large enough to be distinguished from background.

Motivated by these excesses, a search conducted in the 300 fb$^{-1}$ data set for 
$\ell \ell + \gamma \gamma$ consistent with anomalous $Zh$ production yields 
$\sim 3$ events whose $m_{\ell \ell \gamma \gamma}$ accumulate at 300~GeV, 
further suggesting the presence of a new state decaying to $Zh$ but not 
substantially increasing the significance of the excess. 
Given that these signals in the $Zh$ final state are consistent with a pseudoscalar 
Higgs at low $\tan \beta$, both collaborations consequently extend their inclusive 
diphoton resonance searches to close the gap in coverage between the endpoint of the 
SM-like Higgs search at 150 GeV and the beginning of the KK resonance search at 500 GeV. 
Upon unblinding the analysis, they detect a signal consistent with the production and 
decay of a 300~GeV particle decaying to pairs of photons with 
$\sigma \cdot {\rm Br} \sim 7$ fb at 14 TeV. 
The lack of events in dijet-tagged categories indicate that there is no meaningful 
associated production, bolstering the case for a new pseudoscalar.

Attempts to interpret the resonance in terms of the MSSM are stymied by the resolution 
of Higgs coupling measurements in the 300 fb$^{-1}$ data set. 
For a pseudoscalar in the MSSM at low $\tan \beta$ with $m_A = 300$~GeV, the generic 
expected deviation in the $hbb$ coupling is of order $\sim 5-20\%$, with much smaller 
deviations in $h \gamma \gamma$, $htt$, and $hVV$. 
The precision of Higgs coupling measurements at 300 fb$^{-1}$ only serve to bound 
$\tan \beta < 4$ in the MSSM. Both the high-luminosity run of the LHC and the 
ILC become high priorities for establishing the discovery of the new state and 
triangulating measurements of Higgs couplings at both 125 GeV and 300~GeV.
 
At the high luminosity run of the LHC, the signal reaches $5 \sigma$ significance in the $bb \ell \ell$ final state by 1000 fb$^{-1}$ of integrated luminosity, sufficient to announce the discovery of a new state. The excess in the $\tau^+ \tau^- \ell \ell$ channel grows to several $\sigma$, consistent with a production cross section times branching ratio of $\sim 1$ fb, while the excess in the diphoton final state at 300~GeV also reaches $5 \sigma$ significance by the end of the full 3000 fb$^{-1}$. However, the experimental and theoretical errors on the discovery-level channels are sufficiently large that interpretation based on direct coupling measurements of the new state remains challenging. Interpreted in the context of a Type II 2HDM, the best fit to the production and decay rates favors $\tan \beta \simeq 2, \cos (\beta - \alpha) \simeq -0.012$. This is in mild tension with the MSSM interpretation, for which the tree-level prediction at $\tan \beta \simeq 2$ is closer to $\cos (\beta - \alpha) \simeq -0.04$, but without sufficient experimental resolution to provide meaningful discrimination. Although the errors on the Higgs coupling measurements at 125 GeV improve to $\Delta g_{hbb} \sim 10\%$, still no significant deviation from Standard Model predictions is observed. Finally, after the full 3000 fb$^{-1}$ are analyzed, a collection of 10 $4\ell$ events consistent with $H \to ZZ \to 4 \ell$ are reconstructed around 450 GeV -- sufficient to hint at the presence of an additional CP-even scalar but insufficient to establish discovery. Searches for a resonance in $t \bar t$ prove inconclusive.

A lepton collider such as the ILC or TLEP can explore the electroweak symmetry breaking
sector in detail.
For example,
at the $\sqrt{s} = 250$ GeV run of an ILC, the coupling measurements of the 125 GeV Higgs improve to $\Delta g_{hbb} \sim 5\%$ without observing deviations from SM predictions, increasing tension with the MSSM interpretation. No direct production of the new state is kinematically possible. However, at $\sqrt{s} = 500$ GeV, the pseudoscalar is expected to be kinematically available in both $b \bar b A$ and $Ah$ associated production. 
Indeed, after 500 fb$^{-1}$ are on tape, $b \bar b A$ associated production is observed 
at the level of a small handful of events consistent with $\sigma(b \bar b A) \sim 0.1$ fb. 
Two $Zhh$ events are observed consistent with $e^+ e^- \to Ah \to Zhh,$ but are difficult to distinguish from the SM di-Higgs background given the low statistics. 
By $\sqrt{s} = 1$ TeV the $bbA$ signal increases consistent with a cross section of 
$\sim 1$ fb, leaving several hundred $b \bar b A$ events on tape and substantially 
improving the direct coupling measurements of the pseudoscalar. Most importantly, the ILC operating at $\sqrt{s} = 1$ TeV discovers additional states in the Higgs sector. The first is a 370 GeV charged Higgs with cross section of order $\sim 10$ fb. The primary discovery mode is $H^+ H^-$ Drell-Yan pair production in the $t \bar t b \bar b$ final state. The mass splitting between the charged Higgs and the pseudoscalar are again in tension with tree-level MSSM predictions for the mass spectrum. Reaching closer to the kinematic threshold, the ILC discovers the hinted-at CP-even scalar at 450 GeV through $HA$ associated production in the final state $t \bar t Z h$ with a cross section of several femtobarn. 

In addition, the improvement of Higgs coupling measurements at 
$\sqrt{s} = 500$~GeV indicates 
a small persistent tension with SM predictions at the level of $\Delta g_{hbb} \sim 2\%$.
While the departure is not statistically significant, the smallness of $\Delta g_{hbb}$ 
is in tension with conventional MSSM predictions. 
In the context of a Type II 2HDM, the combination of measurements of the light SM-like Higgs at 125 GeV, the pseudoscalar at 300~GeV, the charged Higgs at 370 GeV, and the second CP-even scalar at 450 GeV imply an extended Higgs sector that is closer 
to the alignment limit than implied by supersymmetric decoupling alone, with the best-fit point in a Type II 2HDM ultimately corresponding to $\tan \beta = 2, \cos (\beta - \alpha) = -0.0122$, as well as with mass relations between scalars that are discrepant from the standard MSSM predictions. 
This ignites fervent exploration of non-standard corners of MSSM Higgs parameter 
space,  as well as  other natural theories of extended electroweak symmetry breaking. 
On the experimental side, a high-luminosity $e^+ e^-$ collider could reduce
the uncertainty on $\Delta g_{hbb}$ to less than a percent, turning the 
tension into a discovery of new physics.

A high-energy proton collider can also continue exploration of the extended Higgs sector by producing a large sample of heavier Higgs scalars. In this example, although the heavy CP-even scalar $H$ primarily decays to $t \bar t$ pairs, it also exhibits rarer decay modes such as $H \to ZA$ and $H \to H^\pm W^\mp$ that are kinematically squeezed but nonetheless observable provided the large number of $H$ bosons produced at a high-energy $pp$ machine. More generally, a high-energy proton collider has the potential to discover additional Higgs bosons that lie well beyond the reach of the LHC and ILC.


\subsection{WIMP Dark Matter }
\label{sec:dmatter}

Though the presence of dark matter in the universe has been
well-established, little is known of its particle nature or its
non-gravitational interactions.  A vibrant experimental program
is searching for a weakly interacting massive particle (WIMP), denoted as
$\chi$, and interactions with standard model particles via some
as-yet-unknown mediator.  

WIMPs appear in many theories of physics beyond the standard
model ({\it e.g.} SUSY),  or other theories which posit a rich dark
sector complete with dynamical self-interactions and striking features
at colliders~\cite{Dienes:2012yz}. For other examples, see
Refs.~\cite{Chang:2010en,Agrawal:2011ze,Fortin:2011hv,Hooper:2007qk,Beltran:2010ww,Ellis:2006vu}.

However, this section focusses on a phenomenological
approach, searching directly for  WIMPs rather than on other
states which may appear in the theory. Specifically, this section describes the sensitivity of  searches for pair-production of WIMPs at particle colliders, 
$pp\rightarrow \chi\bar{\chi}$ at the LHC or $e^+e^-\rightarrow
\chi\bar{\chi}$ at a lepton collider  via some  unknown mediator.

\begin{figure}
\centering
\includegraphics[width=0.5\linewidth]{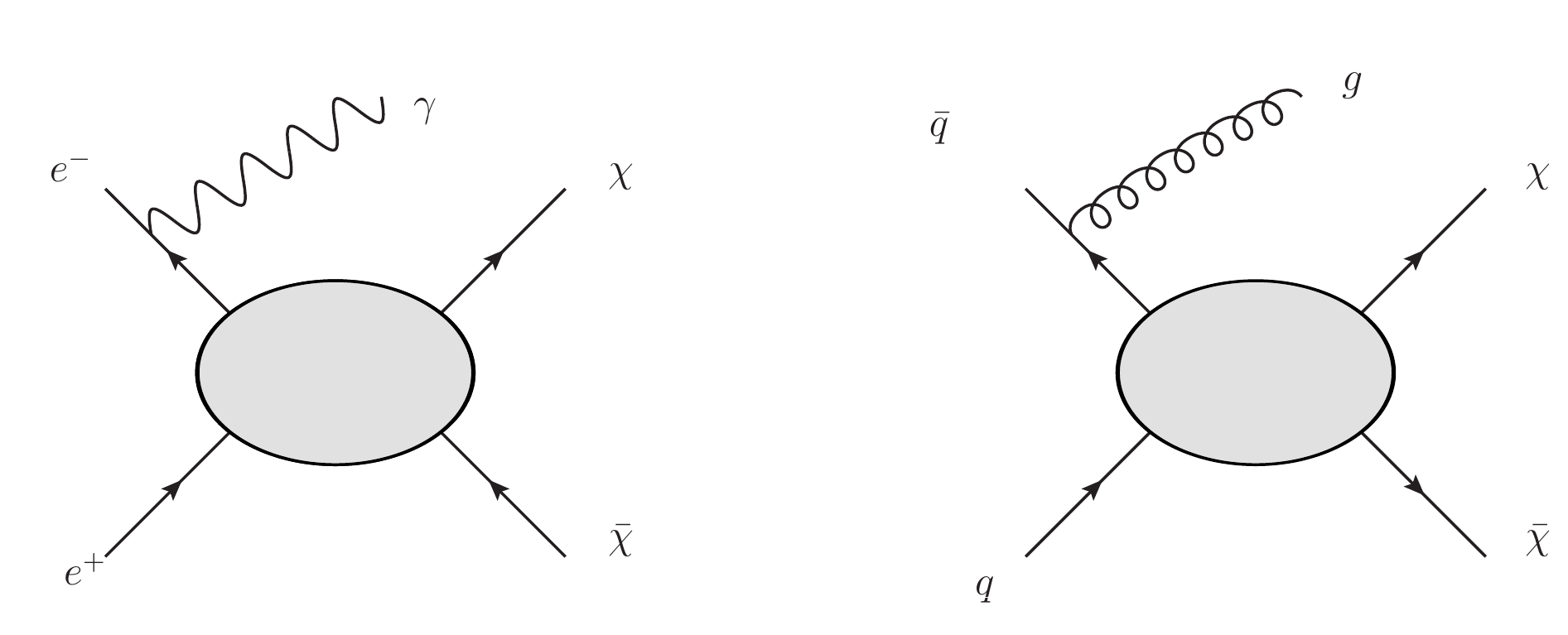}
\caption{ Pair production of WIMPs ($\chi\bar{\chi}$) in $e^+e^-$
  collisions (left), or $pp$ collisions (right), both via an unknown
  intermediate state, with initial-state radiation of a standard model particle.}
\label{fig:dmdiag}
\end{figure}

 If the mediator is too heavy to be
resolved, the interaction can be modeled as an effective field theory
with a four-point interaction, otherwise an explicit  model is needed
for the heavy mediator. As the final state WIMPs are invisible to the
detectors, the events can only be seen if there is associated
initial-state radiation of a standard model
particle~\cite{Beltran:2010ww,Fox:2011pm,Goodman:2010ku}, see Fig~\ref{fig:dmdiag}, recoiling against the dark matter pair.

In this section, we describe the sensitivity of future $pp$ and
$e^+e^-$ colliders in various configurations to  WIMP pair production
using the mono-jet final state (in the $pp$ case) or mono-photon final
state (in the $e^+e^-$
case). We consider both effective operators and one example of a real, heavy
$Z'$-boson mediator.

\subsubsection{Searches at $pp$ colliders}

The LHC collaborations have reported limits on the cross section of
$pp\rightarrow  \chi\bar{\chi}+X$ where $X$ is a
hadronic jet~\cite{atlasjet,cmsjet}, photon~\cite{atlasphoton,cmsphoton}, and
other searches have been repurposed to study the cases where $X$ is a
$W$~\cite{monow} or $Z$ boson~\cite{atlaszz,monoz}. In each case,
limits are reported in terms of the mass scale
$M_\star$ of the unknown interaction expressed in an effective field
theory~\cite{Beltran:2008xg,Beltran:2010ww, Fox:2011pm,Goodman:2010ku,
  Shepherd:2009sa,Cao:2009uw,Goodman:2010yf,Bai:2010hh,Rajaraman:2011wf}
though the limits from the mono-jet mode are the most powerful~\cite{dmcombo}.

In Ref.~\cite{dmextrap}, the sensitivity of possible future
proton-proton colliders is studied in various configurations (see
Table~\ref{tab:facdm}) to WIMP pair production using the mono-jet final
state. Both effective operators and one example of a real, heavy
$Z'$-boson mediator are considered.

The analysis of jet$+\missET$ events uses a sample of events with one
or two high $p_\mathrm{T}$ jets and large $\missET$, with angular cuts
to suppress events with two back-to-back jets (multi-jet
background). The dominant remaining background is $Z\rightarrow
\nu\bar{\nu}$ in association with jets, which is indistinguishable
from the signal process of $\chi\bar{\chi}$+jets. The estimation of the background at large $\missET$ is problematic in
simulated samples, due to the difficulties of accurately modeling the
many sources of $\missET$.  The experimental results, therefore, rely
on data-driven background estimates, typically extrapolating the
$Z\rightarrow \nu\bar{\nu}$ contribution from $Z\rightarrow\mu\mu$
events with large $Z$ boson $p_\mathrm{T}$.  These results use estimates are anchored in experimentally reported
values~\cite{atlasjet,cmsjet} of the background estimates and signal
efficiencies (at $\sqrt{s}=7$ TeV, $\mathcal{L}=5$ fb$^{-1}$, $\missET
> 350$ GeV), and use simulated samples to extrapolate to higher
center-of-mass energies, where no data is currently available.  At the
higher collision energies and instantaneous luminosities of the
proposed facilities, the rate of multi-jet production will also be
higher, requiring higher $\missET$ thresholds to cope with the
background levels and the trigger rates, see Ref.~\cite{dmextrap} for details.

\begin{table}
\centering
\caption{Details of current and potential future $pp$ colliders,
  including center-of-mass energy ($\sqrt{s}$),  total integrated
  luminosity ($\mathcal{L}$), the threshold in $\missET$, and the
  estimated signal and background yields. From Ref.~\cite{dmextrap}.}
\label{tab:facdm}
\begin{tabular}{lrrrrrr}
\hline\hline
$\sqrt{s}$ [TeV] & $\missET$ [GeV] & $\mathcal{L}$ [fb$^{-1}$] & $N_{D5}$ & $N_{\textrm{bg}}$ \\
\hline
7 & 350 & 4.9 & 73.3 & $1970\pm 160$\\
14 & 550 & 300 & 2500 & $2200\pm 180$\\
14 & 1100 & 3000 & 3200 & $1760\pm 143$\\
33 & 2750 & 3000 & 8.2$\cdot10^4$ & $1870\pm 150$\\
100 & 5500 & 3000 & 3.4$\cdot10^6$& $2310\pm 190$\\
\hline\hline
\end{tabular}
\end{table}

\begin{figure}[h]
\centering
\includegraphics[width=0.24\linewidth]{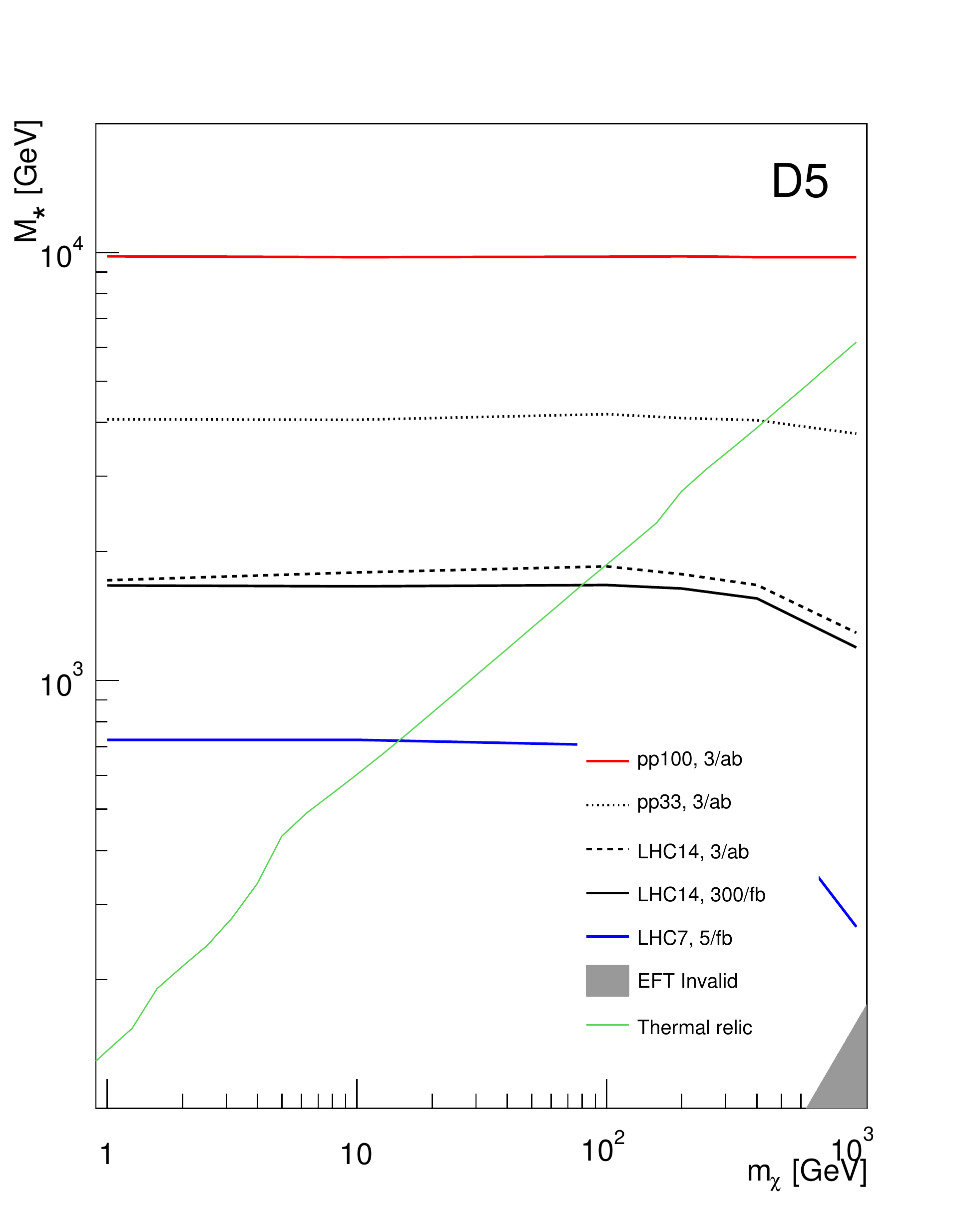}
\includegraphics[width=0.24\linewidth]{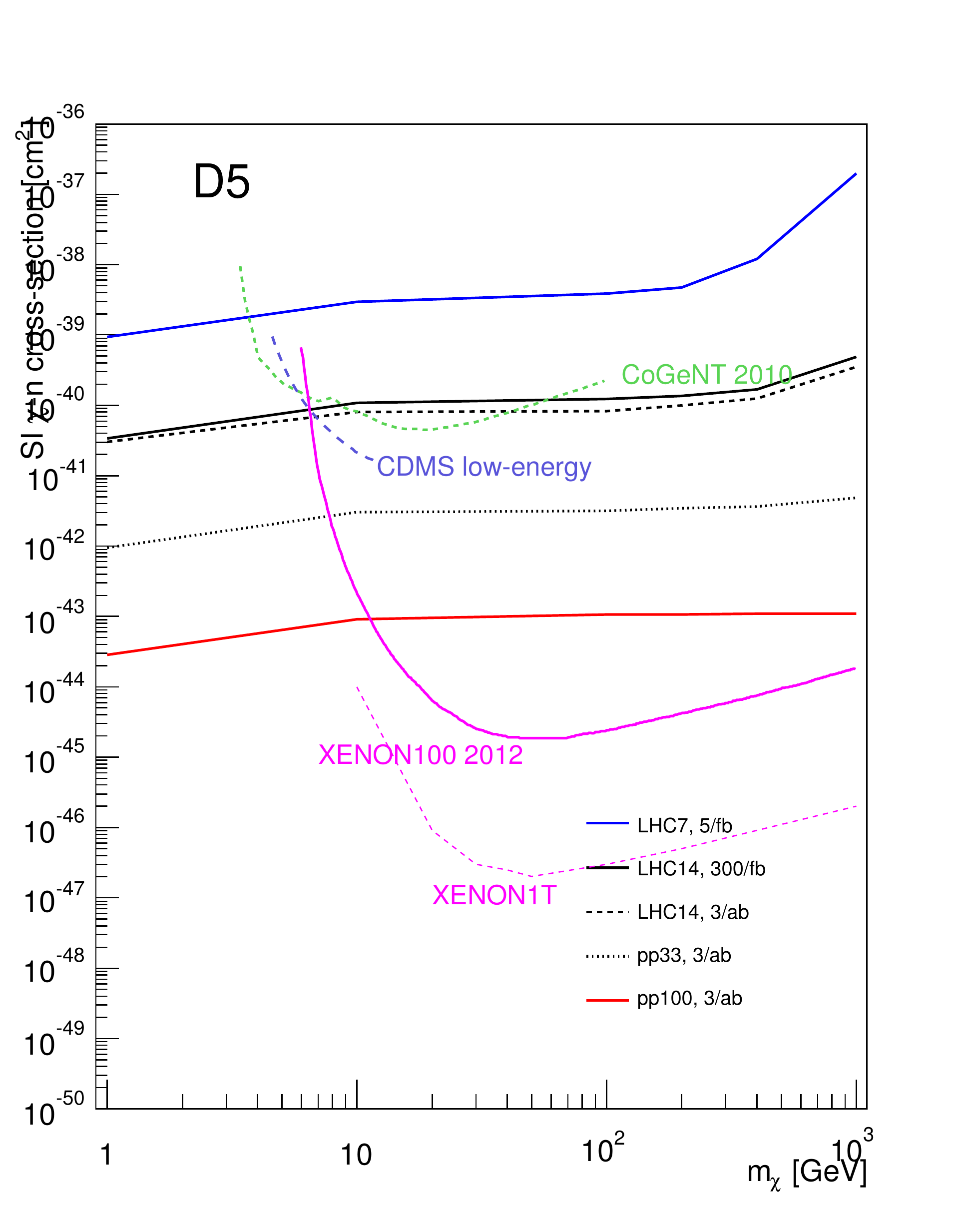}
\includegraphics[width=0.24\linewidth]{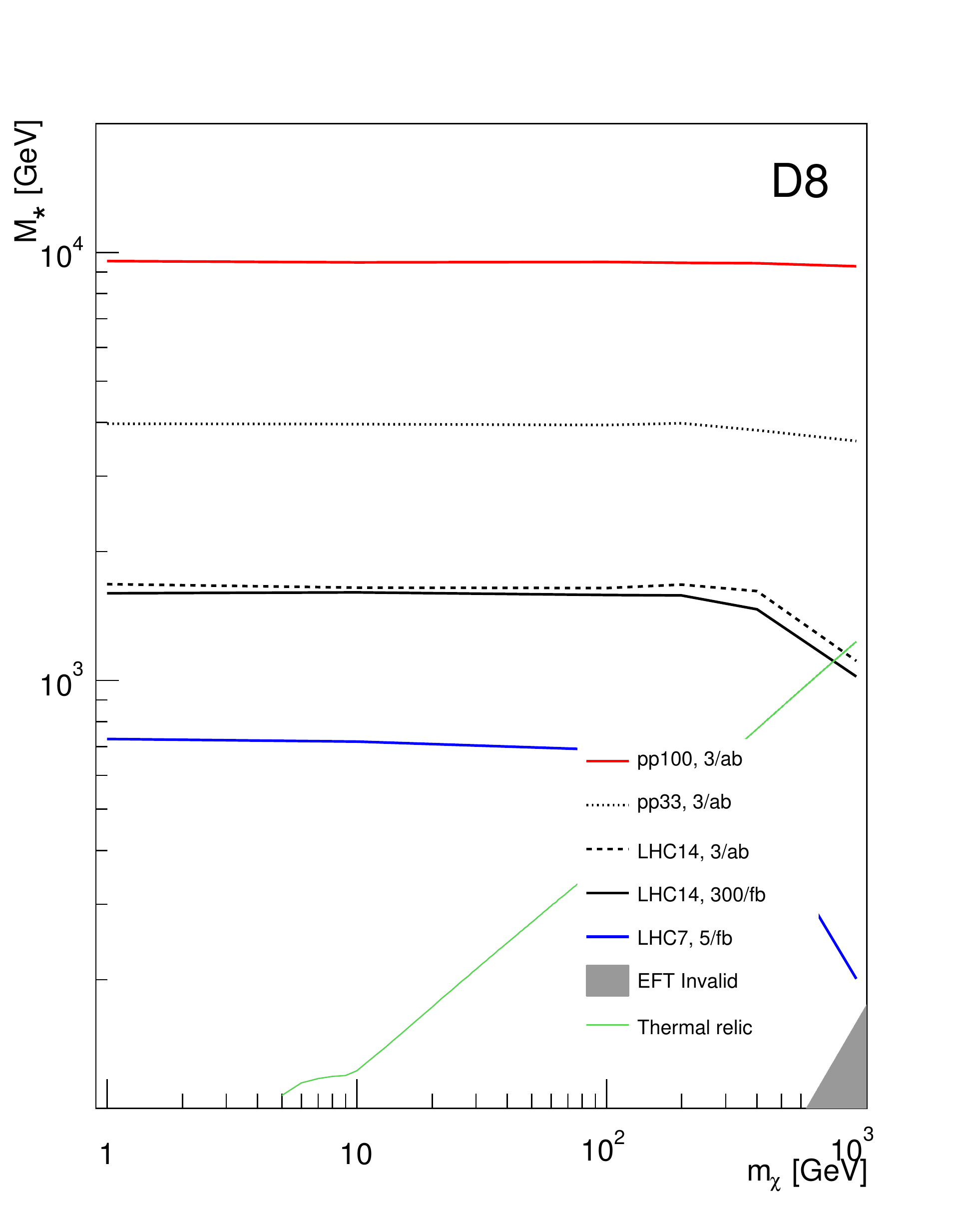}
\includegraphics[width=0.24\linewidth]{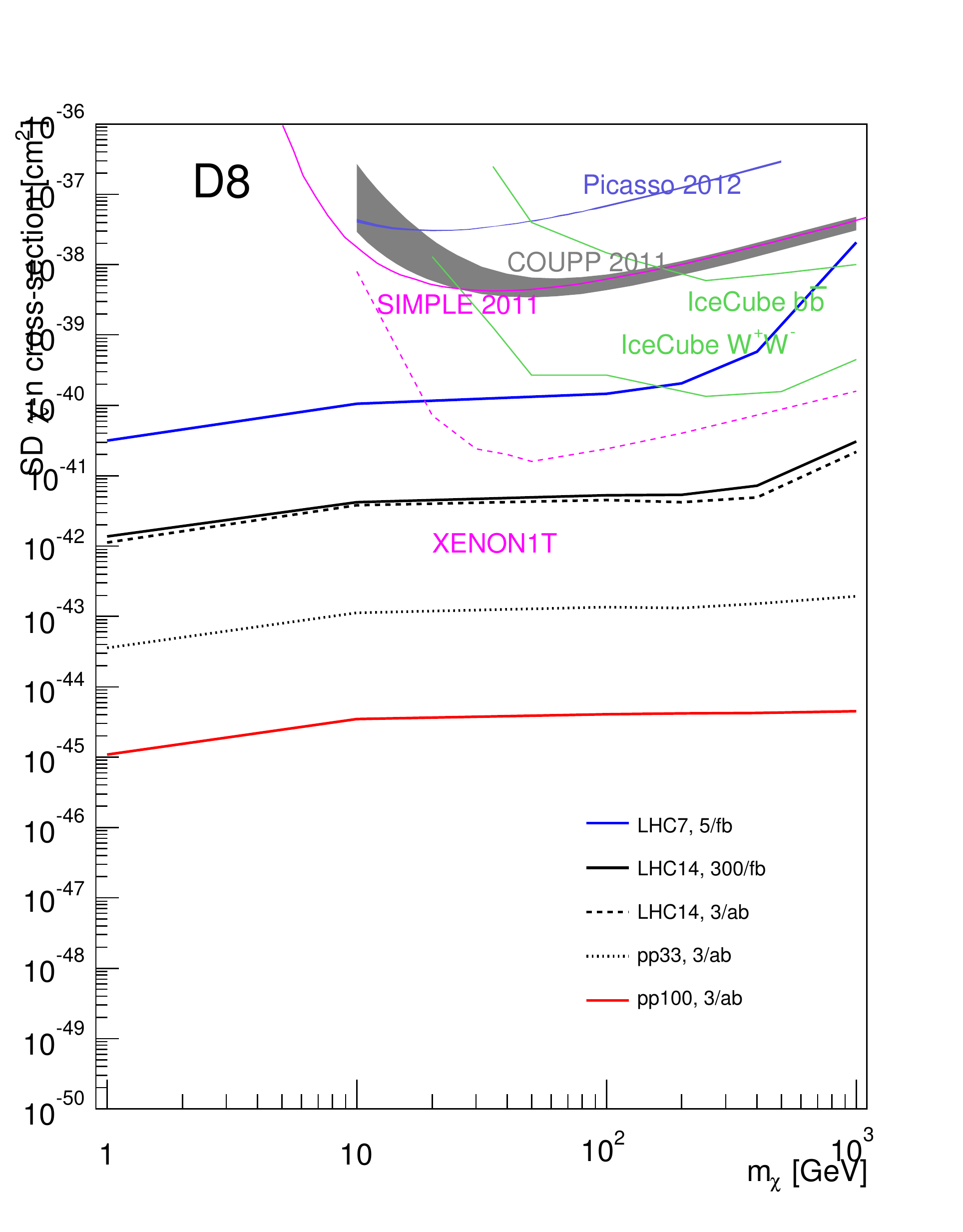}
\caption{Limits at 90\% CL in $M_\star$ (left) and in the spin-independent
  WIMP-nucleon cross section (right) for different facilities using
  the D5 or D8 operator as a function of $m_\chi$.  From Ref.~\cite{dmextrap}.}
\label{fig:d5d8}
\end{figure}

Given the expected background and uncertainties, limits can be
calculated on contributions from new sources, which can be translated
into  limits on
$M_*$, see Fig.~\ref{fig:d5d8}.  These are then translated in limits on the
$\chi$-nucleon cross section.

\begin{figure}[h]
\centering
\includegraphics[width=0.4\linewidth]{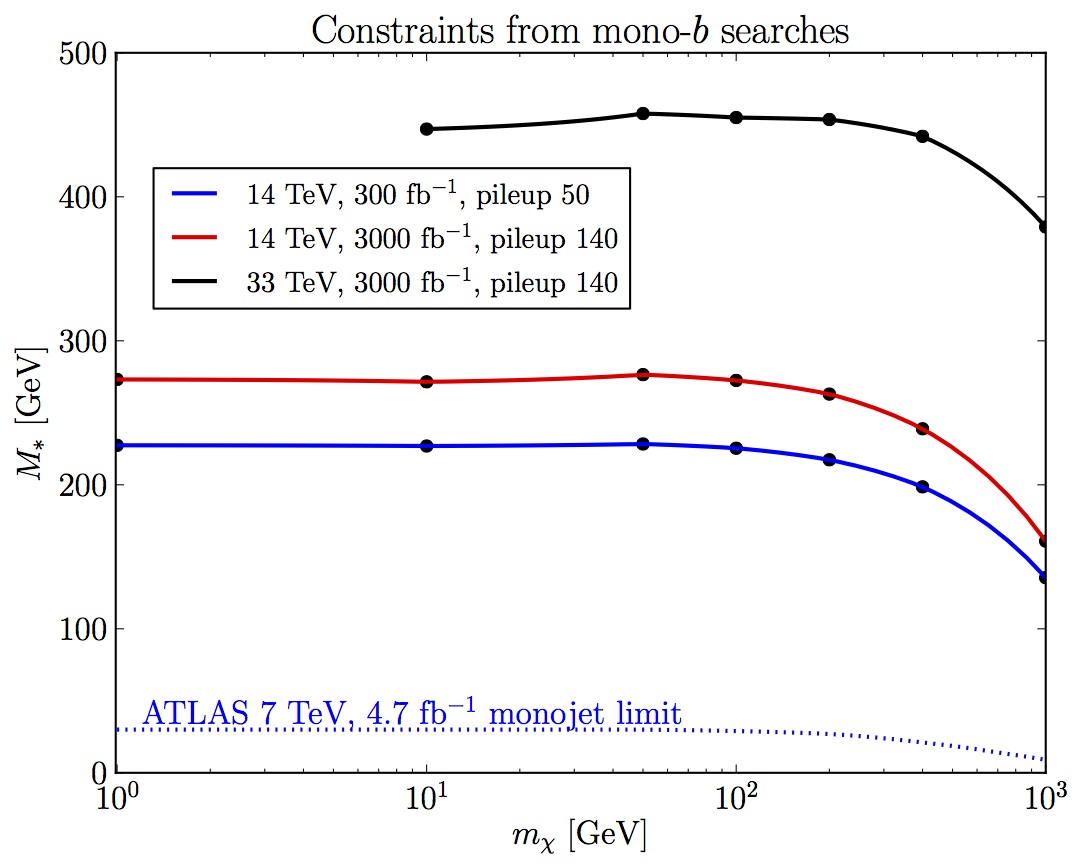}
\includegraphics[width=0.4\linewidth]{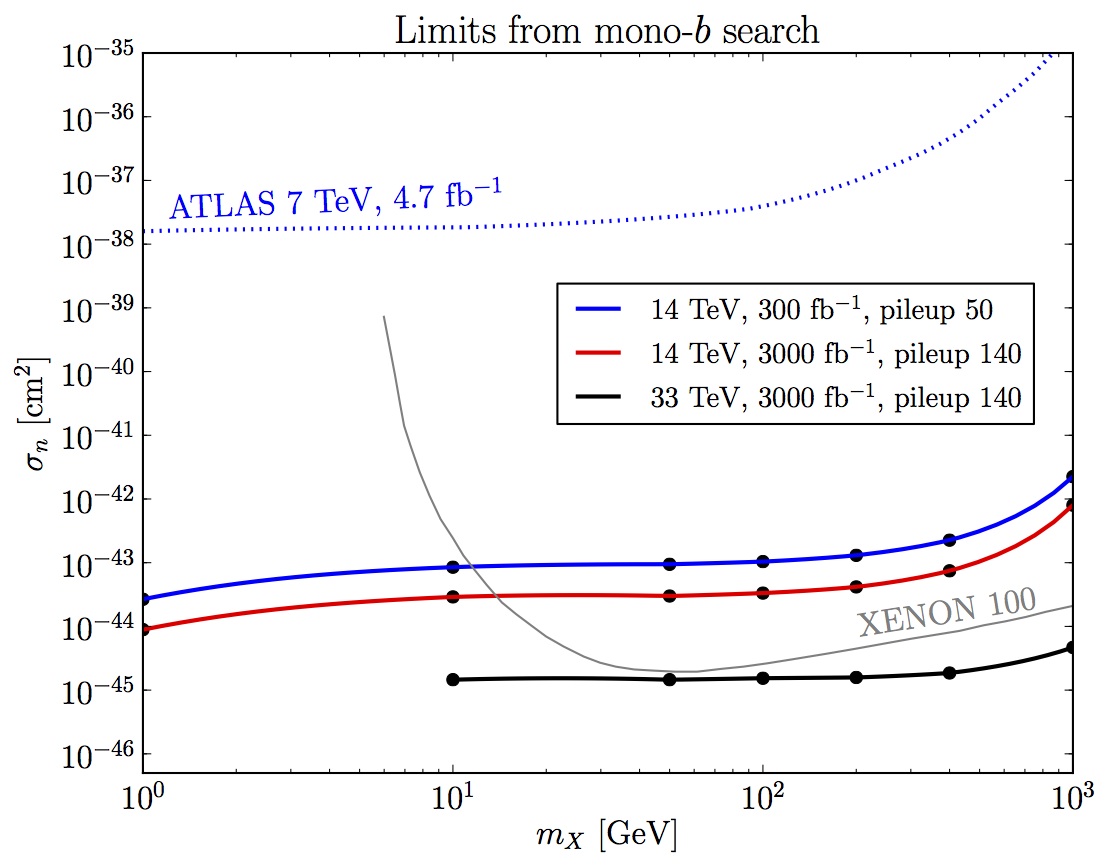}
\caption{Limits at 90\% CL in $M_\star$ (left) and in the spin-independent
  WIMP-nucleon cross section (right) for different facilities when
  requiring a $b$-quark in the final state, as a function of $m_\chi$.  From Ref.~\cite{Artoni:2013zba}.}
\label{fig:monob}
\end{figure}

Despite the kinematic and PDF suppression for producing third
generation quarks, it was shown~\cite{Lin:2013sca}, that for the scalar operator,
searches for $\chi\bar{\chi}+b$ and $\chi\bar{\chi}+t\bar{t}$ final
states offer enhanced sensitivity due to the large suppression of
background and therefore lower missing energy thresholds.

\begin{figure}[ht]
\centering
\includegraphics[width=0.32\linewidth]{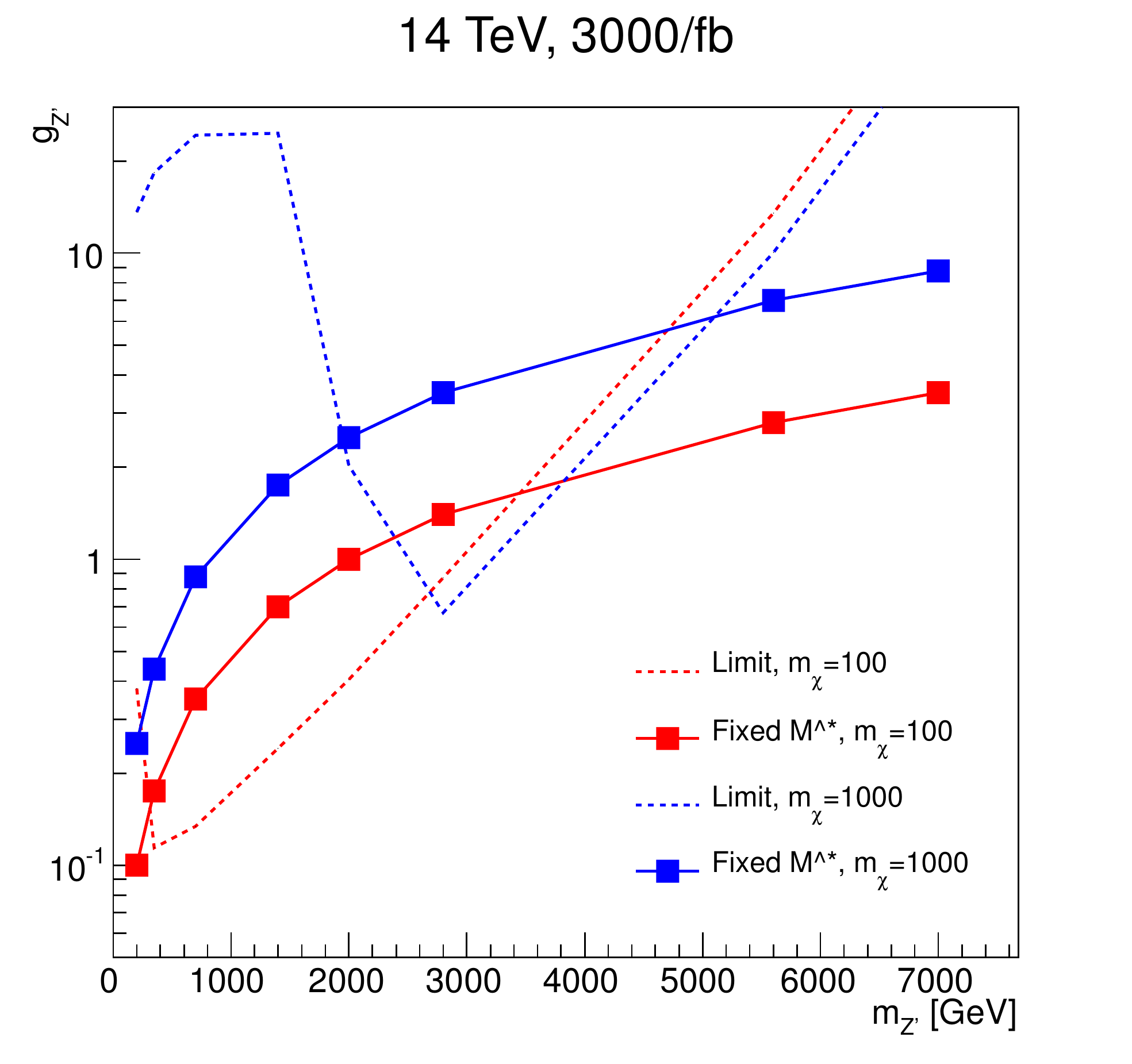}
\includegraphics[width=0.32\linewidth]{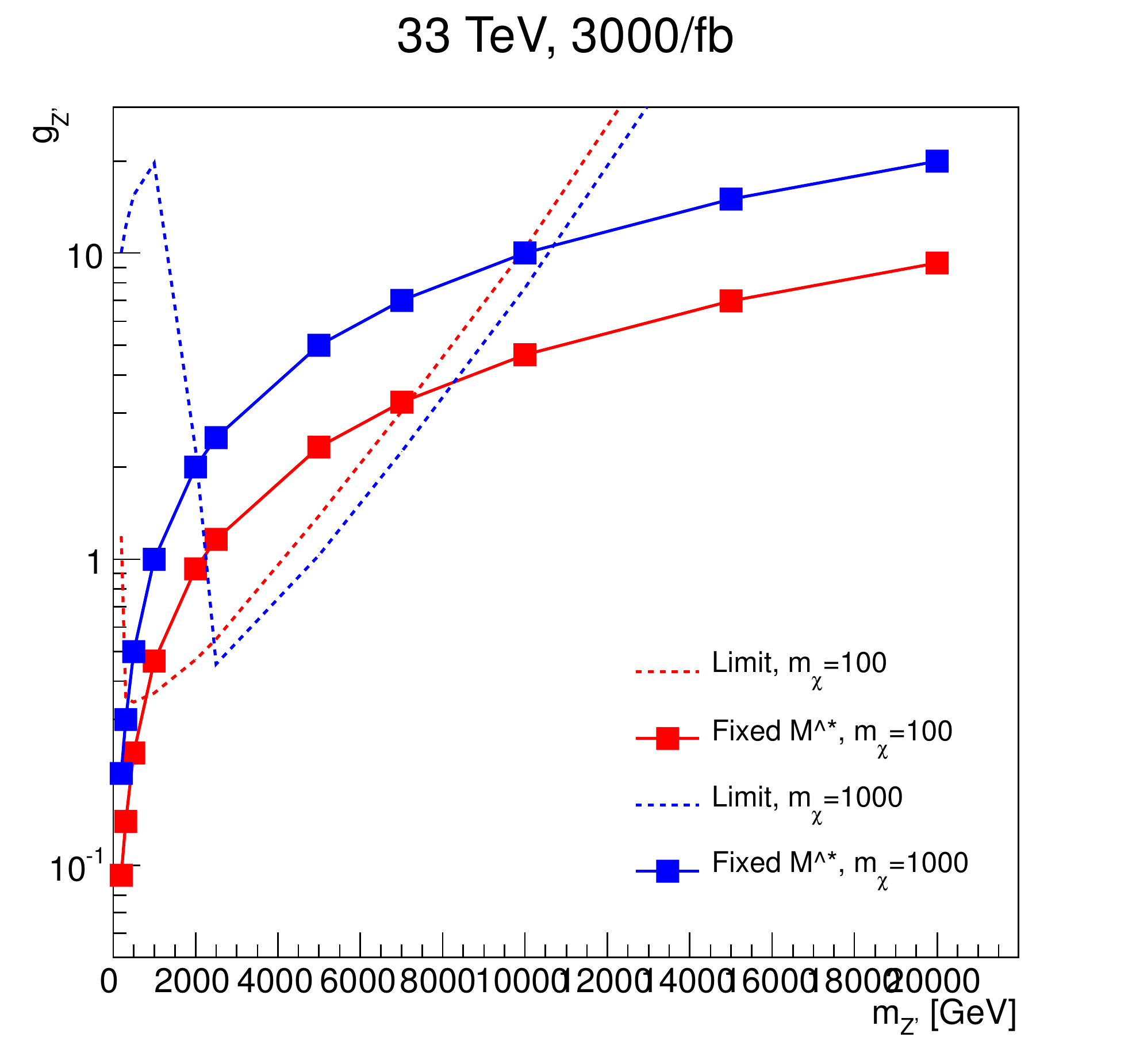}
\includegraphics[width=0.32\linewidth]{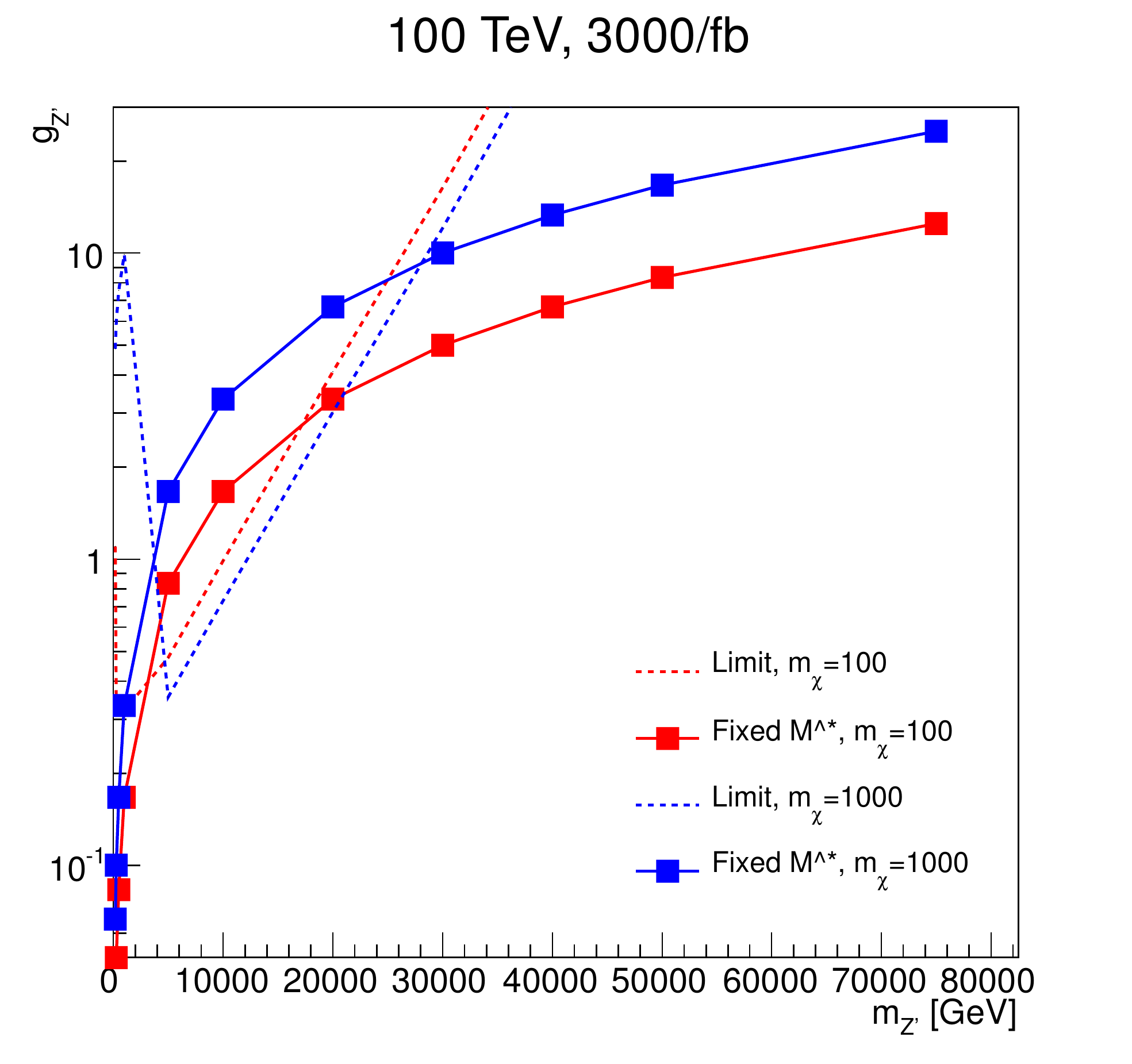}
\caption{ Sensitivity of various $pp$ facilities to a dark matter pairs produced through a real
  $Z'$ mediator. In each case, expected limits on the coupling
  $g_{Z'}$ versus $Z'$ mass for two choices of $m_\chi$ as well as  the values of $g_{Z'}$ which
  satisfy $g'/m_{Z'}=1/M_*$, where $M_*$ are limits from a
  lower-energy facility. From Ref.~\cite{dmextrap}}
\label{fig:dm_zp}
\end{figure}

The EFT approach is useful when the current facility does not have the
necessary center-of-mass energy to produce on-shell mediators.  The
next-generation facility, however, may have such power.    The sensitivity of the proposed facilities
to a model in which the heavy mediator is a $Z'$ which couples to
$\chi\bar{\chi}$ as well as $q\bar{q}$~\cite{zprime,Petriello:2008pu,Gershtein:2008bf} is discussed. The coupling of the $Z'$ is a free parameter in this theory, but
particularly interesting values are those which correspond to the
limit of previous facilities on $M_*$. That is, an EFT model of the
$Z'$ interaction has $\frac{1}{M_*} = \frac{g_{Z'}}{M_{Z'}}$ fixing the relationship between $g_{Z'}$ and
$M_{Z'}$. Figure~\ref{fig:dm_zp} shows the expected limits in terms of
$g_{Z'}$ on the
$Z'$ model at the variety of $pp$ facilities under consideration. The $g'$ expected limits can be compared to the curve with
$g_{Z'} = \frac{M_{Z'}}{M_*}$.

\subsubsection{Searches at lepton colliders}

The same mechanism which allows $pp$ colliders to be sensitivie to the
coupling of the initial-state quarks to WIMP pairs allows $e^+e^-$
colliders to proble the couplings of electrons to WIMP pairs, see
Fig~\ref{fig:dmdiag}.  The couplings of WIMPs to leptons could be
mediated by different operators with different suppression scales than
the WIMP-quark (gluon) couplings. Therefore $e^+e^-$ colliders will
add important complementary information to the WIMP picture~\cite{Bartels:2012ex,Chae:2012bq, Dreiner:2012xm}.

\begin{figure}[ht]
\centering
\includegraphics[width=0.45\linewidth]{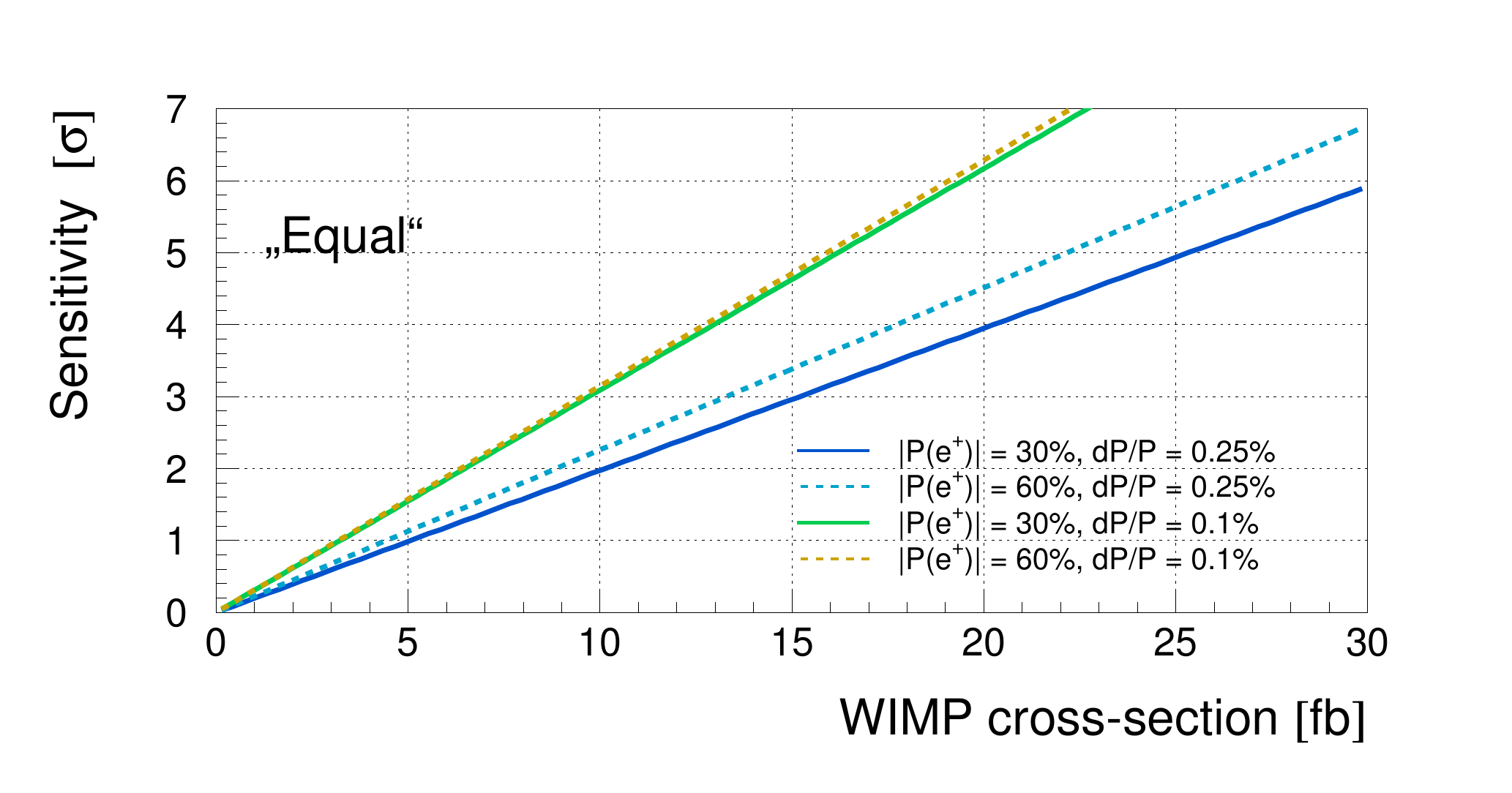}
\includegraphics[width=0.45\linewidth]{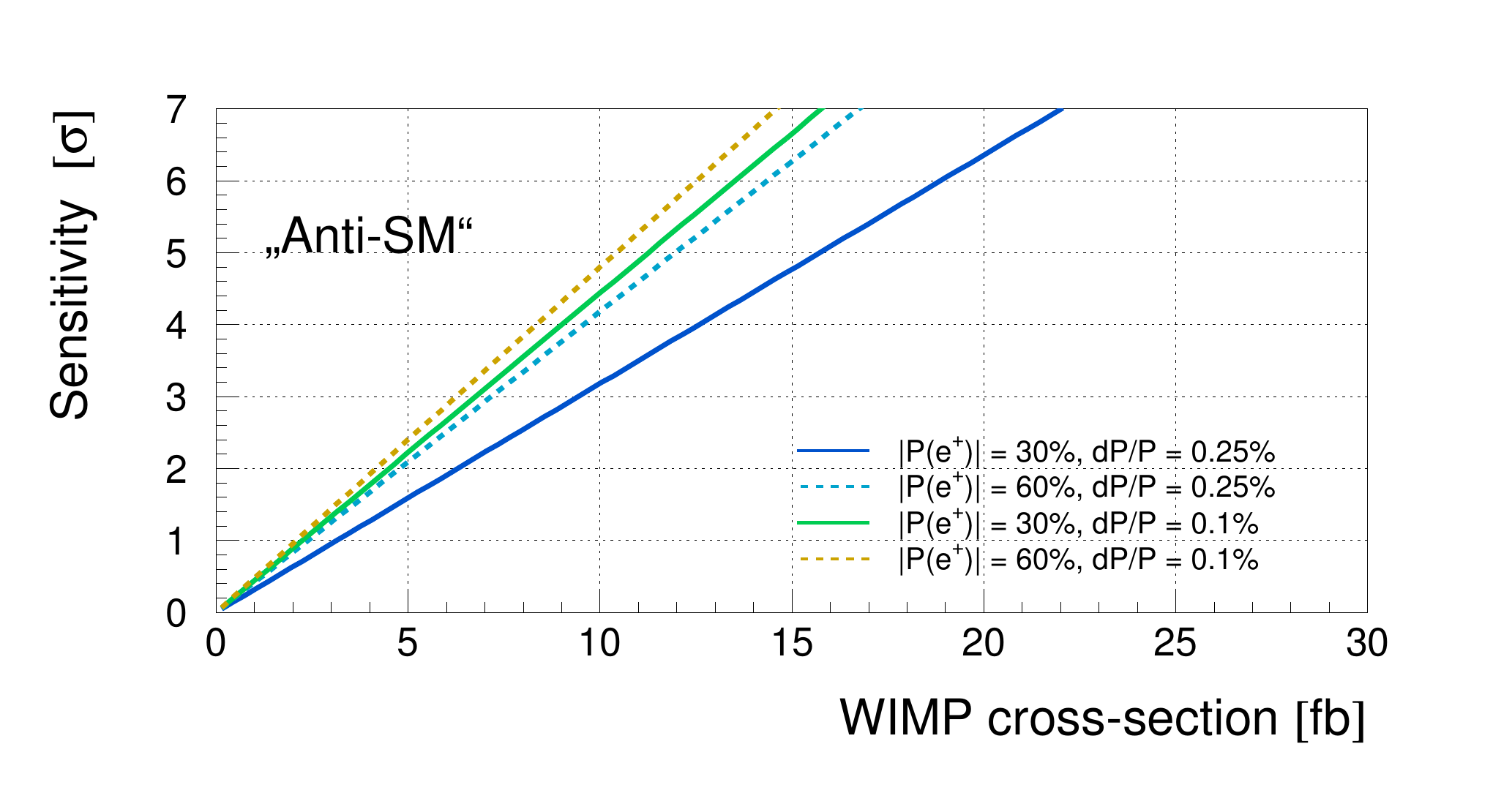}
\caption{Sensitivity as a function of  WIMP pair-production cross
  sections, for two beam polarization options and two uncertainty
  scenarios at the ILC. From Ref.~\cite{Bartels:2012ex}}
\label{fig:ilc_dm_lim}
\end{figure}

The final state is a high-$p_T$ photon with missing momentum due to
the invisible $\chi$ pair.  The dominant background is production of
neutrino pairs via a $Z$ boson, with a photon from initial state
radiation.  The sensitivity reaches up to nearly $\sqrt{s}/2$.

Studies at lepton colliders offer two important advantages compared to
similar studies at $pp$ machines. First, the polarization of the
initial state may be controlled, which gives power to distinguish
between the WIMP signal and the backgrounds, which may have distinct
polarization-dependent couplings.

Following the analysis of Ref.~\cite{Bartels:2012ex}, three coupling
scenarious are considered: 

\begin{itemize}
\item {\it equal}:  couplings are
independent of the helicity of the initial state,
\item {\it helicity}: couplings conserve helicity and parity, and
\item {\it anti-SM}: WIMPs couple only to right-handed electrons
  (left-handed positrons)
\end{itemize}

\noindent
where the final case has the greatest power to disentangle the SM
backgrounds from WIMP production. The relative sensitivity of two of
these scenarios is shown in Fig~\ref{fig:ilc_dm_lim}.

\begin{figure}[ht]
\centering
\includegraphics[width=0.23\linewidth]{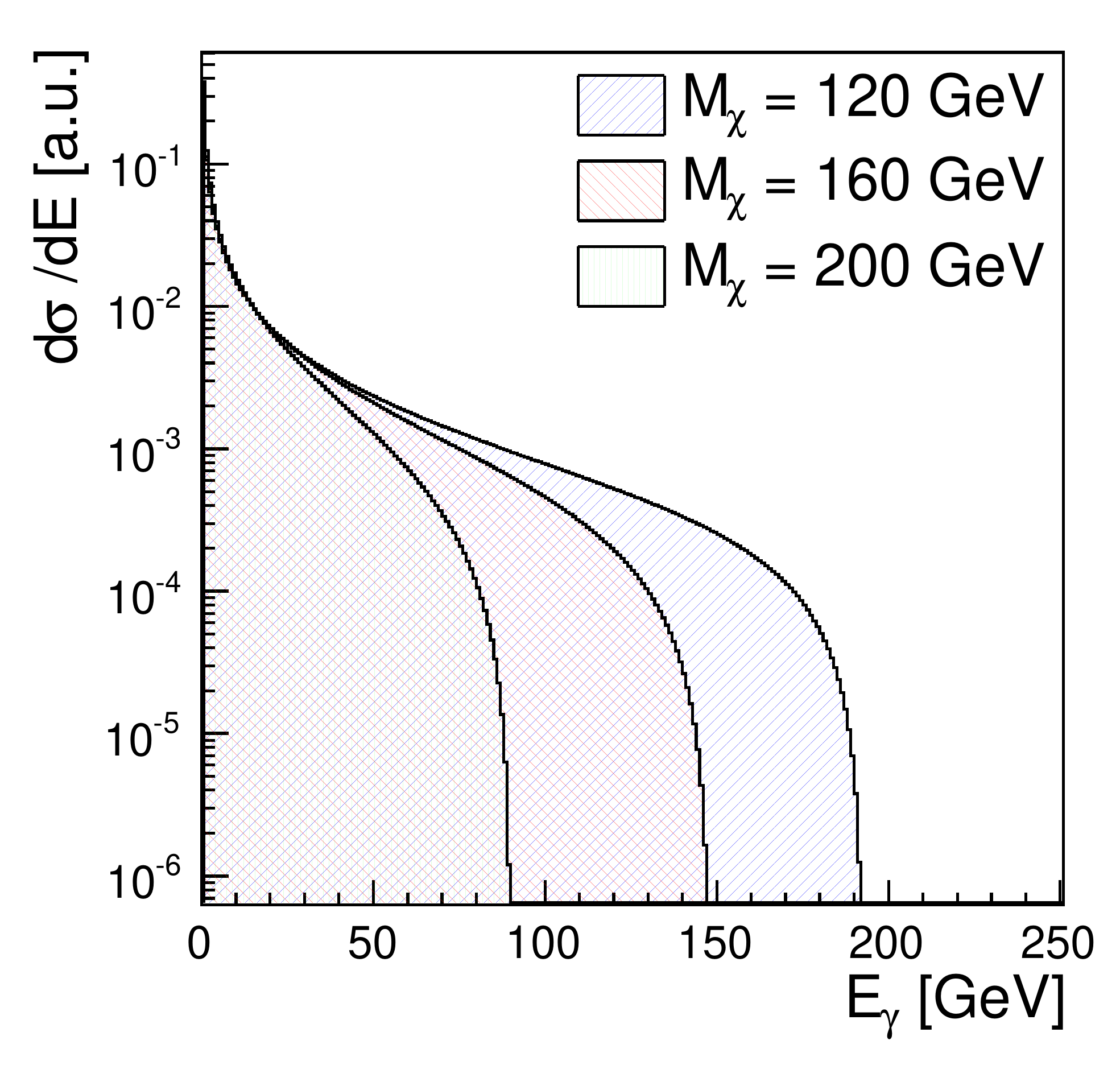}
\includegraphics[width=0.23\linewidth]{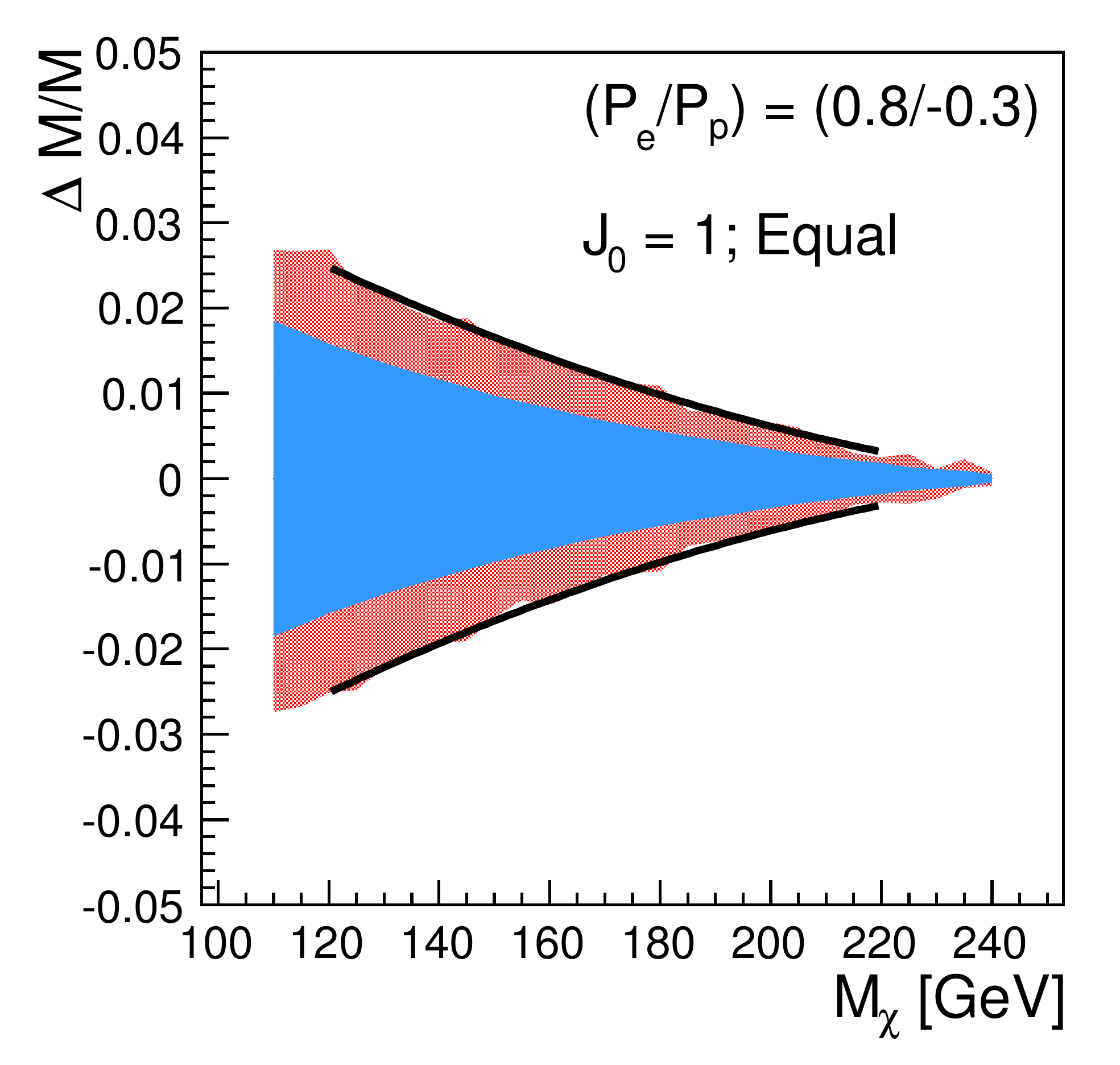}
\includegraphics[width=0.23\linewidth]{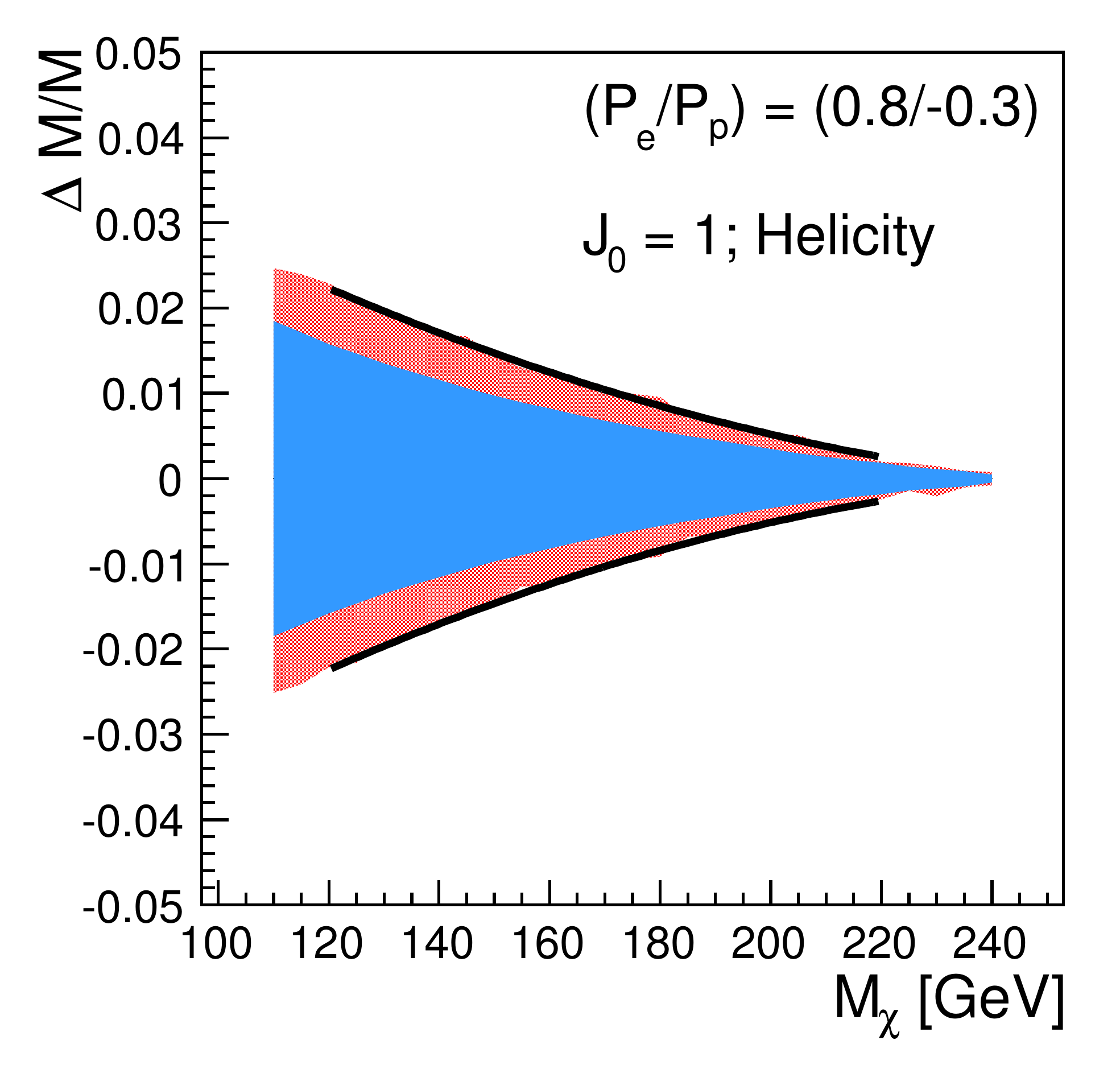}
\includegraphics[width=0.23\linewidth]{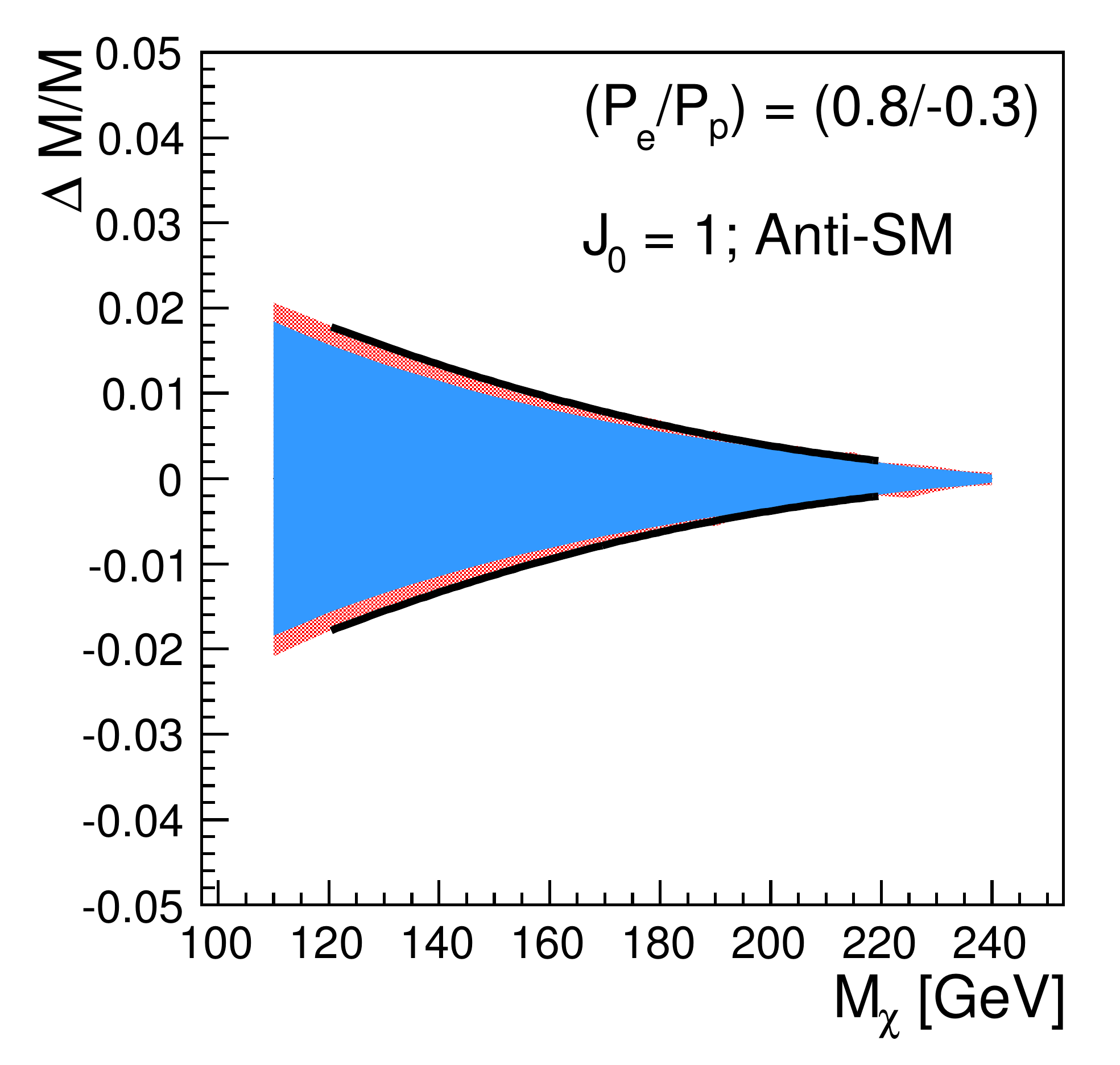}
\caption{ Left, dependence of the photon energy spectrum on the dark
  matter mass, $m_\chi$ at the ILC. Right, expected relative uncertainty on
  $m_\chi$ as a function of $m_\chi$ for three coupling 
  scenarios. From Ref.~\cite{Bartels:2012ex}}
\label{fig:ilc_mchi}
\end{figure}

The second major advantage of a lepton collider is its sensitivity to
the WIMP mass through its effect on the observed photon total energy,
see Fig~\ref{fig:ilc_mchi} for an ILC study. In addition, the shape of the photon energy spectrum and the ratio of cross-sections      measured with different beam polarisation configurations give access to the helicity      structure of the WIMP-lepton interaction and its dominant partial wave~\cite{Bartels:2012ex},      which are equivalent to pinning down the operator of the interaction.

 Such studies were possible at LEP, but the small
integrated luminosity of the dataset and lack of control over beam
polarization results in a significant decrease in sensitivity.

\subsubsection{Connections to Cosmic and Intensity Frontiers}

\begin{figure}[h]
\centering
\includegraphics[width=0.3\linewidth]{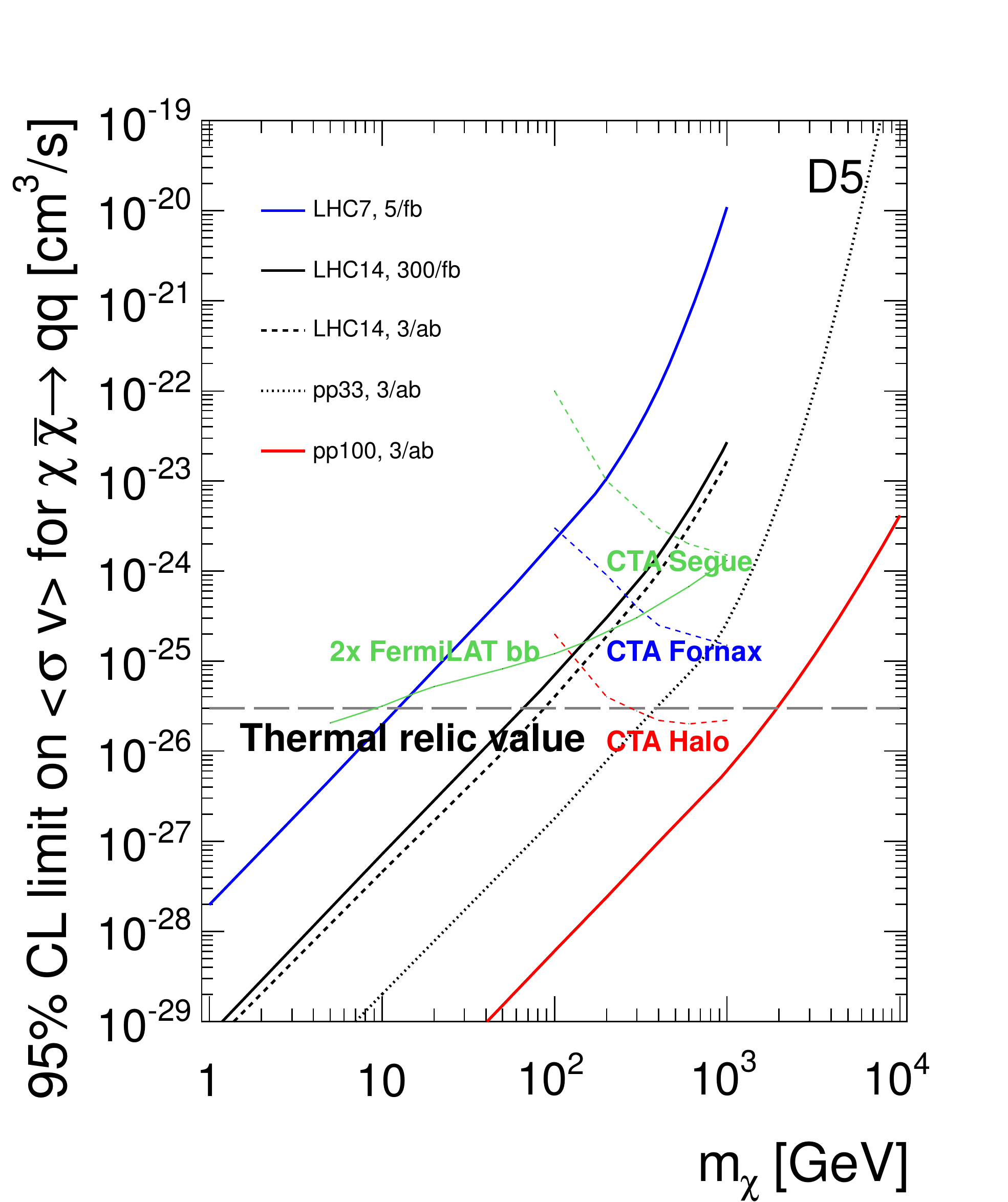}
\includegraphics[width=0.3\linewidth]{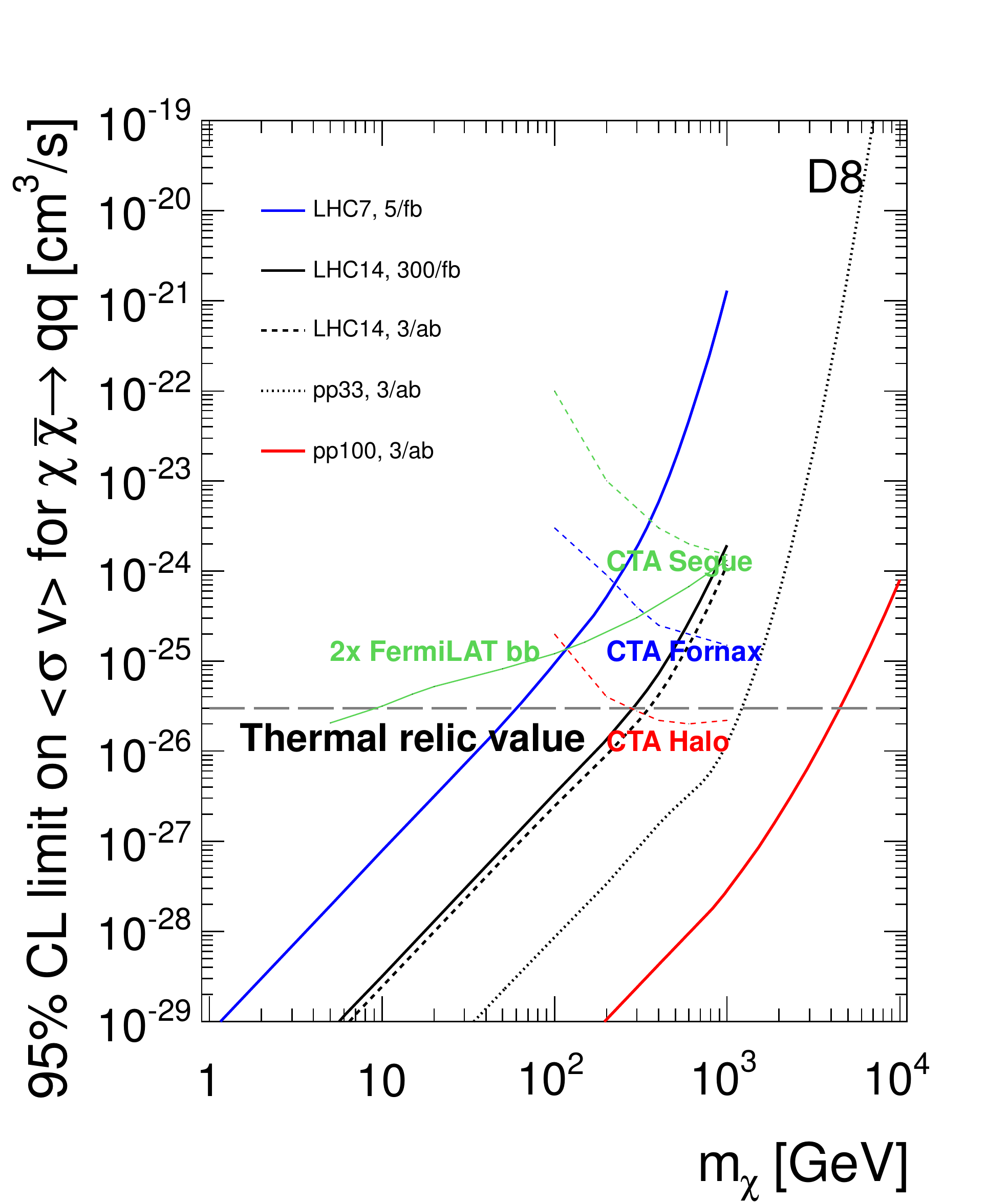}
\caption{Limits at 95\% CL on WIMP pair annihilation for different
  facilities using the D5 (left) or D8 (right) operator as a function of $m_\chi$. From Ref.~\cite{dmextrap}.}
\label{fig:indirect}
\end{figure}

The search for WIMPs via their interactions with the standard model is
clearly an area where the energy frontier overlaps with the cosmic
frontier, where there are dedicated direct-detection experiments
searching for recoil interactions $\chi+n\rightarrow \chi+n$. We have
compared the collider sensitivity to these direct-detection
experiments by translating the collider results into limits on the
$\chi-n$ interaction cross section.  In addition, the results may be
translated to compare with indirect detection experiments, which probe
WIMP annihilation into standard model particles,
$\chi\bar{\chi}\rightarrow XX$.  In Fig~\ref{fig:indirect}, we map $pp$ sensitivities to WIMP pair annihilation
cross-section limits. Predictions are compared to Fermi-LAT limits
from a stacking analysis of Dwarf galaxies~\cite{fermi}, including a
factor of two to convert the Fermi-LAT limit from Majorana to Dirac
fermions, and to projected sensitivities of CTA~\cite{CTA}.

At the ILC, WIMPs can be probed even if the WIMP-lepton coupling is so small that the      reverse process, namely the cosmic annihilation process $\chi\bar{\chi} \rightarrow f \bar{f}$ includes only a small fraction  of $e^+e-$ pairs.

These searches also probe models which are commonly considered to be
the domain of the intensity frontier, such as extensions of the Standard Model modifying neutrino-quark interactions~\cite{Friedland:2011za}.

\subsection{New gauge bosons}
\label{sec:resonances}
\subsubsection{$Z'$}

\newcommand{\zpr}{\ensuremath{Z'}}
\newcommand{\zp}{\ensuremath{Z'}}
\newcommand{\n}{\nonumber \\}
\newcommand{\commt}[1]{{\color{blue} \bf  [#1]}}
\newcommand{\gzp}{\ensuremath{\Gamma_{Z'}}}
\newcommand{\ffb}{^{f \bar f}}
\newcommand{\pol}{\ensuremath{\mathcal{P}}}
\newcommand{\mzp}{\ensuremath{M_{Z'}}}
\newcommand{\uprm}{\ensuremath{U(1)'}}
\newcommand{\x}{\ensuremath{\times}}        
\newcommand{\sto}{\ensuremath{SU(2) \x U(1)}}   
\newcommand{\eeql}[1]{\label{#1}\eeq}
\newcommand{\refl}[1]{(\ref{#1})}
 \renewcommand{\arraystretch}{1.2}
 \def\tev{\ ,{\rm TeV}}
\def\fbi{\,{\rm fb}^{-1}}

Additional colorless vector gauge  bosons (\zp ) occur in many extensions of the Standard Model (SM),
in part because it is generically harder to break additional abelian $U(1)^\prime$ factors than  non-abelian ones\footnote{For  reviews, see~\cite{Langacker:2008yv,Cvetic:1995zs,Hewett:1988xc,Leike:1998wr}.  Specific properties are reviewed in~\cite{Cvetic:1997wu,Erler:2009jh,Langacker:2009im,Nath:2010zj,Jaeckel:2010ni}. }.
Although \zpr s can occur at any scale and with couplings ranging from extremely weak to strong, we concentrate here on TeV-scale masses with couplings not too different from electroweak,  which might therefore  be observable at the LHC or future colliders. 

In this section, we summarize both the discovery reach and the potential of measuring the properties of new vector gauge bosons at future facilities. Following the notation in~\cite{Langacker:2008yv}, we define the couplings of the SM and additional gauge bosons to fermions by

\[-L_{NC}=eJ^\mu_{em}A_\mu+g_1J_1^\mu Z^0_{1\mu} + g_2 J^\mu_2 Z_{2 \mu}^0, \ 
J_\alpha^\mu=\sum\limits_i \bar f_i \gamma^\mu [\epsilon_L^{\alpha i}P_L + \epsilon_R^{\alpha i} P_R] f_i. \]

In this report, We will focus on several well known examples, listed in Table~\ref{tab:benchmark}. 
\begin{table}[htb]
  \centering
    \begin{tabular}{|c|c|c|c|c|c|c|c|}
    \hline
          & $\chi$ & $\psi$ & $\eta$ & LR & BL & \multicolumn{2}{|c|}{SSM}\\
    \hline
    $D$ & $2\sqrt{10}$ & $2\sqrt{6}$ & $2\sqrt{15}$ & $\sqrt{5/3}$ & 1 & \multicolumn{2}{|c|}{1} \\ \hline
 \multirow{2}{*}{$\hat \epsilon_L^q$} & \multirow{2}{*}{--1}    & \multirow{2}{*}{1}     & \multirow{2}{*}{--2}    & \multirow{2}{*}{--0.109} & \multirow{4}{*}{1/6} & $\hat \epsilon_L^u$ & $\frac 1 2 -\frac 2 3{\rm sin}^2\theta_W$\\ \cline{7-8}
         &   &   &   &   &   & $\hat \epsilon_L^d$ & $-\frac 1 2 +\frac 1 3{\rm sin}^2\theta_W$\\ \cline{1-5} \cline{7-8}
    $\hat \epsilon_R^u$ & 1     & --1    & 2     & 0.656  &  & $\hat \epsilon_R^u$ & $- \frac 2 3 {\rm sin}^2\theta_W$\\ \cline{1-5} \cline{7-8}
    $\hat \epsilon_R^d$ & --3    & --1    & --1    & --0.874 &  & $\hat \epsilon_R^d$ & $ \frac 1 3 {\rm sin}^2\theta_W$\\ \hline
    \multirow{2}{*}{$\hat \epsilon_L^l$} & \multirow{2}{*}{3}     & \multirow{2}{*}{1}     & \multirow{2}{*}{1}     & \multirow{2}{*}{0.327}  & \multirow{3}{*}{--1/2} & $\hat \epsilon_L^\nu$ & $\frac 1 2 $\\ \cline{7-8}
           &   &   &   &   &   & $\hat \epsilon_L^e$ & $-\frac 1 2 + {\rm sin}^2\theta_W$\\ \cline{1-5} \cline{7-8}
    $\hat \epsilon_R^e$ & 1     & --1    & 2     & --0.438 &  & $\hat \epsilon_R^e$ & $ {\rm sin}^2\theta_W$\\ \hline
    \end{tabular}%
  \label{tab:benchmark}%
\caption{Benchmark models and couplings, with $\epsilon^i_{L,R} \equiv \hat \epsilon^i_{L,R} /D$.}
\end{table}%


\begin{figure}[h!]
\subfigure{
\includegraphics[width=139pt]{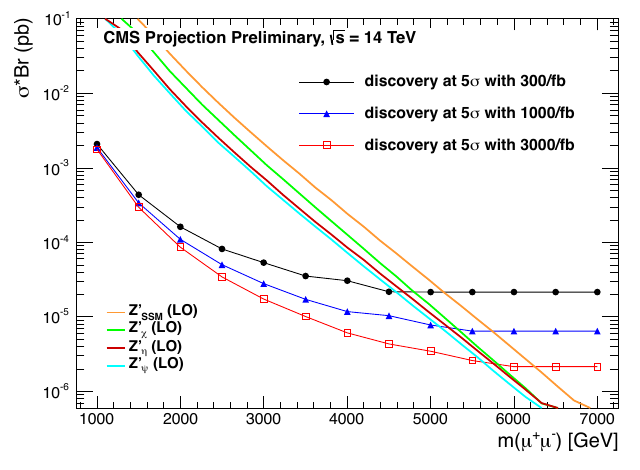}}
\subfigure{
\includegraphics[width=139pt]{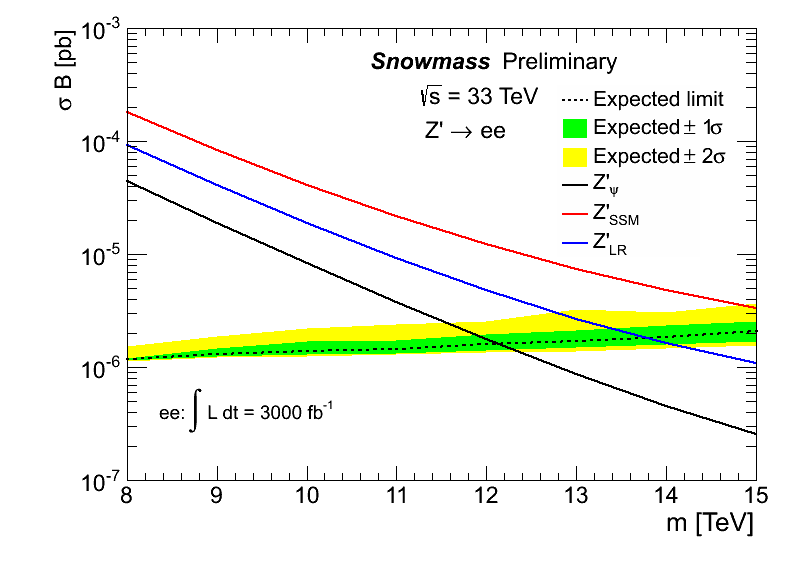}}
\subfigure{
\includegraphics[width=139pt]{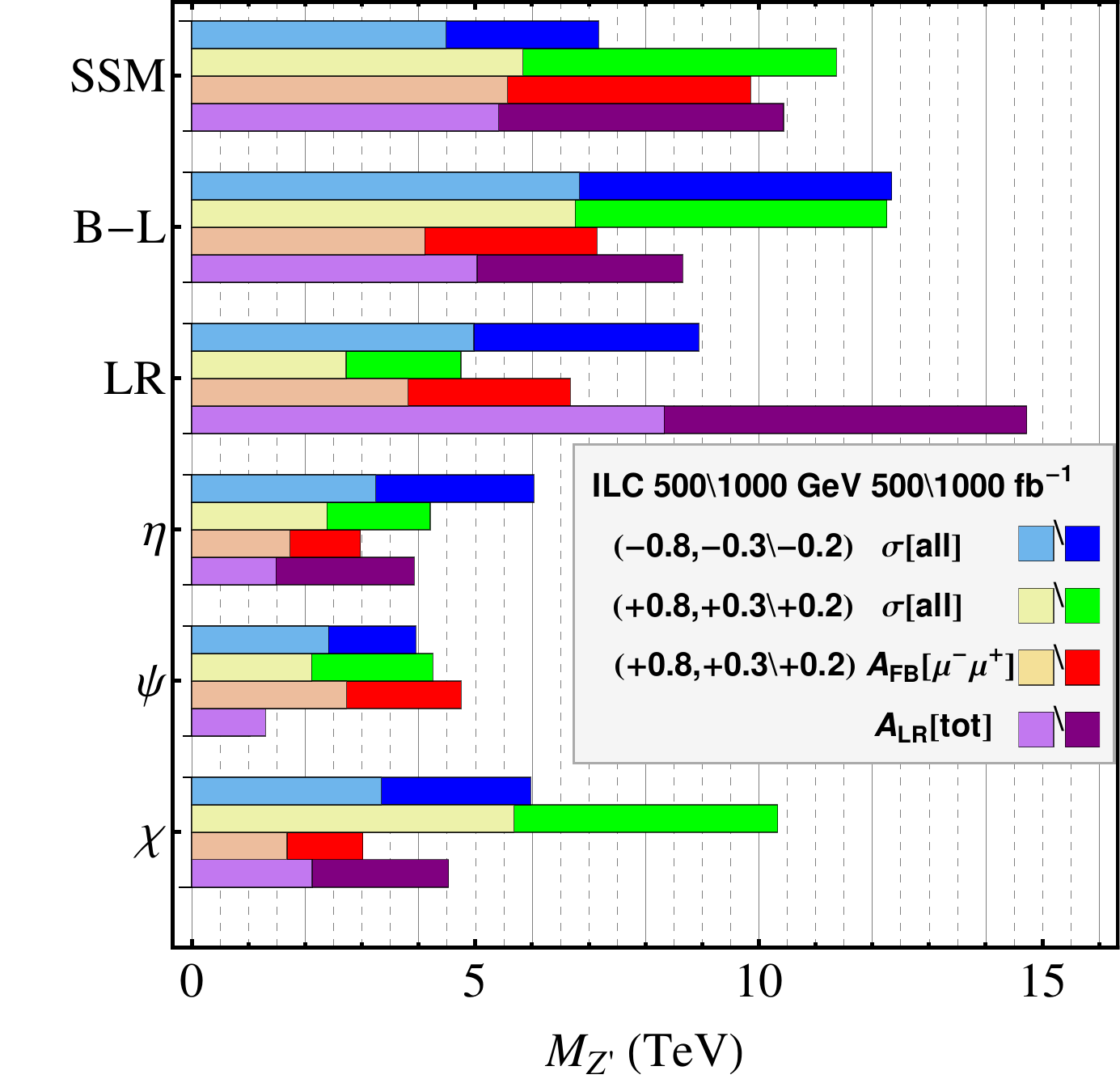}}
\caption[]{ Reaches for \zp \ at colliders. Left and middle panel: the reach at the LHC \cite{cmswp} and HL-LHC \cite{Hayden:2013sra}. Right Panel: the reach at the ILC \cite{Han:2013mra,Freitas:2013xga}. }
\label{fig:Zprime_reach}
\end{figure}

Hadron colliders are great for searching for \zp. Such searches typically look for a resonance peak in the lepton pair invariant mass distribution. Due to its simplicity and importance, it is usually among the earliest analyses to be carried out at hadron colliders.  The reach at the LHC and HL-LHC are presented in the left and middle panel of Fig.~\ref{fig:Zprime_reach}.   In particular, LHC Run 2 can discover \zp \ up to about 5 TeV, while the HL-LHC  and a 33 TeV hadron collider can extend that reach to about 7 TeV and 12 TeV, respectively. Previous studies \cite{Diener:2009vq}  indicate that a 100 TeV VLHC can ran up to $M_\zp \sim 30$ TeV.  

High energy lepton colliders can search for \zp \ by observing its interference with the Standard Model $Z$ and photon.  As an example, the ILC reaches for several \zp \ models are presented in the right panel of Fig.~\ref{fig:Zprime_reach}. In particular, in addition to the total rate, the asymmetry observables (defined later) are very powerful in many cases. We see that it can go beyond the capabilities of the 14 TeV LHC. For example, this is the case for $Z'_{\rm B-L}$ and $Z'_{\rm \chi}$. The CLIC reach is significantly higher \cite{clicwp}.

\begin{figure}[h!]
\subfigure{
\includegraphics[width=139pt]{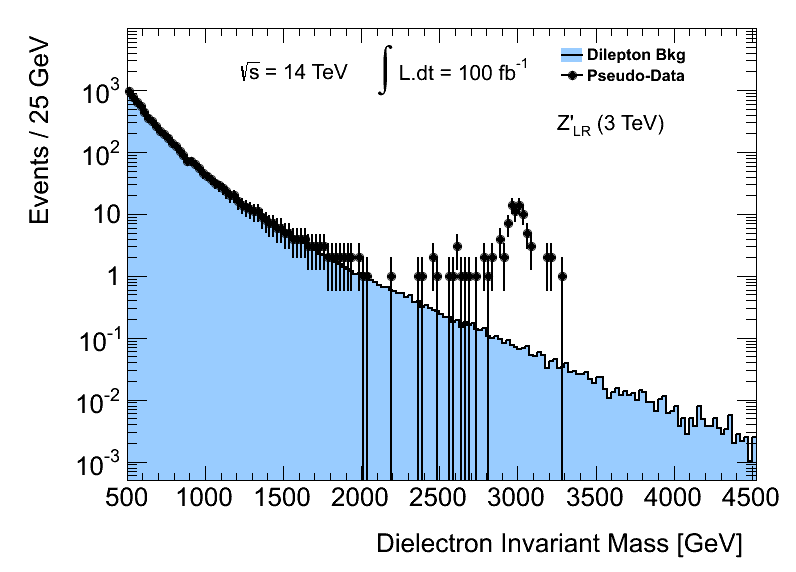} }
\subfigure{
\includegraphics[width=139pt]{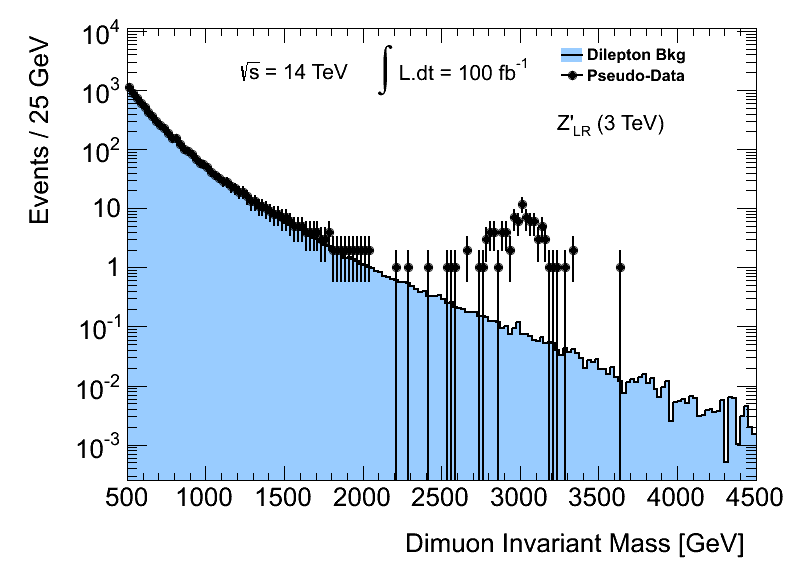}}
\subfigure{
\includegraphics[width=139pt]{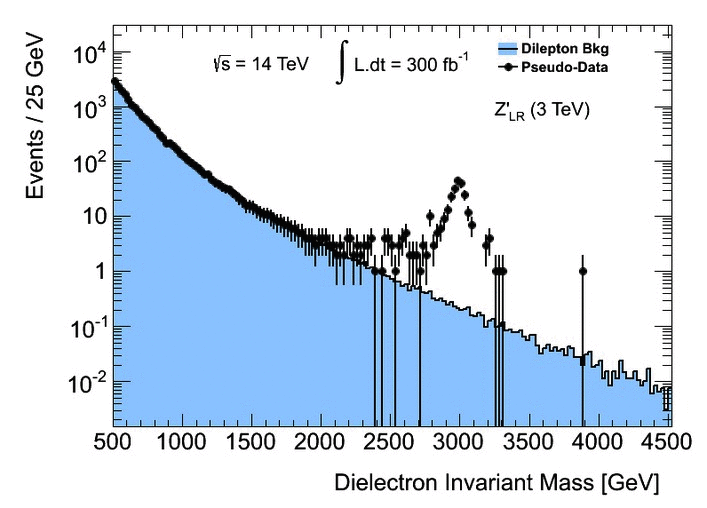}}
\caption[]{A \zp discovery story at the LHC \cite{Hayden:2013sra}. Left: Drell-Yan backgrounds and the emerging signal for a LR $Z'$ at 3 TeV,  for $e^+e^-$ pairs  after 100 fb$^{-1}$. Middle: Drell-Yan backgrounds and the emerging signal for a LR $Z'$ at 3 TeV,  for $\mu^+\mu^-$ pairs  after 100 fb$^{-1}$. While the  muon line shape is much broader due to resolution effects, the observation would be definitive, confirming evidence of a discovery. Right: The Drell-Yan backgrounds and signal for a LR $Z'$ at 3 TeV,  for $e^+e^-$ pairs  after 300 fb$^{-1}$. }
\label{fig:Zprime_story}
\end{figure}
A $Z^\prime$ discovery would be spectacular at the LHC and will be a focus of attention as the energy doubles in 2015. Such an observation will come steadily for invariant masses above a few TeV/c$^2$ and will accumulate over years. While acceptances are good for both electrons and muons (better than 80\% and independent of pileup, mass resolutions are quite different between electrons ($\Delta M/M \propto$ a percent) and muons ($\Delta M/M \propto$ 10\%) and precision measurements will eventually rely on the former. But the observation of a signal in both channels would be definitive and so the muon states will be an important part of a discovery story for a new vector resonance.

 a possible evolution of a 3 TeV $Z′$ discovery at the LHC in electron pair final states. A potential signal will begin to emerge with the first half year of data in 2015 (at about 30 fb$^{-1}$ with a few 10s of events.  By itself, such a small bump could be overlooked as a background fluctuation. But a broader, similarly-populated enhancement would have started to emerge in the $\mu^+\mu^-$ invariant mass distributions and this would be a major focus by the end of Run 2. The left panel of Fig.~\ref{fig:Zprime_story} shows shows the nature of such a signal in electron pairs by the end of Run 2, while the middle panel of Fig.~\ref{fig:Zprime_story} shows shows the same object as it would appear in muon pair combinations. By the end of LHC Run 2 with 100 fb$^{-1}$, a discovery would be declared but without much information available for deciphering its source. The right panel of Figure \ref{fig:Zprime_story} shows how the next run (corresponding to the Phase 1 upgrades) of 300 fb$^{-1}$ could begin the process of discriminating a dynamical source. Of course after 3000 fb$^{-1}$ at the HL-LHC precision measurements of such a new state could be made.


\begin{figure}[h!]
\subfigure{
\includegraphics[width=209pt]{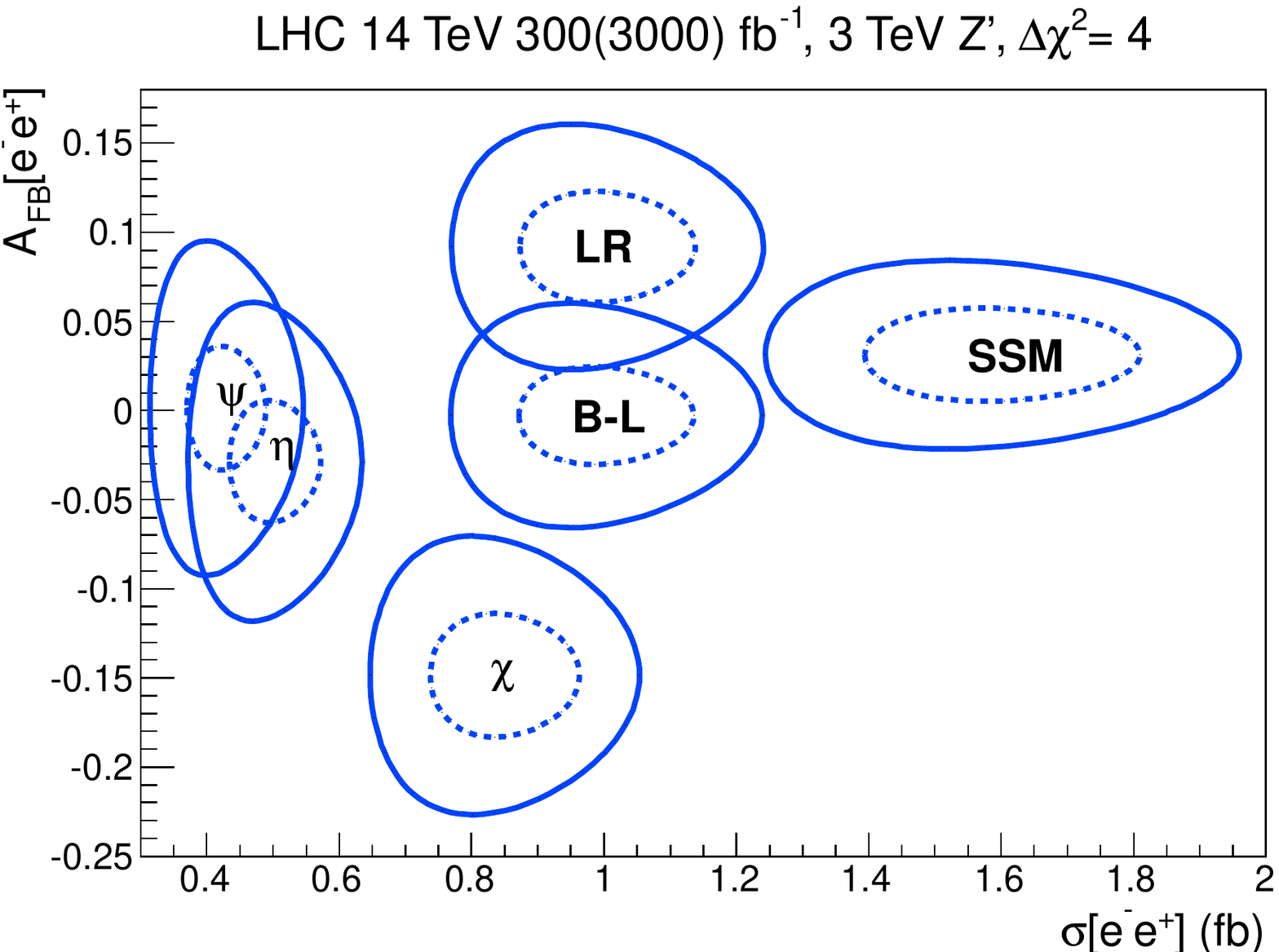} }
\subfigure{
\includegraphics[width=209pt]{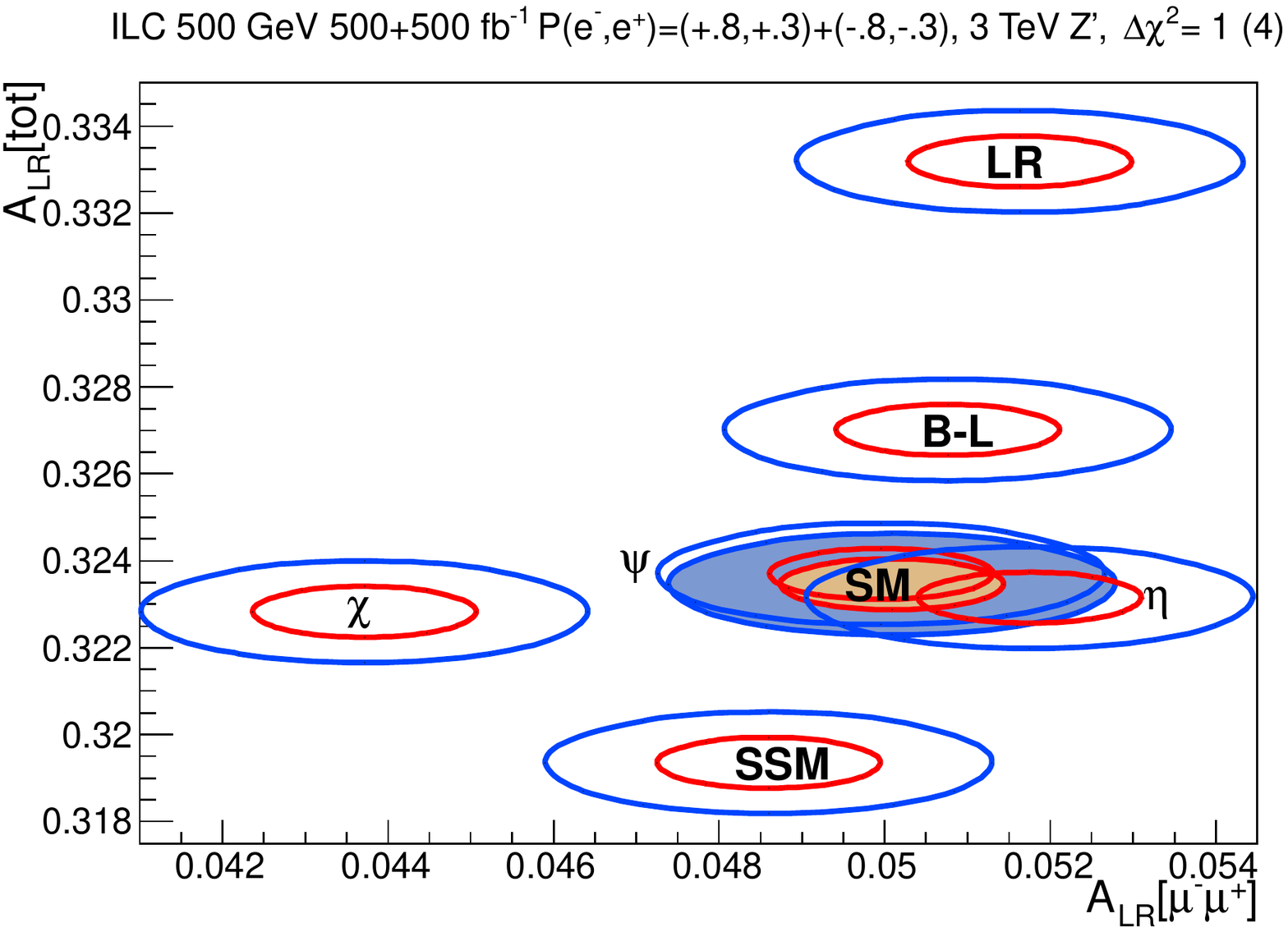}}
\caption{Distinguishing \zp \ models at colliders \cite{Han:2013mra,Freitas:2013xga}. {\it Left panel}:  $\Delta \chi^2 =4 $ contours of the simulated forward-backward asymmetry versus cross section for the benchmark models at LHC Run 2 (solid) and HL-LHC (dashed).
 {\it Right panel}: Right panel: $\Delta\chi^2=1$(red) and $\Delta\chi^2=4$(blue) contours of polarization asymmetry in dimuon final state and all di-fermion final states (excluding $e^-e^+$ and $\nu\nu$) at the ILC.  }
\label{fig:LHCmmfit}
\end{figure}

If a \zp  \ has been discovered, the immediate next step would be to measure its properties as much as we can.  There have been studies on this topic, for example \cite{delAguila:1993ym,DelAguila:1993rw,DelAguila:1995fa,Diener:2011jt,Godfrey:2005pm,Carena:2004xs}.
The useful observables are $\sigma_{\rm prod} \times $BR in various channels and the total width. Many \zp \ candidates are chiral. To reveal this nature of their couplings, it is useful to consider the forward backward asymmetry  variable. In addition,  we can also include the left right asymmetry variable at lepton collider with polarized beams.  
As a concrete example, we consider a benchmark with $M_{Z'} =3 $ TeV, which is within the discovery reach of the LHC Run 2. 
The predicted value as well as experimental precision for the $\sigma_{\rm prod} \times $BR($Z' \to$ dilepton ) and $A_{\rm FB}$ ($A_{\rm LR}$) at the LHC (ILC) are shown in the left panel of  Fig.~\ref{fig:LHCmmfit}. We can see that combining the measurements at  a Hadron collider and lepton collider can be very valuable in distinguishing different models. For example, $\zp_{\rm LR}$ and $\zp_{\rm B-L}$ cannot be clearly distinguished at LHC Run 2. HL-LHC can start to discern their differences. On the other hand, ILC with polarized beams can clearly tell them apart.

 Discovery of a \zp \ leads to many new implications which can lead to further searches at colliders. There should be (at least) an associated Higgs with the \zp. Discovering this new Higgs would be much harder than discovering the \zp, similar to the discovery of $W/Z$ vs the Higgs in the Standard Model.  The understanding of the nature of \zp \ couplings, even if a partial one, will give us insight about its embedding in the high scale (UV) and more fundamental theory. Such UV completions of \zp \ usually leads to additional predictions.  A \zp \ with the Standard Model fermions could be anomalous, in which case there have to be new fermions that may be produced by colliders.  If a \zp \ is consistent with the one from the Left-Right symmetric model, there should also be additional heavy resonances, such as $W_R'$ and exotic Higgses, with similar masses. \zp \ can also play an important role in the dynamics of electroweak symmetry breaking, and decaying into SM gauge bosons will give us a smoking gun signal in this scenario.

In the context of supersymmetry, \zp \ can play an important role, such as the solution of the $\mu$ problem and the mediator of the supersymmetry breaking. 
\zp \ decaying into superpartners can be an important discovery channel. 

In addition to $Z^\prime$ minimal gauge couplings to the Standard Model
fermions discussed here, gauge-invariant anomalous (magnetic moment type)
couplings with the known fermions could also be present. The dilepton final states, like $e^+e^-$ and $\mu^+\mu^-$, are still the
most clear channels. The reach of this scenario at hadron colliders have been presented in \cite{Chizhov:2013hua}. For example, with integrated luminosity of 3 ab$^{-1}$, pp collider at 14(33) TeV can discover such a $Z^\prime $ up to 6(13) TeV. 

\subsubsection{New hadronic resonances}
\begin{figure}[h]
\subfigure{
\includegraphics[width=209pt]{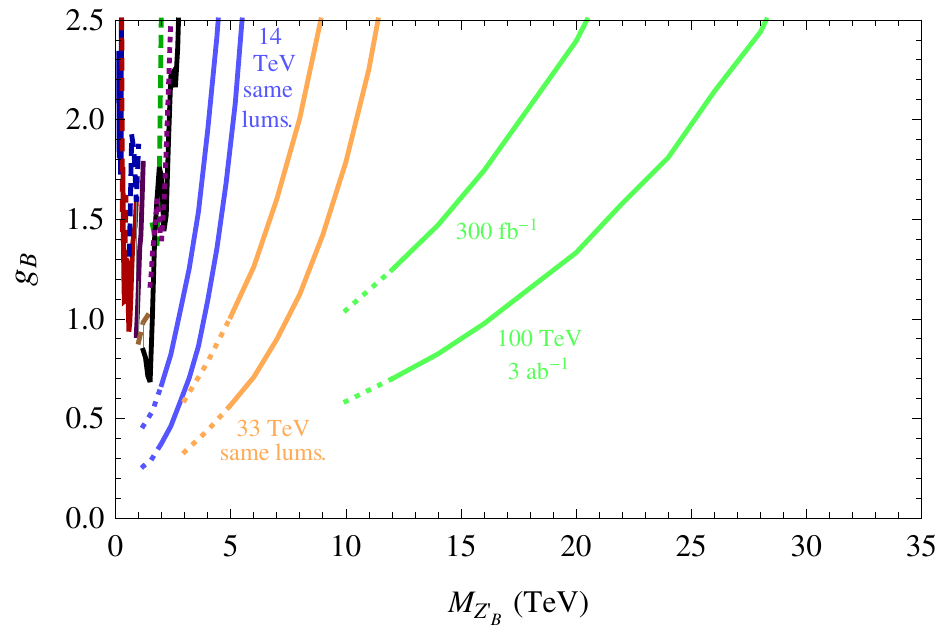} }
\subfigure{
\includegraphics[width=209pt]{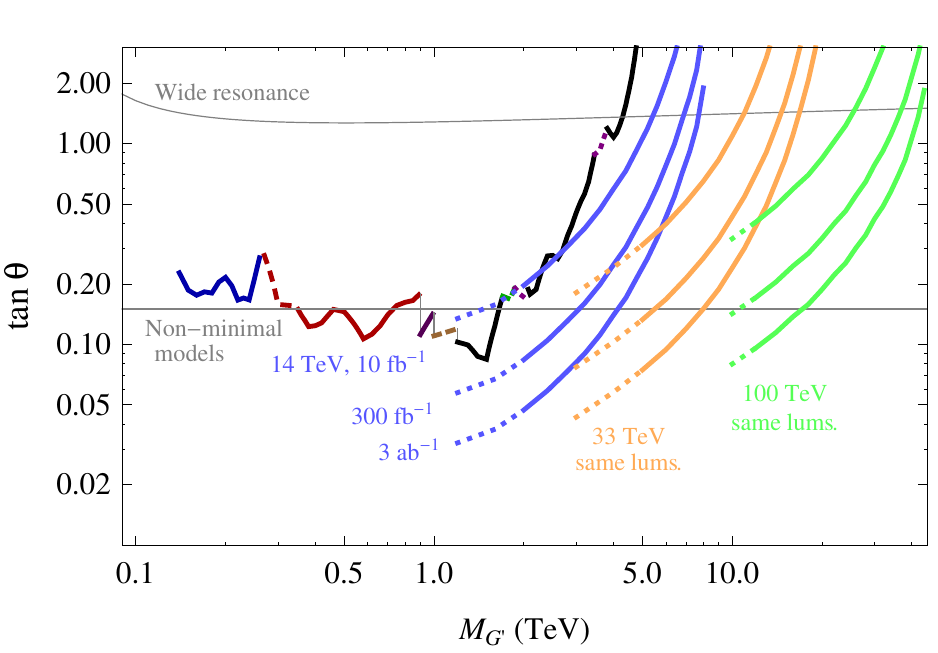}}
\caption[]{Hadronic resonance discovery sensitivity at hadron colliders. Left panel: $\zp_B$. Right panel: Octet coloron. }
\label{fig:LHC_had}
\end{figure}

\begin{figure}[h]
\subfigure{
\includegraphics[width=209pt]{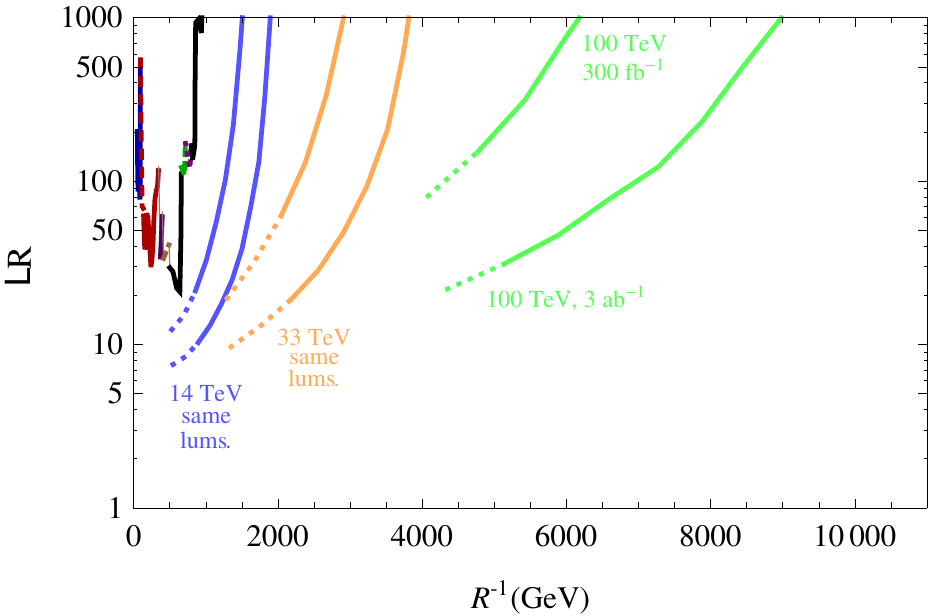} }
\subfigure{
\includegraphics[width=209pt]{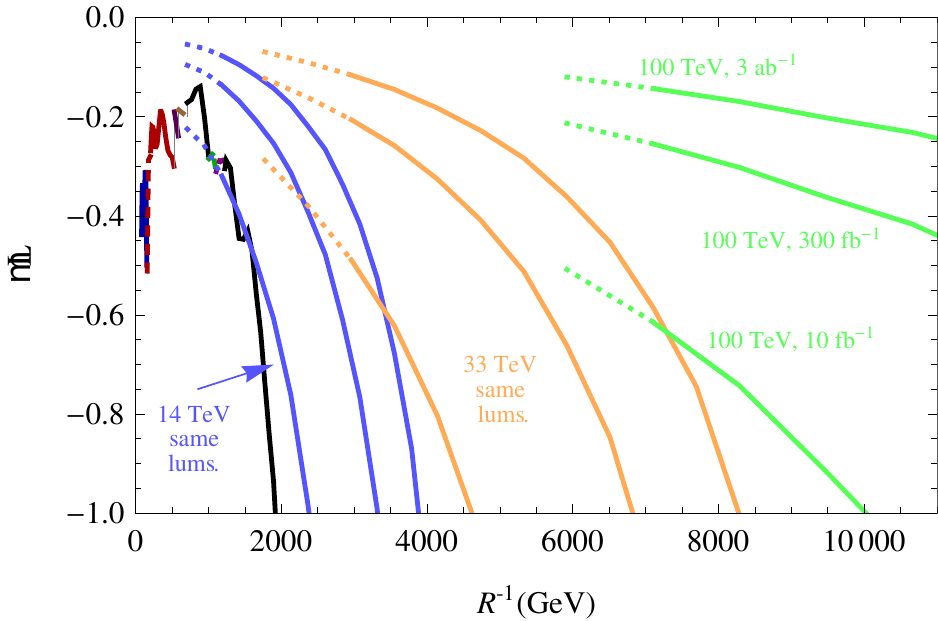}}
\caption[]{Left panel, the discovery reaches for KK-gluon in minimal UED model at hadron colliders. Right panel, the discovery reaches for  KK-gluon in next-to-minimal UED model at hadron colliders. }
\label{fig:LHC_UED}
\end{figure}

Hadron colliders are also ideal for searching for new leptophobic resonances by looking for a peak in the dijet invariant mass distribution.  Aside from serving as a standard candle for understanding experimental issues such as jet energy resolution, these searches are strongly motivated in theories with a new U(1) baryon number gauge symmetry, coloron models, and models of Universal Extra Dimensions (UED).  The discovery reach in the coupling--mass plane~\cite{Dobrescu:2013cmh} for LHC and HL-LHC, a 33 TeV $pp$ collider, and a 100 TeV $pp$ collider is shown in Fig.~\ref{fig:LHC_had}   for a $Z'_B$ colorless vector resonance (left panel) and  a $G'$ color-octet vector resonance (right panel)~\cite{Yu:2013wta}.  Similarly, Fig.~\ref{fig:LHC_UED} shows the discovery sensitivity  for the level-2 Kaluza Klein gluon in minimal UED models  (left panel)  and next-to-minimal UED models (right panel)~\cite{Kong:2013xta}. Higher energy machines clearly extend the reach for dijet resonances to higher masses, for example allowing the discovery of relatively strongly coupled $Z'_B$ bosons (colorons) progressing from 4.5 (6.5) TeV with the LHC to 5.5 (7.5) TeV with the HL-LHC, to 11.5 (16) TeV for a 33 TeV collider and as high as 28 (40) TeV for the VLHC.  High luminosity at these machines is also critical in order to probe couplings as small as $g_B \sim 0.25$ and $\tan \theta \sim 0.08$, depending on the multijet trigger threshold.


%
%
%
%
%

\subsection{Discovery in Jets $+$ MET: `Simple' Supersymmetry}
\label{sec:story1}

As discussed in the introduction,
perhaps the best motivated and most successful framework for
physics beyond the standard model is supersymmetry (SUSY).
In almost all SUSY models, the colored superpartners (gluino and squarks)
are significantly heavier than the lightest supersymmetric particle (LSP),
which is stable and appears in the detector as missing energy.
This is due to the fact that the superpartners get large contributions
to their mass from quantum corrections, and the colored particles get the
largest contributions.
A very general search strategy at hadron colliders is therefore to 
look for production of gluinos and squarks.
These will decay to jets, possibly leptons, and missing energy.
Superpartners of leptons (sleptons), 
as well as the partners of the $W$, $Z$, and Higgs (electroweak-inos)
are generally lighter than the colored superpartners, and lead
to decays with leptons in the final state.
These provide a clean signal, but the lepton signal
is highly model-dependent.
A very general search strategy is therefore to search for jets plus
missing energy.
Detailed study has shown that this is the most sensitive search for
many well-motivated SUSY models, {\it e.g.}~`minimal supergravity.'

The observation of any excess of missing energy immediately
raises the question of whether the missing energy is due to
stable dark matter particles being produced.
This connection is much more general than SUSY:
WIMP dark matter motivates a stable particle that can be produced
at colliders, and the hierarchy problem motivates `partner' particles
with the same gauge quantum numbers as standard model particles.
Colored partners generally decay to the dark matter particle,
and can therefore give a signal in jets plus missing energy.
Examples of such models include `universal' extra dimension models
where the standard model fields propagate in an extra dimension,
and `little Higgs' models with $T$-parity.
A discovery in the jets plus missing energy channel is therefore 
potentially the first step producing and studying dark matter in the laboratory,
as well as establishing new symmetries of nature.
It is therefore recognized as one of the most important searches for new 
physics at the LHC, with both ATLAS and CMS each performing several independent
searches using different background rejection strategies.

\paragraph{LHC sensitivity:}
The upcoming LHC run 2 (14~TeV $300/$fb)
has tremendous potential for discovery in this channel.
The strongest current bounds come from the recently-completed LHC run 1,
and the reach is significantly improved due to the increased center-of-mass
energy of run 2.

\begin{figure}
\centering
\includegraphics[width=0.5\linewidth]{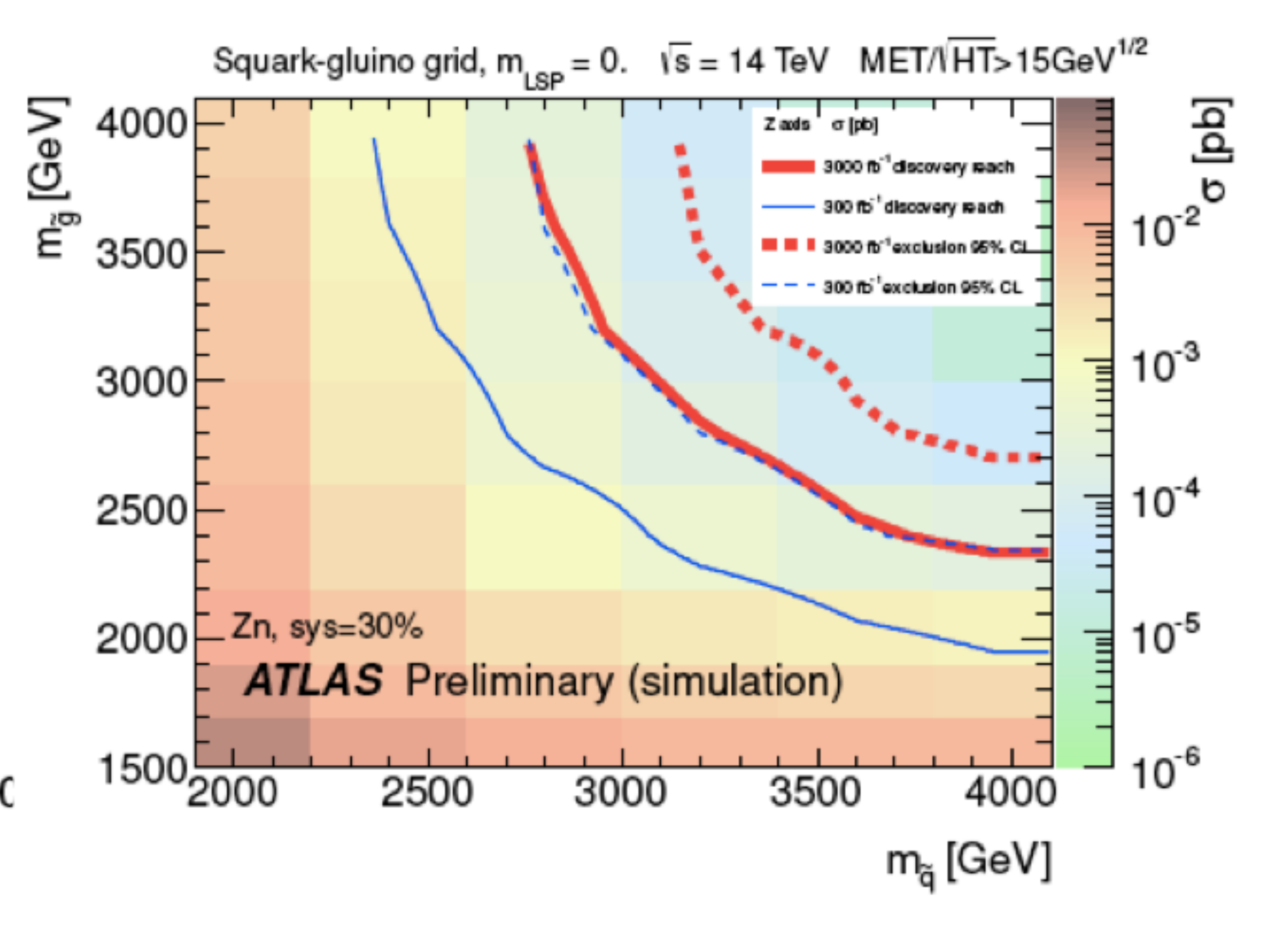}
\caption{Estimated reach of ATLAS run 2 for squarks and gluinos. CMS has similar sensitivity.}
\label{fig:story1atlas}
\end{figure}

\begin{figure}
\centering
\includegraphics[width=0.45\linewidth]{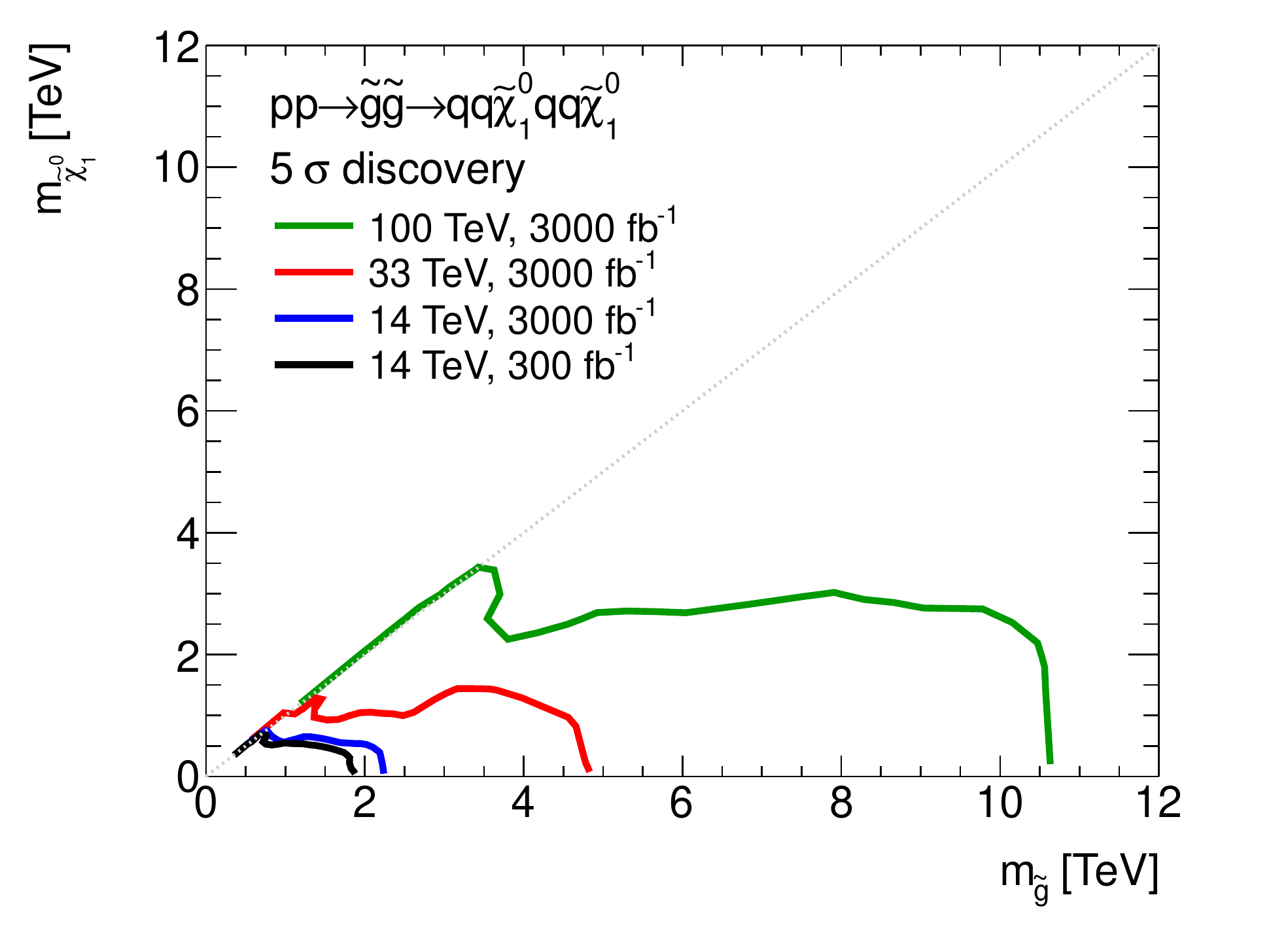}
\includegraphics[width=0.45\linewidth]{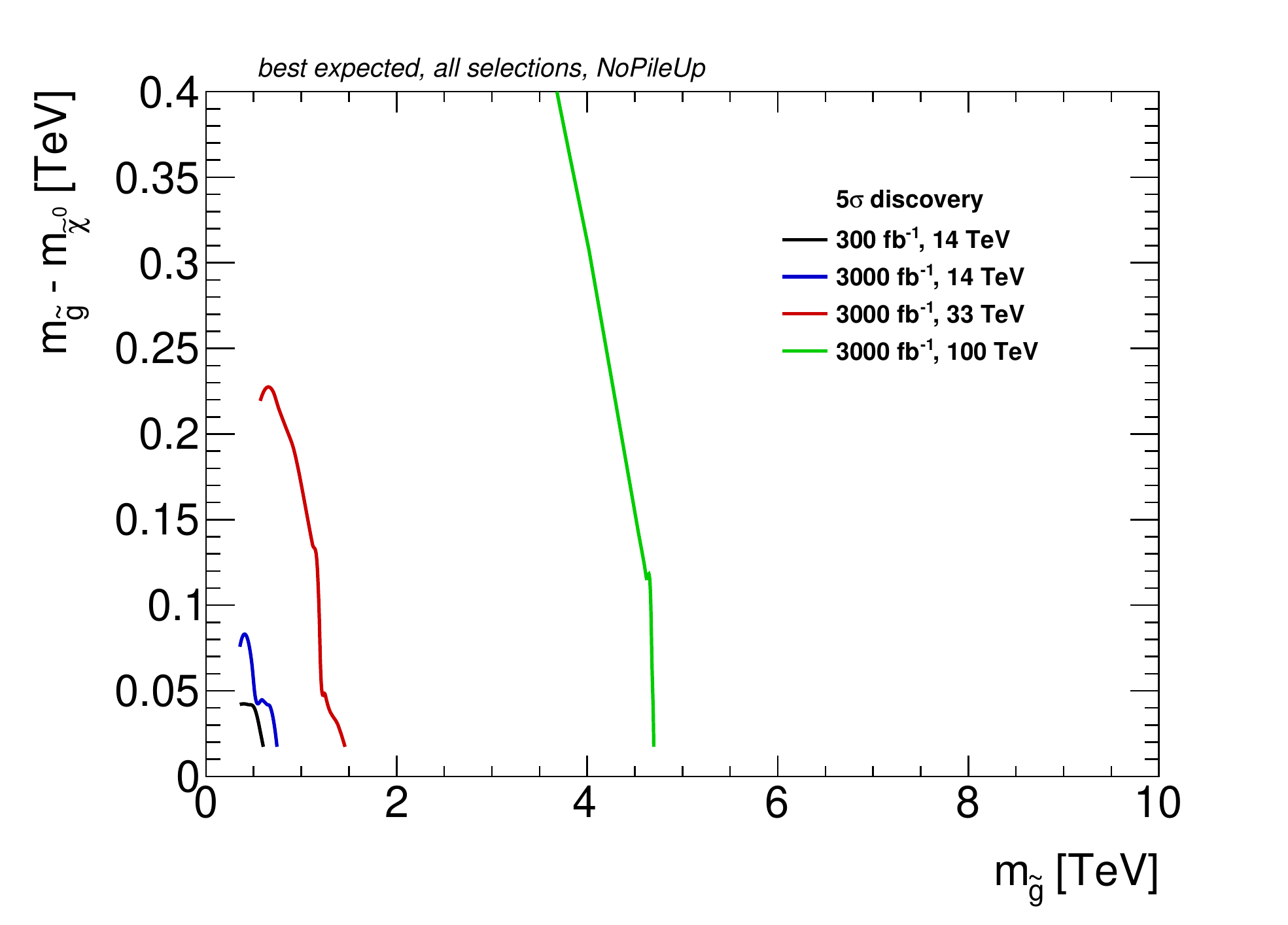}
\caption{Estimated reach for gluinos using DELPHES simulation \cite{squark-gluino-scan}. Only gluino and LSP are assumed to be light. Left panel shows the result of multijet search, right panel - monojet search optimized for squeezed spectrum.}
\label{fig:story1simple}
\end{figure}

It is impossible to discuss the reach for SUSY without caveats and assumptions.
One approach is the simplified model approach, which focuses on a subset
of the particles and allows exploration
of a wide range of kinematics for new physics,
but does not include the effects of the many decay modes present
in realistic models.
The reach of these searches for simplified models has been studied 
in the the ATLAS and CMS WhitePapers \cite{atlaswp,cmswp},
and also using the Delphes Snowmass LHC detector
\cite{squark-gluino-scan}.
The reach is shown in Fig.\ref{fig:story1atlas} and \ref{fig:story1simple}.
Based on these results, we can expect squarks and gluinos with masses up to 
around 2 TeV to be visible at LHC run 2, provided there is no significant
dilution due to other decays.

\begin{figure}
\centering
\includegraphics[width=0.9\linewidth]{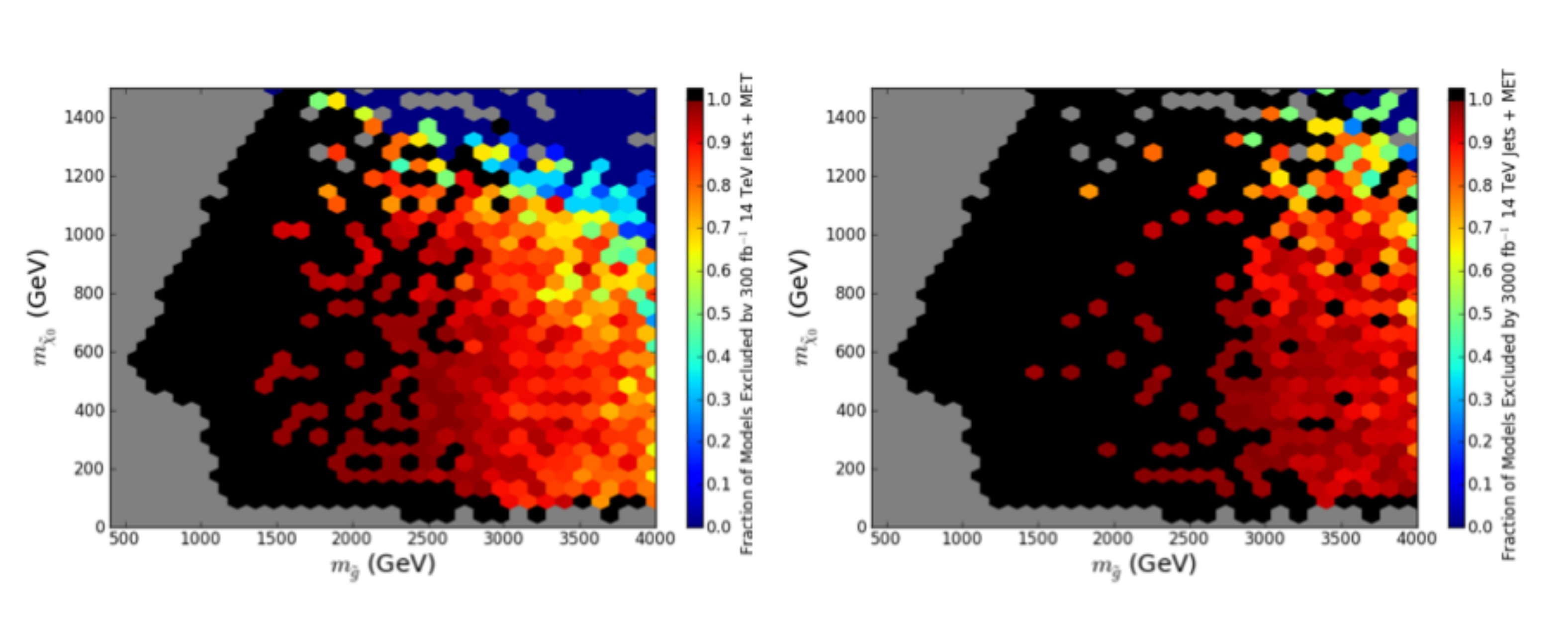}
\caption{Projections for pMSSM model coverage efficiency \cite{Cahill-Rowley:2013yla} shown in gluino-LSP pane for 14 TeV LHC and integrated
luminosity of 300/fb (left) and 3000/fb (right)}
\label{fig:story1pmssm}
\end{figure}

Another way to assess SUSY reach is to study scans over complete models.
This approach is taken in the 
`phenomenological minimal supersymmetric standard model' 
(pMSSM)  \cite{Cahill-Rowley:2013yla, Cahill-Rowley:2013gca}.
Here 19 independent superpartner masses are independently 
uniformly scanned.
This scan shows that of the models that are not excluded at the
LHC with 300/fb, $75\%$ are in $95\%$ confidenct level reach of the HL-LHC,
see Fig.~\ref{fig:story1pmssm}.
Similar conclusions are reached when surveying other complete models, i.e.  SO(10) 
SUSY GUT models \cite{Anandakrishnan:2013pja}

Another important caveat to the LHC sensitivity is Majorana vs Dirac nature of gauginos. 
The naturalness bounds on gaugino masses are dramatically relaxed for Dirac gaginos \cite{Fox:2002bu,Kribs:2012gx}.
For example, gluinos as heavy as 3-4 TeV may not result in significant fine-tuning. One of the consequences 
of high gluino masses is the suppression of squark production by almost two orders of 
magnitude \cite{kribsWP}. 

%

\begin{figure}
\centering
\includegraphics[width=0.5\linewidth]{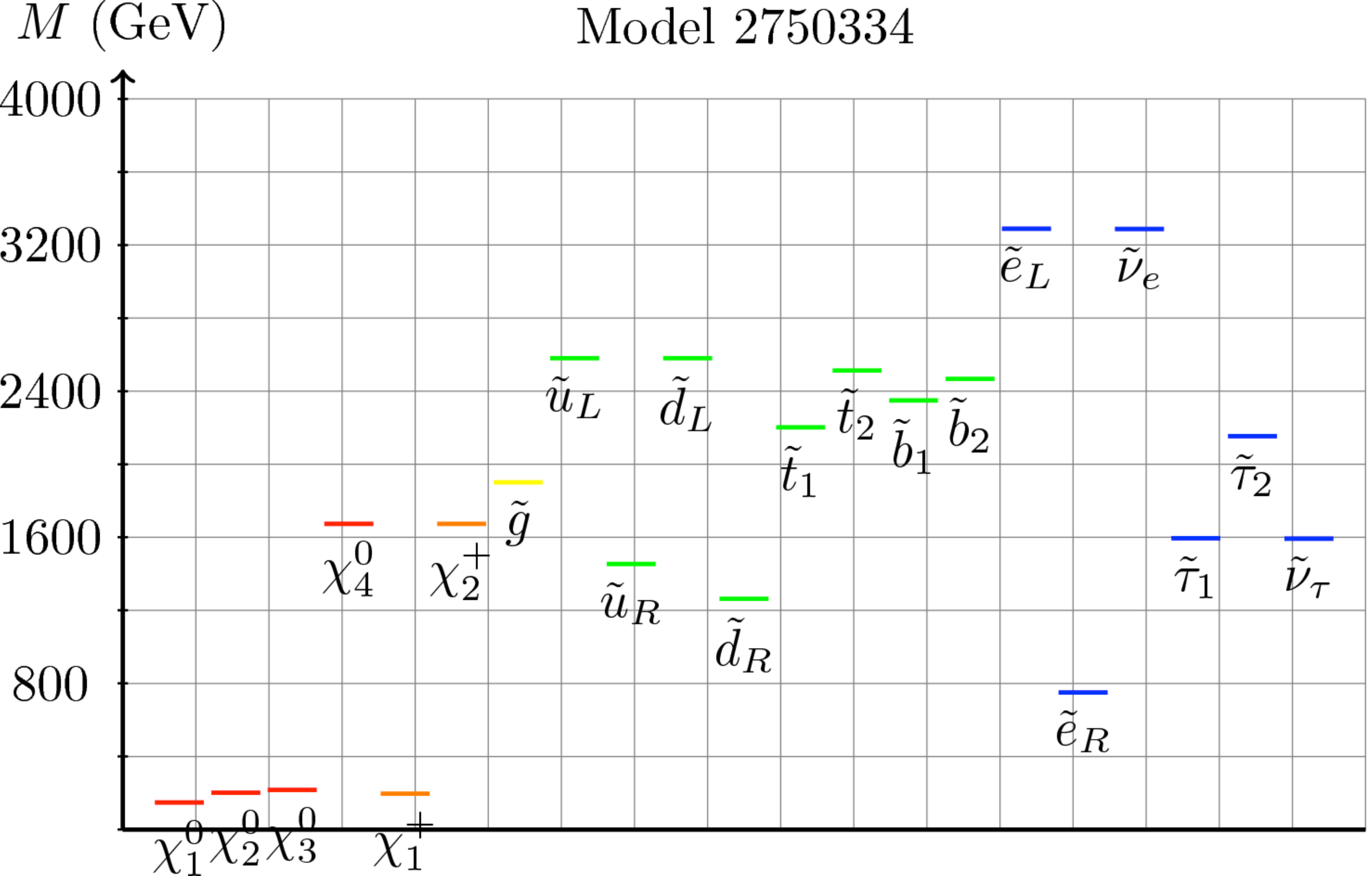}
\caption{Spectrum of the pMSSM model used for discovery scenario.}
\label{fig:story1spectrum}
\end{figure}

\paragraph{An example model}

To illustrate the potential impact of a discovery in this channel, 
we discuss a scenario based on model 2750334 of the pMSSM scan 
\cite{Cahill-Rowley:2013gca}.
The spectrum of the model is given in Fig.~\ref{fig:story1spectrum}.
Complete details of the model can be found in \cite{Cahill-Rowley:2013gca}.
This model has light neutralinos and charginos clustered around 200~GeV;
the lightest neutralino is a mixture of bino and Higgsino
(`well-tempered Bino-Higgsino'),
and consititues a viable dark matter candidate.
The lightest squark has a mass of 1.3~TeV, and evades
searches at LHC run 1.

In this model, the LHC14/300 will discover new physics in the
jets plus MET channel with high significance.
At the same time, no other signal of new physics would be observed.

The simplest phenomenological explanaion of this excess is 
production of a new colored particle, followed by decay
to jets plus a stable neutral particle accounting for
the missing energy.
Because this is the most sensitive channel for SUSY,
this is clearly the leading interpretation of the signal 
that must be further explored.
But there are also other possibilities to be considered.
For example models where all the standard model 
particles propagate in extra spatial dimensions 
(`universal extra dimensions') 
have extra-dimensional excitations for all the standard model particles
that give rise to similar singals.
This motivates both detailed study of 
the jets plus MET excess as well as searches for other 
particles predicted in these models.

Using kinematic variables such as MT2, one can get an estimate of the
mass difference between the colored particle and the stable neutral particle.

It is difficult to get additional information about the spectrum,
because the energy distributions are sensitive mainly to the difference
between the produced colored particle mass and the mass of the 
stable particle at the end of the decay.
Information about the rate is also difficult to interpret because
production is due to an unknown number of similar states,
and there may be multiple decays.
In this case, the energy distribution and rate are not a good match with what is
expected in the simplest SUSY models (which generally have addtional
degenerate squarks and/or a lighter gluino), leading one to suspect
that there may be additional colored superpartners.

In addition, there is no sign so far of the
electroweak-inos and sleptons that must be present if the signal
is SUSY. These evade searches because this part of the spectrum is highly
compressed and slepton production cross section is small.

The HL-LHC (14~TeV with $3000/$fb) extends the squark discovery reach to 
approximately 2.5 TeV. For our example model it might be possible to 
claim evidence of more then one strongly interacting state.

Lepton collider reach for new particles extends up to masses of $\sqrt{s} / 2$.
Therefore ILC with 500 GeV would be able to precisely measure the
masses and spins of the gauginos and sleptons, as well as the branching fractions in their
transitions \cite{Baer:2013vqa}. The presence of these additional expected partners, as well as their
spins, would be direct evidence for supersymmetry, which relates fermions
and bosons. This would allow us to estimate the composition of the electroweak-ino
mass eigensstates. This is an important step in connecting collider measurements with the
dark matter relic abundance (see below).

The selectron $\tilde{e}_R$ would not be directly produced at the 500 GeV ILC, 
but its existence may be inferred by a
significant enhancement of the neutralino production cross-sections due to
t-channel selectron exchange, which could be identified by its characteristic
polarisation dependence. Measuring the polarised cross-sections to few percent precision
would allow to extract the selectron $\tilde{e}_R$ mass to a few 10 GeV. In general, the t-channel   
enhancement is observable for selectron masses up to a few TeV. The other sleptons in the
example scenario would not be found at the 500 GeV ILC,
and mass limits on the lightest slepton would be at around 250~GeV.
These bounds are nominally weaker than those from the LHC, but
unlike the LHC limits, they are essentially free of loopholes from
complicated decays \cite{Berggren:2013vna}.
This would suggest that the sleptons are not important for the thermal
relic density of the LSP, and estimates of the relic abundance from
collider data would give values consistent with the observed relic
density.

At this point, it would be very clear that supersymmetry has been
established, and the sleptons are the last major missing piece of
the puzzle. An ILC upgrade, or CLIC, or a muon collider would be 
strongly motivated to search for these. In our example model,
a 3 TeV CLIC \cite{clicwp, cliccdr} would easily discover an 800 GeV $\tilde{e}_R$,
but 1.6 TeV $\tilde{\tau}_1$ and heavier sleptons would remain out of reach.

The higher mass colored superpartners can only be searched for
at a higher energy hadron machine, either a 33~TeV LHC or a
VLHC (see Fig. \ref{fig:story1simple}).


\subsection{ SUSY with a light stop }
\label{sec:lightstop}

One of the essential elements of any solution to the naturalness problem is a top-partner, which is responsible for temper the quantum corrections to the Higgs mass generated by the top quark. In SUSY, the superpartner of the top quark, stop, plays this role. Therefore, search for the stop is directly connected to the test of naturalness.

\begin{figure}[h]
\begin{center}
\subfigure{
\includegraphics[width=209pt]{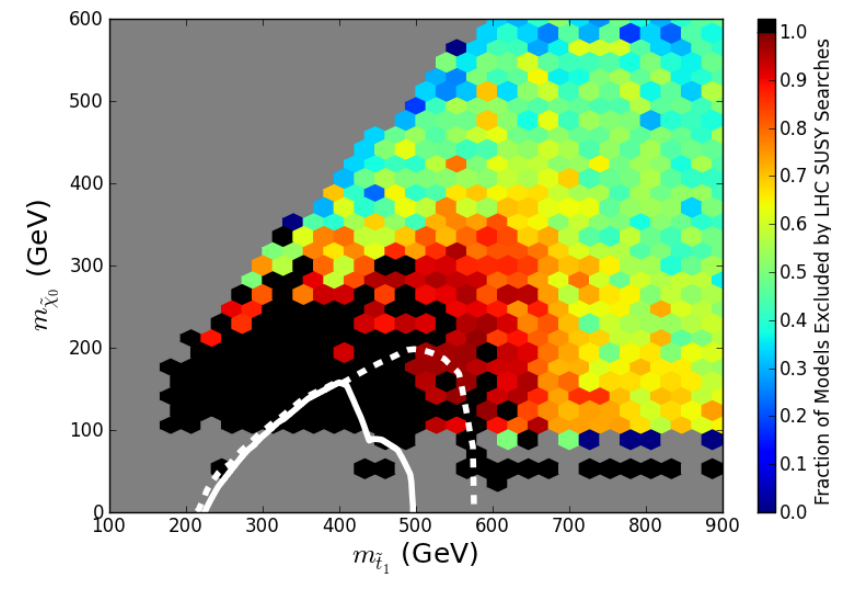} }
\end{center}
\caption[]{ Limits on the pMSSM parameter space \cite{Cahill-Rowley:2013yla} from current LHC stop searches. }
\label{fig:LHC_stop_8TeV_pMSSM}
\end{figure}
\begin{figure}[h]
\subfigure{
\includegraphics[width=229pt]{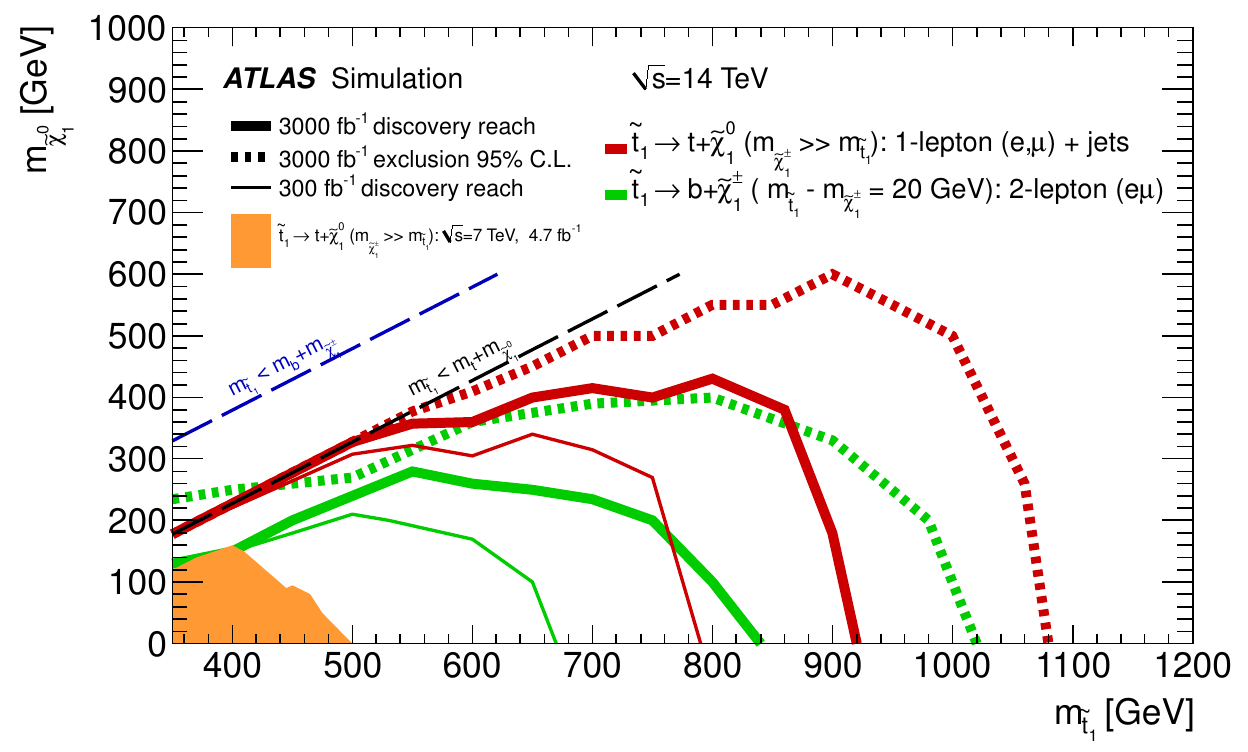} }
\subfigure{
\includegraphics[width=209pt]{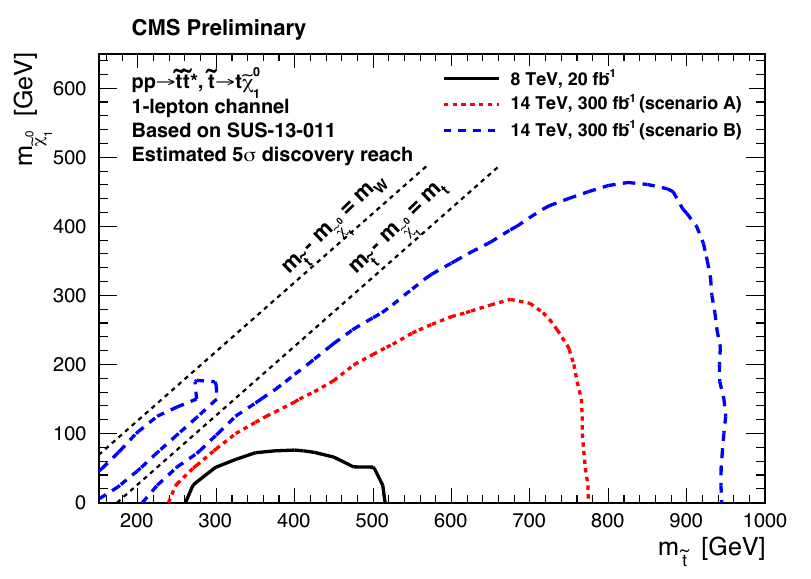} }
\caption[]{ ALTAS \cite{atlaswp} and CMS \cite{cmswp} projections of reaches for stop in direct pair production LHC Run 2 and HL-LHC.}
\label{fig:LHC_stop}
\end{figure}

The simplest stop decay channel is $\tilde{t} \to t + $ LSP,  giving rise to signature $t\bar{t}+\missET$. LHC run 1 has made significant progress in exploring the relevant parameter region of light stop. At the same time, there are still large portion of model space left unconstrained, as demonstrated in the pMSSM scan \cite{Cahill-Rowley:2013yla} shown in Fig.~\ref{fig:LHC_stop_8TeV_pMSSM}.
This
channel is going to be one of the foci of the LHC Run 2 program.  
The reach is estimated in ATLAS and CMS white papers \cite{atlaswp,cmswp}, as shown in Fig.~\ref{fig:LHC_stop}

If an excess is observed in this channel in LHC Run 2,  it may be evidence for
SUSY with a light stop quark. During  the subsequent run and the operation of  HL-LHC, the significance of this excess will grow and reach discovery level.  This would be a major discovery which marks the beginning of an new era. It gives solid evidence to naturalness, and hints at many new particles to be discovered in the coming decades.
In addition to the stop discovery, the presence of $t \bar t + $MET signal implies the presence of a stable neutral particle. This would the first collider signal for dark matter as well! 

The immediate goal after this discovery would be to measure the properties of the new particle, and check it is consistent with that of the stop. The most important properties include its mass and couplings. With some simplifying  assumptions which can be checked later, we can get an initial  estimate of the stop mass just from the measured production cross section. 

The initial discovery would also tell us that stop has a significant coupling to top and the LSP so that $\tilde{t} \to t + $ LSP is a main decay mode. Indeed, this would also be one of  simplifying assumption which allows us to estimate the stop mass using production rate. At the same time, the stop can have a rich collection of decay channels to  charginos and neutralinos. Measuring them will paint a full picture of stop couplings.  Many of these channels will be subdominant, and discovering them require large statistics. HL-LHC is indispensable in accomplishing this task.

To confirm the initial estimates of the stop properties, more detailed measurements of properties need to be carried out. Indeed, there can be other new physics scenarios, for example the Universal Extra Dimension (UED), which can have signals very similar to SUSY. Therefore, during the period after discovery, there will be competing interpretations. To distinguish them, model independent measurements of spin and mass are necessary. Such measurements are difficult, since we can not fully reconstruct the momentum of LSPs. Precise measurement of subtle features of kinematical distributions will be necessary. High statistics at the level of HL-LHC will great enhance our capability of carrying out these measurements.

\begin{figure}[h!]
\subfigure{
\includegraphics{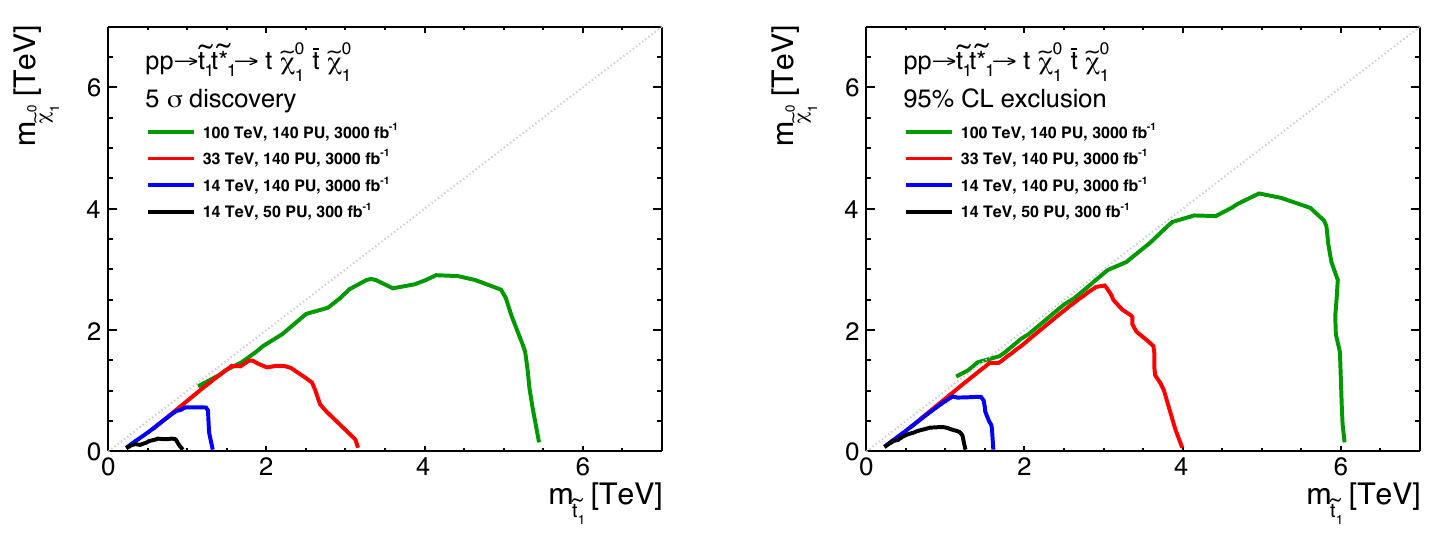} }
\caption[]{ Reaches for stop-neutralino simplified model using the single lepton channel \cite{Cohen:2013zla}.  The left [right] panel shows discovery reach [$95\%$ CL exclusion]. }
\label{fig:vlhc_stop}
\end{figure}

The most interesting coupling of stop is probably with the Higgs boson. Confirming its consistency with SUSY prediction would be a directly proof of the stop's crucial role in solving the fine-tuning problem. To directly probe this coupling, one would have to observe the $pp \to \tilde{t} \tilde{t}^* h$ process. However, this process has an extremely low rate at 14 TeV LHC. It can only be reached at the VLHC with $E_{\rm CM} = 100$ TeV. At the same time, a robust test of the divergence cancellation can be performed by testing the ``SUSY-Yukawa sum rule"~\cite{Blanke:2010cm}, a relation among stop and sbottom masses and mixing angles which is tightly connected with this cancellation. A meaningful test of the sum rule requires precise measurements of the masses and mixing angles, which could be performed at a future lepton collider.

Higher energy proton colliders are necessary to significantly extend the reach of stops. This can be clearly seen in Fig.~\ref{fig:vlhc_stop}, which presents the reach on a stop-neutralino simplified model using the single lepton channel \cite{Cohen:2013zla}. In particular, an 100 TeV hadron collider can probe stop up to at least 6 TeV. 
The improvement in stop reach significantly enhanced our capability of testing the naturalness of the Higgs mass, as a significant part of the  fine-tuning is proportional to $m_{\tilde{t}_1}^2$. 

We note that, from Fig.~\ref{fig:LHC_stop_8TeV_pMSSM}, very light stop below 250(500) GeV are still not ruled out. They are within the reach of 500(1000) GeV ILC. Heavier stop would be reachable at CLIC and/or high energy muon collider. If stop can be produced at lepton collider, their properties can be studied in great detail. 

The presence of a light stop implies the presence of a full set of superpartners not far away from the TeV scale. Indeed, naturalness also implies certain electroweak-inos, in particular the Higgsino,  should be within a couple hundred GeV. This would be case in which a high energy lepton collider can play a crucial role in putting together a full picture of the new physics immediately above the weak scale. As discussed above, model independent measurements of spin, mass and couplings of superpartners are challenging at pp-colliders and require at least high luminosity. At the same time, high energy lepton colliders can provide a much cleaner environment. Moreover, it is much easier to reconstruct the kinematics.  In addition, it could also answer definitively another question of equal importance: the identity of the LSP dark matter.  

\begin{figure}[h!]
\subfigure{
\includegraphics[scale=0.5]{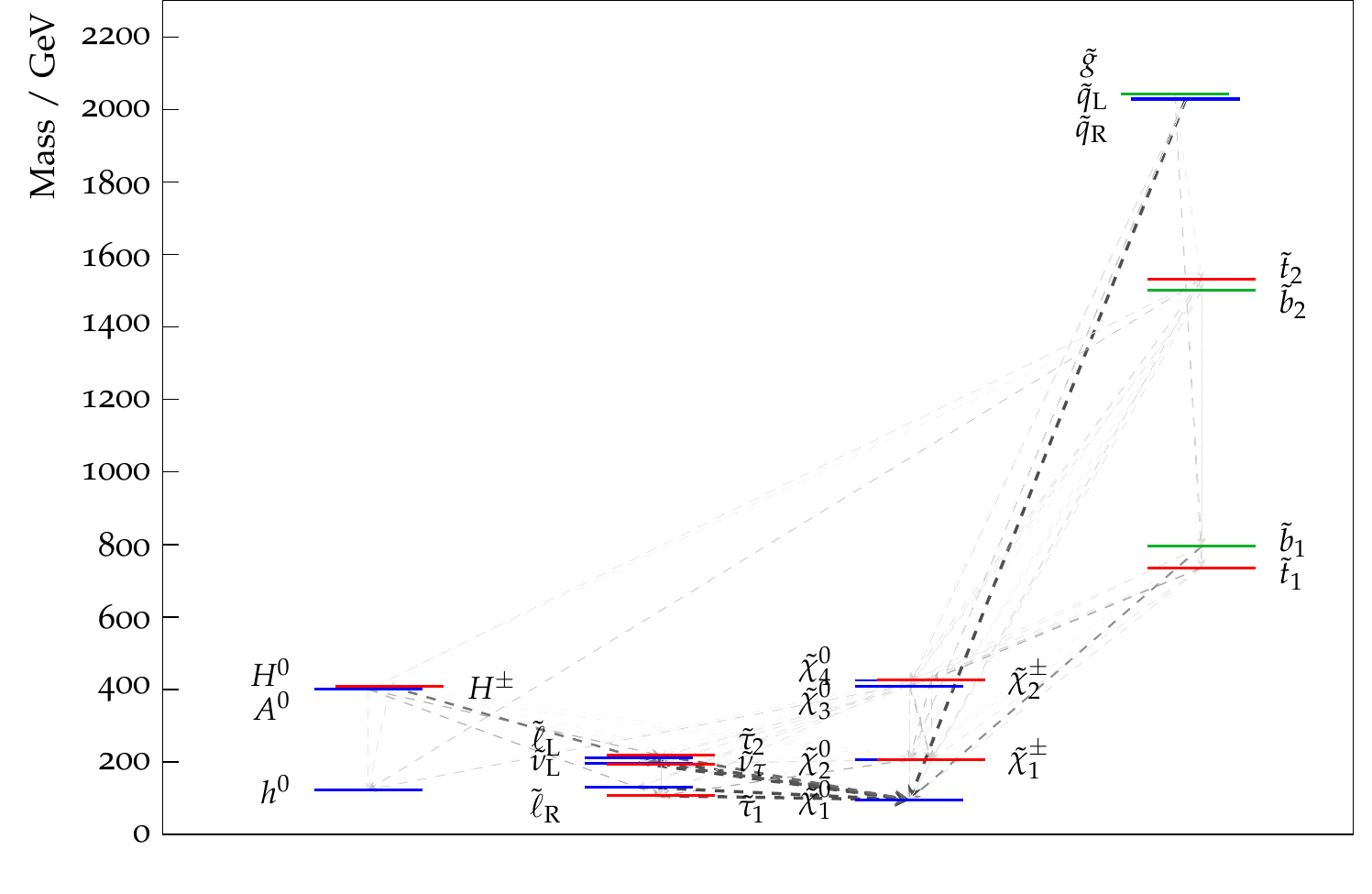} }
\subfigure{
\includegraphics[scale=1.1]{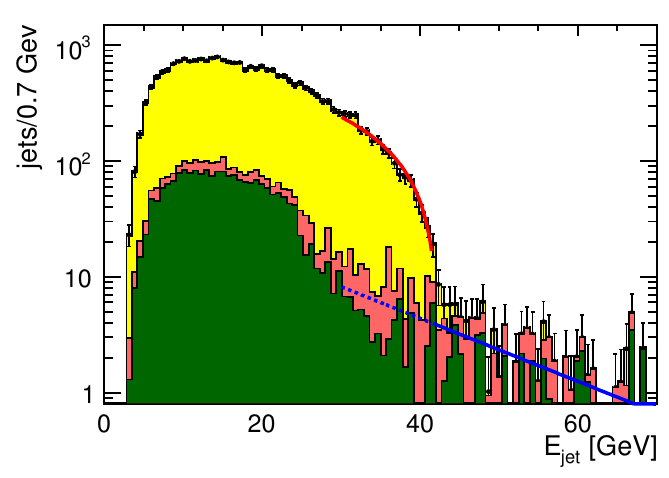} }
\caption[]{ Left panel: spectrum of the benchmark model in the $\tilde{\tau}$-coannihilation region. Right: $\tilde{\tau}_1$ spectrum (yellow) and background (SM: red, other SUSY: green), with end-point fit. }
\label{fig:ILC_LHC_staubenchmark}
\end{figure}

A great example of such scenarios is explored in the joint
ILC-LHC study of the stau co-annihilation model \cite{Berggren:2013hda}. In addition to
low fine-tuning, the neutralino in that model accounts for the 
observed amount of the Dark Matter in the Universe. The mass
spectrum and allowed transitions in this model are shown in the left panel of Fig.~\ref{fig:ILC_LHC_staubenchmark}.  

The discovery sensitivity of this model has been studied for several LHC analyses: two analyses heading for stop and sbottom discovery, and one tuned to the discovery of the electroweak particles. All analyses are already limited by the systematic uncertainties at the LHC with 14 TeV and 300 fb$^{-1}$. We nevertheless believe that 10 times higher luminosity will also help to reduce the systematic uncertainties on the number of background events, as these are partly caused by low statistics in the control regions used to determine the background from data.
The searches for the stop are based on a selection with one or zero leptons and tagging of b quarks. The existence of a stop will lead to an excess in data with a significance of 3.5~$\sigma$, if the systematic uncertainty can be controlled up to 15$\%$. A 5-$\sigma$ discovery could only be obtained, if the systematic uncertainties were pushed below 5\%, which we consider too optimistic. 
The electroweak particles are searched for with an analysis based on a same-sign di-lepton selection. An excess of 3~$\sigma$ could be achieved here, if the systematic uncertainty drops to 15\%, but we expect that this analysis will have a larger uncertainty due to the large background of di-boson production, which is less well known.

At the 500 GeV ILC sleptons and lighter gauginos are accessible,
and their mass and quantum numbers will be measured with high precision. In particular, the $\tilde{\tau}_1$ mass could be measured to $0.2 \%$ and $\tilde{\tau}_2$ mass to $3 \%$ \cite{Baer:2013vqa}. The production cross section can be determined to $4 \%$, and the polarization of $\tau$ leptons could be measured to $< 10\%$. 
By measuring tau polarization one can measure higgsino fraction of the
lightest neutralino. These measurements in the stau sector are of particular importance  to predict the relic density in this scenario. In a very similar scenario  it has been shown that within the MSSM-18, the precision of the ILC  measurements allows a prediction of the relic density at a similar level  as today's determination based on Planck data.  The masses of the third generation squarks               enter through loop-effects in the gaugino sector. Exploiting the full tool-kit  of the ILC, including data-taking at the production thresholds,                          
the achievable precision on the gaugino observables allows inference of the third  generation squark masses at the level of 25$\%$ to 50$\%$~\cite{Bharucha:2012ya}.  This provides valuable first information on whether the stops and sbottoms  could be accessible at a 1 TeV ILC upgrade or at a 3 TeV CLIC, and give hints to HL-LHC data analyses.  At the 1 TeV ILC, the heavier gauginos and heavy higgs bosons also become  accessible for direct production.Beyond the ILC at 1 TeV, CLIC would have access to direct pair production of the lighter stops and sbottoms and allow precision characterization  of these particles which are so closely related to naturalness.
 The plethora of precision measurements at $e^+e^-$ colliders will allow for precise determination of SUSY model parameters, and will help to confirm or rule out different proposed SUSY breaking mechanisms.

\subsection{Discovery in Leptons+MET}

In many scenarios of supersymmetry breaking, the color-neutral
superpartners such as electroweakinos (gauginos, higgsinos) and sleptons can be significantly
lighter than the colored states (gluino and squarks). In this case,
the production of electroweakinos may be the initial supersymmetry
discovery mode.  The typical signature involves production of $W$,
$Z$, and $h$-bosons when the electroweakinos decay to the stable LSP, giving
signatures with charged leptons (from $W$- and $Z$-boson decays),
b-jets (from $h$ decay) and
missing transverse momentum (from the invisible LSPs).

There are persuasive arguments for why some of the charginos and
neutralinos could be light \cite{Baer:2013faa,Baer:2013ava}.
For example, while the fine-tuning
arguments about the upper limits on masses of stop and gluino
 can be relaxed in NMSSM, it is much harder
to avoid the requirement that $\mu$ is small. Therefore, an NMSSM
reality with only light states being charged and neutral higgsinos
can still be ``natural''.

\paragraph{LHC sensitivity.}

The LHC run 2 will greatly extend the reach in searches for
superpartners without strong interactions.

\begin{figure}
\centering
\includegraphics[width=0.45\linewidth]{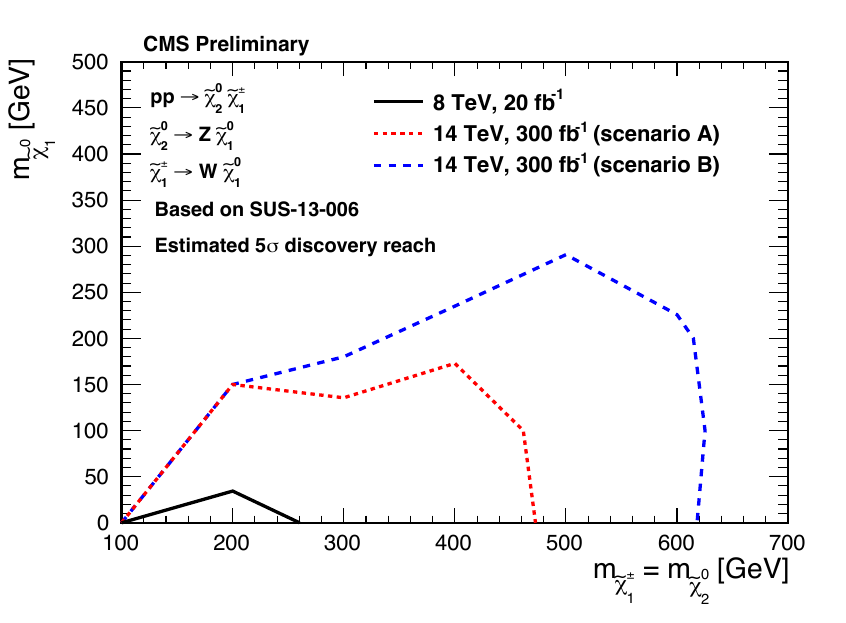}
\includegraphics[width=0.45\linewidth]{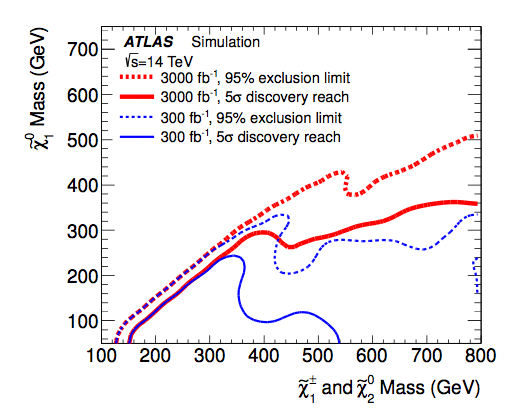}
\caption{Estimated reach for chargino-neutralino production
followed by $\chi_1^\pm \to W^\pm \chi_1^0$
and $\chi_2^0 \to Z \chi_1^0$ with 100$\%$ branching ratio.}
\label{fig:story1ewinoLHCrun2CMS}
\end{figure}

The reach for $\chi_1^\pm \chi_2^0 \to WZ + MET$
has been estimated in the ATLAS and CMS WhitePapers \cite{atlaswp,cmswp},
and has a reach up to $m_{\chi^\pm_1} = 500$--$600$~GeV
(see Fig.~\ref{fig:story1ewinoLHCrun2CMS}).

\begin{figure}
\centering
\includegraphics[width=0.5\linewidth]{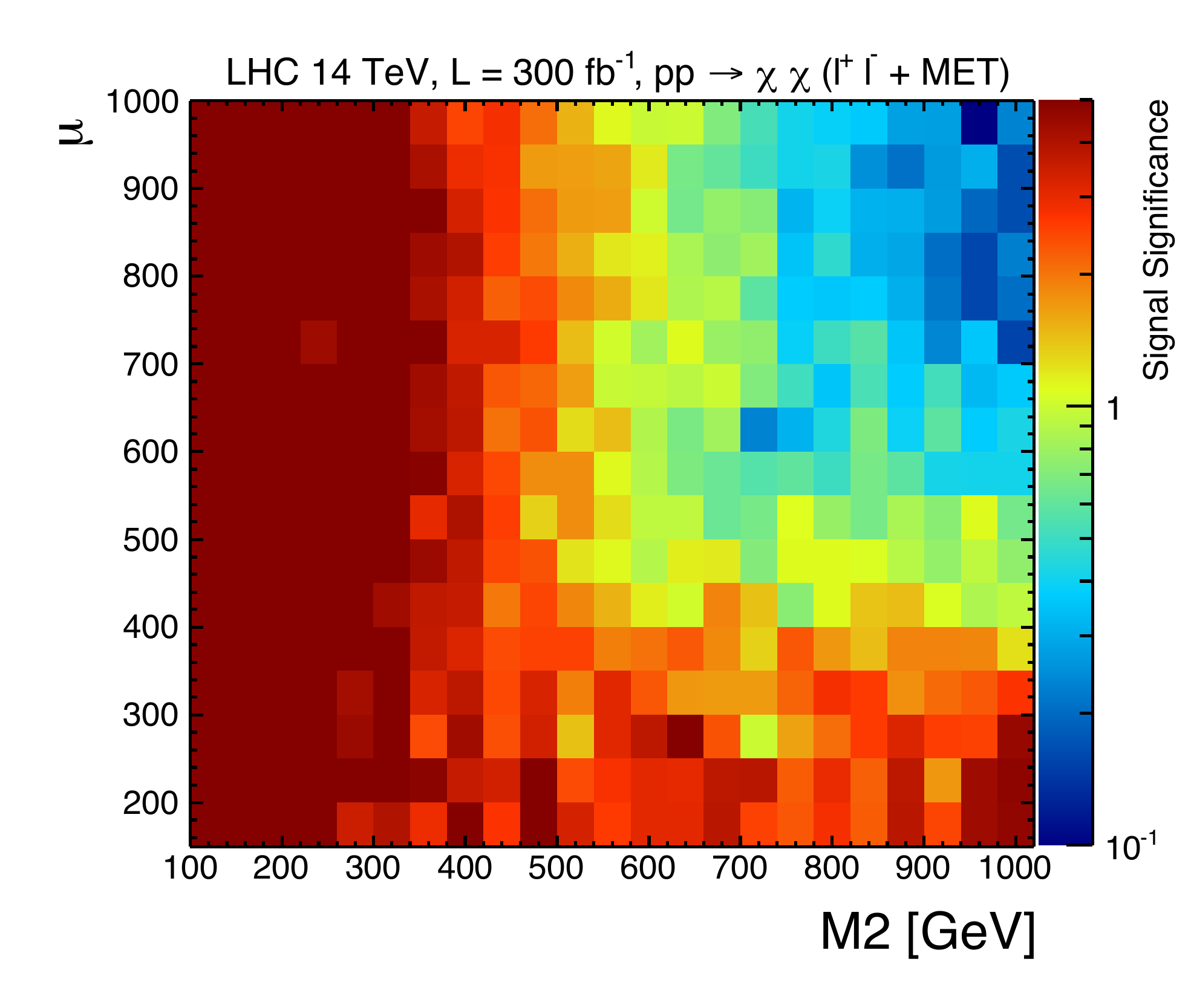}
\caption{Estimated reach of LHC run 2 for chargino production
followed by $\chi^\pm \to W^\pm \chi^0$, 
assuming Bino LSP. }
\label{fig:story1ewinoLHCrun2}
\end{figure}

In realistic models with typical SUSY parameters, the leading
channels resulting in production of detectable final states
involve heavier gauginos production.
Important cases include same sign leptons from $W^{\pm}W^{\pm} + MET$ 
from  $\chi_2^\pm\chi_4^0$
production, followed by $\chi_2^\pm \rightarrow W^\pm \chi_{1,2}^0$
and $\chi_4^0 \rightarrow W^\pm\chi_1^\mp$
 \cite{Baer:2013yha},
opposite sign leptons from $W^+W^- + MET$ 
(see Fig.~\ref{fig:story1ewinoLHCrun2})
from chargino
pair production and subsequent $\chi_1^\pm \to W^\pm \chi_1^0$,
and $W^\pm h + MET$ from chargino-neutralino production
followed by $\chi_{2,3}^0 \to \chi^{0}_1 h$ \cite{Berggren:2013bua}.

\begin{figure}
\centering
\includegraphics[width=0.8\linewidth]{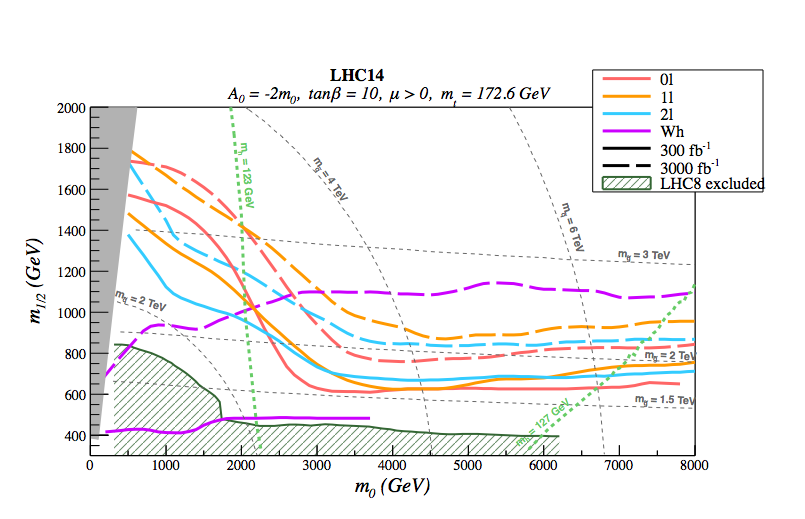}
\caption{Estimated reach of LHC for 300/fb and 3000/fb for mSUGRA model.}
\label{fig:m0mhalf}
\end{figure}

In fact, the channel $W^\pm h + MET \to \ell b\bar{b} + MET$ plays 
increasingly important role \cite{Baer:2013fva} as the integrated 
luminosity of the LHC increases (see Fig. \ref{fig:m0mhalf}).

In the case of very small splitting between gauginos, the final state would be
essentially invisible, but analysis of events in which a jet or photon
recoils from the initial state (see Section~\ref{sec:dmatter}) would
be sensitive at high integrated luminosities. 

Recently, studies have shown that vector-boson-fusion production of
winos, with a final state of two forward jets and missing
transverse energy could be sensitive to models with small splittings
between the electroweakinos with masses of a few hundred
GeV~\cite{ Delannoy:2013ata, Delannoy:2013dla, Dutta:2012xe}.

LHC sensitivity to the EWKino states greatly increases with
integrated luminosity, owing to their relatively low masses and very low production
cross-sections. 

\paragraph{Beyond Run 2 of the LHC}

An excess at the LHC could be studied in detail at the 
HL-LHC, revealing the mass splittings via the dilepton mass edges.
Together with the cross sections and assuming high  higgsino fraction,
a rough estimate of the absolute masses might be possible.

%

The HL-LHC would also extend the sensitivity to colored states from 
about 2 to 2.5 TeV (see Section \ref{sec:story1}), but to make significant gains
in mass reach a higher energy hadron collider will be required. A 33 (100) TeV collider will
be able to push the SUSY squark/gluino discovery reach to 7 (15) TeV \cite{squark-gluino-scan}.


\paragraph{Studies at the ILC}

   At ILC the reaction $e^+e^-\to\tilde{\chi}_1^+\tilde{\chi}_1^-$ 
   should be easily seen above background provided 
   that $\sqrt{s}>2m_{\tilde{\chi}_1^+}$.
   Since the beam energy is precisely known, this allows for extraction 
   of the masses $m_{\tilde{\chi}_1^+}$ 
   and $m_{\tilde{\chi}_1}$ to high precision.
   In addition, mixed production $e^+e^-\to \tilde{\chi}_1^\pm\tilde{\chi}_2^\mp$ 
   can allow access to the heavier
   $\tilde{\chi}_2^\pm$ chargino state even if $\sqrt{s}<2m_{\tilde{\chi}_2^+}$. 
   Also, the ten $\tilde{\chi}_j\tilde{\chi}_{j'}$ 
   reactions should be easy to spot
   provided that $\sqrt{s}>m_{\tilde{\chi}_j}+m_{\tilde{\chi}_{j'}}$. 
   Threshold scans and beam polarization will help
    to differentiate these reactions and to discriminate the Higgsino/gaugino components of each
     $\tilde{\chi}_j$ state which is accessible.
   Even in the case where $e^+e^-\to\tilde{\chi}_1\tilde{\chi}_1$, 
   radiation of initial state photons can be used to
   tag this reaction against $e^+e^-\to\nu\bar{\nu}\gamma$ background.

\begin{figure}
   \centering
   \includegraphics[width=0.6\textwidth]{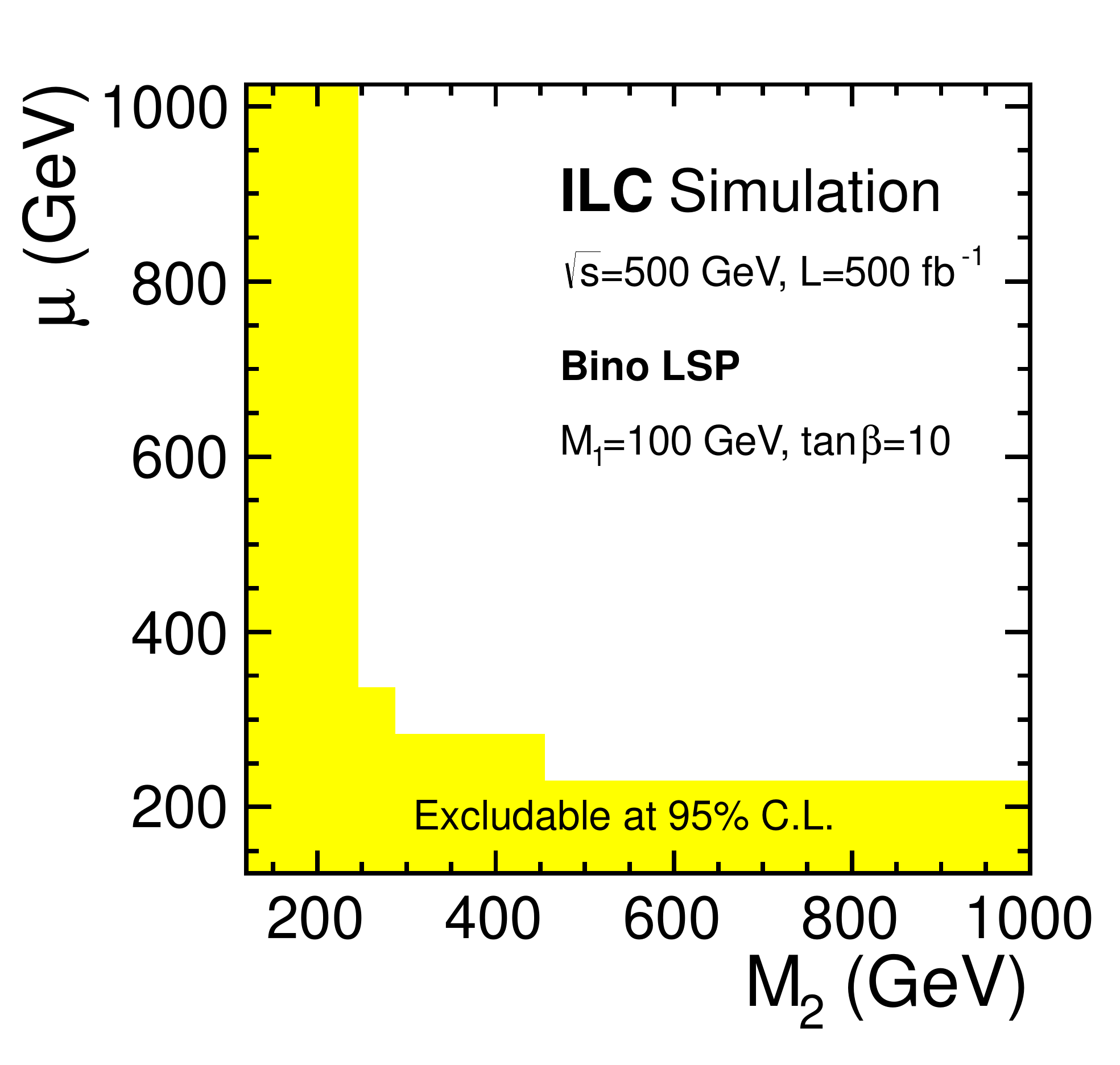}
   \caption{Exclusion reach for electroweakinos at the ILC
     studied with fast simulation in the case of Bino LSP.
     The shaded region (yellow) is the expected exclusion reach at 95\% confidence level.
   }
   \label{fig:ewkino}
   \end{figure}

   A joint LHC/ILC study demonstrates the complementarity of the two 
   machines~\cite{Berggren:2013bua, Baer:2013vqa}.
   In this study, a parameter scan is performed in the context of the MSSM
   across the three-dimensional parameter space in the $M_1$, $M_2$, and $\mu$ variables,
   fixing the Higgs mass to the observed value and $\tan\beta$.
   The expected exclusion reach is shown in Fig.~\ref{fig:ewkino} for the case of the ILC
   with $\sqrt{s}=500$~GeV.

   In the case of Bino LSP, the ILC is essentially sensitive to
   $M_2$ and $\mu$ of up to 250~GeV, which is half the center-of-mass energy;
   the NLSP is accessible under this condition and its decays can be detected.
   The LHC is expected to be able to cover larger $M_2$ values,
   which provide the large mass gaps allowing the detection of the electroweakinos.
   The case of LSP pair production without NLSP can be searched
   at the ILC with a single photon signature from the ISR
   and is covered in Sec.~\ref{sec:dmatter}.

   In the case of Higgsino LSP with $\mu=100$~GeV,
   the LSP and NLSP are accessible regardless of $M_1$ and $M_2$.
   The detection of Higgsino decays, which are typically soft,
   can be used to exclude the entire parameter space with a light Higgsino,
   which is a capability unique to the ILC.

   In addition, the ILC can separate the chargino and neutralino contributions in many cases,
   providing cross-section and mass measurements at the $O(1)$\%
    level~\cite{Suehara:2009bj} 

\subsection{R-parity violating SUSY}
Naturalness suggests that superpartners must be sufficiently light to
produced at the TeV scale.
In particular, it suggests that $m_{\tilde t} < 500$~GeV
and $m_{\tilde{g}} < 1$~TeV,
which larger values requiring tuning to explain the small
observed mass of the Higgs boson.
These `natural' values of the gluino and stop masses are ruled
out in the simplest models by searches at LHC run 1,
motivating the study of SUSY models where the bounds are weaker.

One way that natural SUSY could have escaped detection so far is
that $R$-parity is not conserved.
$R$-parity is a discrete symmetry that guarantees that the
superpartners are produced in pairs, and that the LSP is stable.
Giving up $R$-parity means that missing energy is no longer a generic 
signature of SUSY at colliders.
Dark matter would have to be explained by a particle other than
the LSP.

There are a large number of possible $R$-parity violating (RPV) couplings,
each of which have an extremely rich phenomenology.
These couplings violate baryon and/or lepton number,
and necessarily have a non-trivial flavor structure.
For this reason, there are many constraints on these couplings
coming from flavor physics and baryon and lepton-violating
processes.
However, these constraints generally depend on products of different
RPV couplings, and individual couplings can be large enough to
be relevant for collider physics.
(For a review, see \cite{Barbier:2004ez}.)

In this section, we consider three different discover scenarios.
Two are based on the operators $L_3 Q_3 D_3$ are $U_2 D_1 D_2$,
where the subscripts denote the quark or lepton generation.
These were chosen because they make searches for stops quite challenging.
However, we will see that specialized searches are quite sensitive.
The third scenario explores RPV bilenear terms in the superpotential and 
soft lagrangian, which together allow higgsino decay into W and lepton.
The later scenario, which we will refer to as $bRPV$, 
may be related to the origin of neutrino masses \cite{Porod:2000hv}.

\begin{figure}
\includegraphics[width=0.45\linewidth]{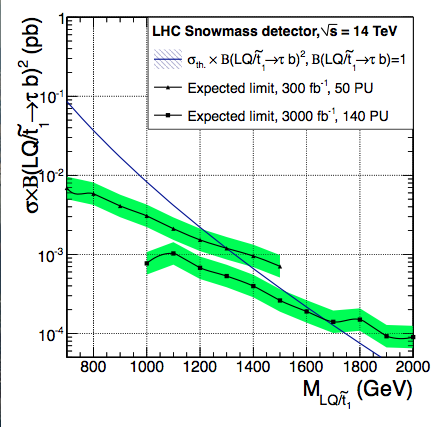}
\includegraphics[width=0.45\linewidth]{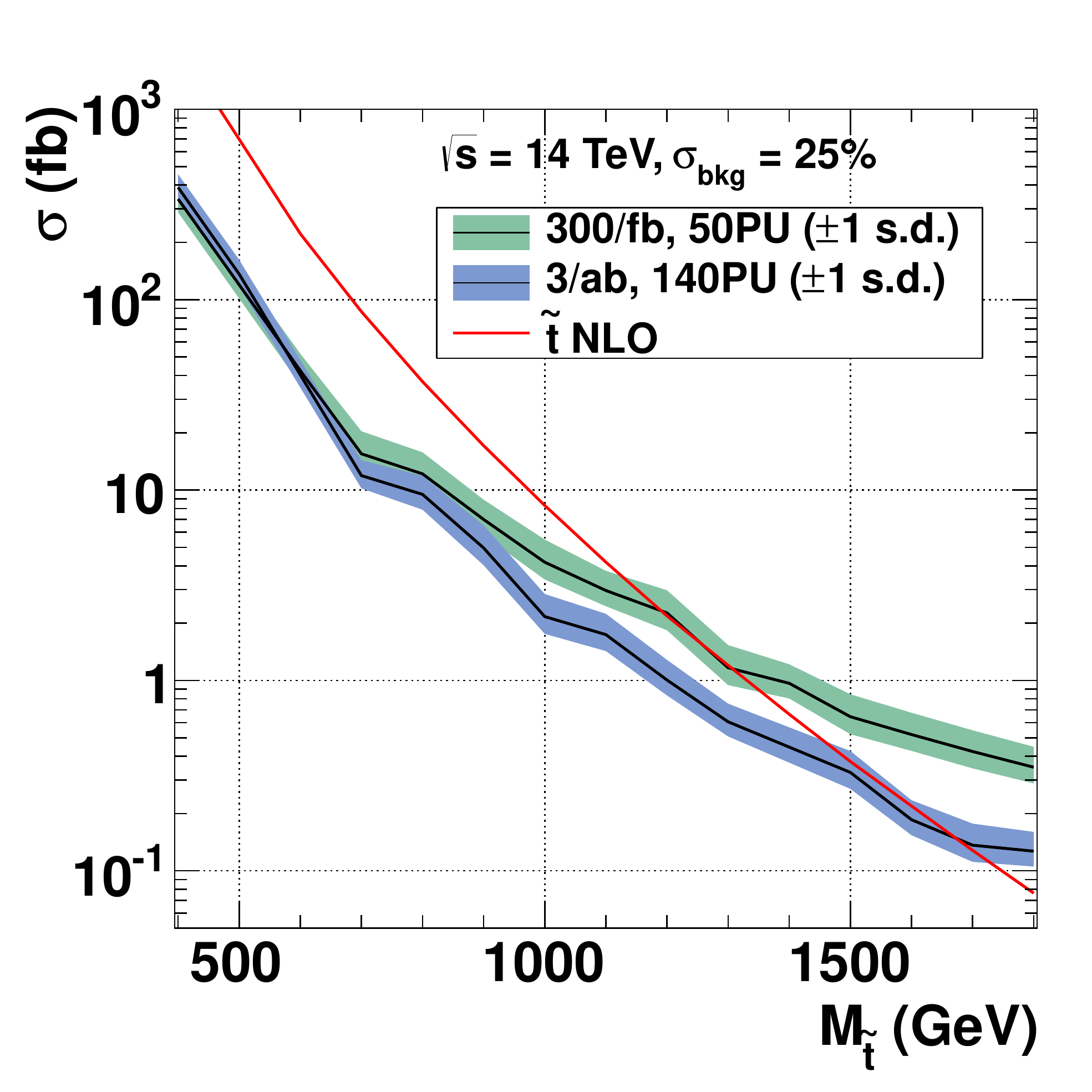}
\caption{Left: limits on stops decaying to $\tau b$ via the $L_3 Q_3 D_3$
operator. Right: limits on stop decaying to $t \chi^0 \to t (jjj)$
via the $U_2 D_1 D_2$ operator}
\label{fig:story4run2reach}
\end{figure}

\paragraph{$L_3 Q_3 D_3$:}
This coupling allows the decay $\tilde{t} \to \tau b$,
so the stop appears as a third-generation leptoquark resonance.
LHC run 2 can probe this mode for stop masses up to 
$1.3$~TeV \cite{RPVwhitepaper} (see Fig.~\ref{fig:story4run2reach}).

RPV SUSY will be the leading interpretation of such a signal.
Another possible interpretation is double higgs production in an extended
Higgs sector followed by $hh \rightarrow (bb)(\tau\tau)$.
This is straightforward to eliminate due to the different kinematics
as well as other decay modes of the Higgs.
Another interpretation is a spin-1 third-generation leptoquark.
This can be distinguished using rate information if the mass of the
produced particle is known, but this is difficult to determine because
of multiple sources of missing energy in the event.

The SUSY interpretation of this signal can be probed in a number of ways.
The sbottom mass is different from the stop mass in general,
but the masses are similar in many models.
This motivates searches for sbottoms, which decay via
$\tilde{b} \to b \nu$ or $t\tau$.
These can be searched for with good sensitivity at the LHC 
\cite{Chatrchyan:2012st}.
Also, the electroweak-inos are expected to be lighter than the stops.
These are expected to have the decays
$\chi^\pm \to tb\nu$, $bb\tau$, or $\chi^0 W^\pm$,
and $\chi^0 \to tb\tau$ or $bb\nu$.
These can be searched for both in direct production and from
stop decays $\tilde{t} \to t \chi^0$ or $b\chi^\pm$.
Another plausible signal is gluino pair production followed by
$\tilde{g} \to t \tilde{t}$ or $b\tilde{b}$, followed by any of the
decays discussed above.
All of these possibilities can be extensively probed at the HL-LHC.

Precision flavor physics also gives a complementary probe of other $LQD$ operators.
Mixing in the $B$ meson system probes $L_3 Q_3 D_1$,
and $B \to X_s \nu\nu$ probes $L_i Q_3 D_2$ and $L_i Q_2 D_3$ for $i = 1, 2, 3$.

The HL-LHC can increase the signifance to $5\sigma$ for the 
1 TeV stops in our scenario.
In the $\mu\tau_h$ channel alone, there are 270 events with a significance of 
$4.4 \sigma$.  
The addition of the $e\tau_h$ channels would give enough to claim discovery.

\paragraph{$U_2 D_1 D_2$:}
In this case, stops can decay via $\tilde{t} \to t \chi^0$
followed by $\chi^0 \to jjj$ from the RPV coupling.
If the stops have a mass of 900~GeV, a search for a lepton 
(from the top decay) with multiple additional 
jets would be sensitive to these at LHC run 2, yielding 9 events above background
and a significance 
of $3.4 \sigma$ (estimated with 50PU and assuming $m_{\tilde{t}} : m_{\chi^0} = 2:1$
\cite{RPVwhitepaper}.

At the same mass, LHC 14 with $3000/$fb and 140PU, would yield 15 events 
(with expected similar background) and a significance of $5.6 \sigma$.
The number of signal events has been suppressed to increase significance,
but looser cuts can give larger samples. 
These may be important to address the question of whether the excess is due to a
misunderstanding of QCD tails.  
For example, the possibility of reconstructing the boosted $\chi^0$ could 
help to resolve this question.

There are a number of associated channels that can be studied in the SUSY
interpretation of this signal.
One is gluino pair production, followed by the decay $\tilde{g} \to jjj$
via virtual squarks, or $\tilde{g} \to t\tilde{t}$ followed by $\tilde{t} \to tjjj$.

Future colliders will also be able to probe this scenario.
Lepton colliders
will be able to probe the electroweak-ino sector essentially without
loopholes for chargino and neutralino masses up to half the center of mass
energy.
In this scenario, the 500~GeV ILC will probe a significant region
of the parameter space, higher energy lepton colliders such as
1~TeV ILC, CLIC, or muon colliders will further extend the reach.
The remaining colored superpartners can be explored only at LHC33 or a VLHC.

\paragraph{$bRPV$:}

\begin{figure}
\centering
\includegraphics[width=0.8\linewidth]{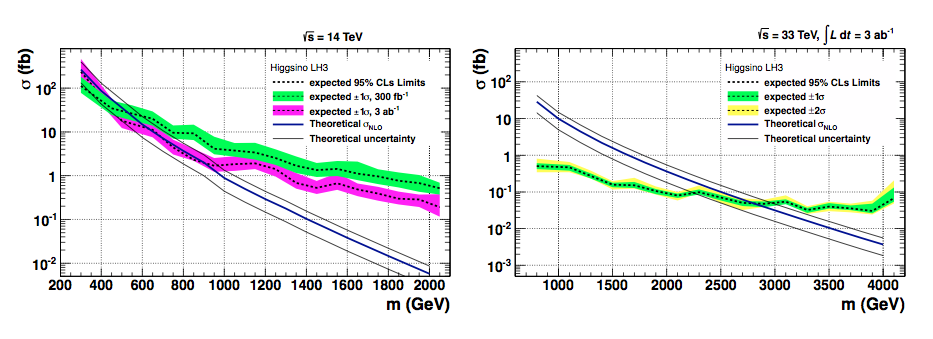}
\caption{Projected LHC sensitivity to the bi-linear RPV SUSY at 14 TeV (left) and 33 TeV (right).}
\label{fig:rpv_LHC_lh}
\end{figure}

At the LHC, higgsino pair production with subsequent decays $\tilde{H} \longrightarrow W\tau$ 
would give rise to excess in multi-lepton production (see left frame of Fig. \ref{fig:rpv_LHC_lh}). 

As an example, let's assume that the higgsino mass is 210 GeV. LHC will establish the excess
with significance above 5 $\sigma$, but the interpretation of the excess at the 14 TeV LHC would 
be highly ambiguous. 

\begin{figure}
\centering
\includegraphics[width=0.8\linewidth]{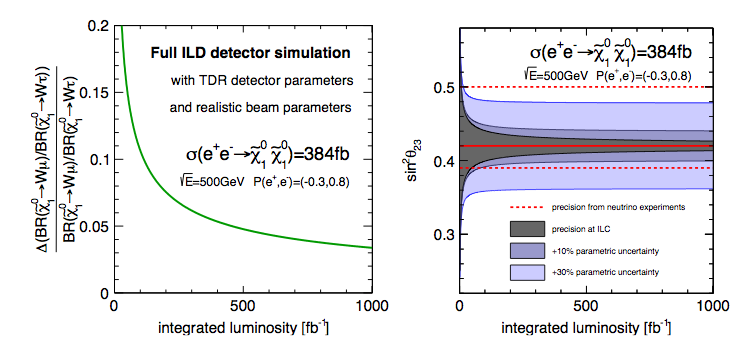}
\caption{Left: Experimental precision on the ratio of $W\mu$ and $W\tau$ branching fractions. Right: Derived 
uncertainty on the $\Theta_{23}$ neutrino mixing angle. }
\label{fig:rpv_ILC_lh}
\end{figure}

At the ILC \cite{List:2013dga,Baer:2013vqa}, however, one would not
 only establish the RPV nature of the 
higgsino, but also detect companion decay into $W\mu$. This would allow to make a very powerful connection
to the neutrino physics: if the R-parity violation is the origin of the neutrino mass, one predicts the 
value of the mixing angle $\Theta_{23}$ (see Fig. \ref{fig:rpv_ILC_lh}).

\paragraph{Further studies}

It's important to note that the sensitivity to such signatures at hadron colliders dramatically improves
with energy. A 33 TeV proton collider will be sensitive to LH scenario up to the 2.5 TeV higgsino masses
(see right frame of Fig. \ref{fig:rpv_LHC_lh}). 

At ILC, RPV decays of SUSY particles mediated either by bilinear or any of the trilinear couplings
can be searched for in a loophole-free and model-independent way
for every sparticle type~\cite{Berggren:2013vna}, as long as its mass is below $\sqrt{s}/2$.

\subsection{Long-lived Heavy Particles}
Massive long-lived particles are predicted by many extensions of the SM 
\cite{Fairbairn:2006gg}. 
In SUSY with $R$-parity considervation the 
next-to lightest SUSY particle (NLSP) may be long-lived due to either approximate mass 
degeneracy in the spectrum
or suppression of the coupling to the LSP due to the higher scale of SUSY breaking.
Small $R$-parity breaking can also result in a long-lived lightest SUSY particle (LSP).
Typical cases include a long-lived stau, chargino, gluino, stop, or sbottom. 
Beyond SUSY, any new fermion added to the standard model must decay through higher-dimension
operators, and is therefore long-lived.
Long-lived particles are therefore a discovery mode for many kinds of new physics.

If the long-lived particle is colored, it will hadronize on the QCD distance scale, 
leading to a color-neutral `$R$-hadron' being observed at longer distances in the detector. 
This $R$-hadron can be either electrically neutral or charged.
The quantum numbers of the lightest $R$-hadron
are uncertain due to unknown non-perturbative QCD dynamics.
$R$-hadrons can undergo nuclear interactions in the detector, changing their charge 
as they traverse the detector. 
%
%
If the long-lived particles are color-neutral and charged, they will lose energy
mostly through ionization and are typically not stopped in the detector.
The speed of such particles can be directly measured using timing information
from the calorimeter and/or muon systems.
Ionization energy deposit is larger than electrons or muons due to the slower
speed, giving another handle on these particles that
can be measured particularly well in silicon tracking detectors.
Combined with a momentum measurement from the radius of curvature of the charged particle, the speed measurement can be used to infer the mass of the particle.

%

\begin{figure}[htbp]
\centering
\includegraphics[width=0.63\textwidth]{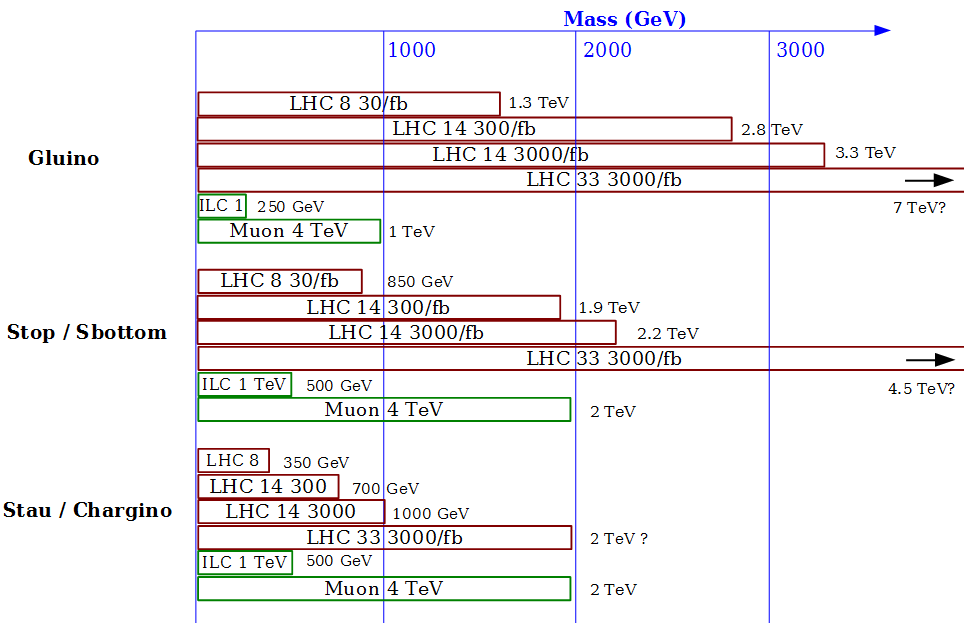}\includegraphics[width=0.37\textwidth]{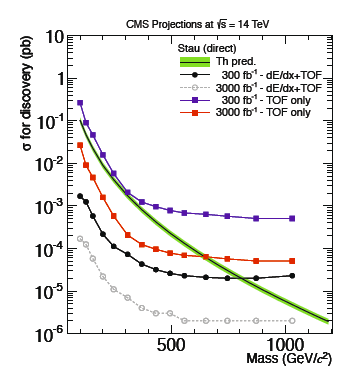}
\caption{Left: Summary of detector-stable partitcle reach at future colliders. Right: CMS study of future LHC reach for long-lived stau, from \cite{CMS:2013xfa}.
\label{fig_reach}}
\end{figure}

ATLAS and CMS have both performed very general searches for long-lived, massive, stable, 
charged tracks, from new particles such as long-lived squarks, gluinos, or staus 
\cite{ATLAS:2012ala, Chatrchyan:2012dxa}. 
The data are compared to models of background derived from data, given measured amounts of timing and ionization mis-measurements based on studies of Z decays. No excess is observed at high mass (large ionization and/or slow time) for any of the searches, so limits are placed on the particles' masses. At ATLAS, the long-lived stau is excluded at 95\% C.L. below 310~GeV, gluinos below 985~GeV, and stop/sbottom below about 600 GeV; similar limits are observed by CMS.

Backgrounds to detector-stable particles are small, given the excellent performance of detectors 
such as ATLAS and CMS. 
Thus the ability to discover these new particles mainly relies on being able to produce them in 
sufficient numbers, so simple extrapolation is sufficient to estimate the reach of future
hadron colliders.
Running at 14~TeV with 300~$fb^{-1}$ of data will enhance the mass reach for 
detector-stable particles by factors of 2--3, to 3 TeV for gluino and 2 TeV for stop/sbottom 
$R$-hadrons, and 1 TeV for staus (see Figure~\ref{fig_reach}). 

Lepton colliders have reach up to $\sqrt{s}/2$, which is generally not competitive with the
bounds from hadron colliders, although they do hold out the possibility of detailed study
of the particles within reach.
A 4 TeV muon collider could study charged stable particles up to about 2~TeV, comparable
to the reach of an upgraded 33~TeV LHC.
Studies of the new particle at the muon collider could probe its mass and spin precisely.

If such a particle is discovered, it will be critical to learn whether it decays,  and its lifetime and decay channels if it does.
Although most long-lived particles escape the detector, a small but significant fraction
stop in the detector due to ionization energy loss and/or nuclear interactions.
ATLAS and CMS have both developed searches for `beam-off' energy deposits in the calorimeter
motivated by this.
Assuming the `generic' model of $R$-hadron interactions, a gluino with mass below 
860~GeV and a stop with mass below 340~GeV are excluded, 
for lifetimes between 10 microseconds and 1000 seconds \cite{ATLAS:sg,CMS:2013xfa}.

The LHC experiments can trap and study long-lived heavy particles up to nearly the mass at 
which they can be discovered and provide reasonable estimates of their lifetimes and decay 
properties, over a large range of potential lifetimes. 
It is expected that this will continue to be true at future colliders, so that long-lived
decays can be studied up to approximately the mass reach for a wide range of particle
lifetimes.
Should a new particle be discovered, specialized detectors could be constructed to trap a 
larger fraction of the particles and optimized to study their decay properties as accurately 
as possible~\cite{Graham:2011ah}. 
Large luminosity, from an upgraded LHC, would be essential for this program of study.

\subsection{Top Partners\label{sec:toppartners}}
 

A natural extension to the standard model would be a new chiral
generation of quarks and leptons.
However, such a
chiral fourth fermion generation would couple to the Higgs boson with a Yukawa
coupling that is proportional to its mass and therefore give a large enhancement
Higgs-boson production through its contributions to the fermion
triangle in gluon fusion. This is
clearly inconsistent with the observed Higgs production cross
section, ruling out a chiral fourth generation of fermions.
 
Vector-like quarks are non-chiral, in that their left- and
right-handed components transform in the same way. Therefore their
mass terms do not violate any symmetry and do not have to be generated
by a Yukawa coupling to the Higgs boson. They couple to the Higgs
boson only through their mixing with standard model quarks. This
mixing is limited to small values by measurements of the $S, T, U$
electroweak precision oblique 
parameters. For such small mixing angles, vector-like quarks are not
expected to affect the gluon fusion production rate of the Higgs boson
significantly and thus are not ruled out by the observed Higgs boson
production cross section.
 
Vector-like quarks are motivated by some solutions to the hierarchy
problem\cite{delAguila:1989rq,Schmaltz:2005ky,Contino:2003ve,Agashe:2004rs,
Agashe:2006at,DeSimone:2012fs,Dawson:2012di}. 
Little Higgs theories predict top-quark partners that cancel
the effects of the top-quark loops on the Higgs boson mass. Models of
compositeness also predict vector-like top partners.  Vector-like
quarks can be weak isospin singlets, such as the charge-2/3 top-quark
partners predicted by little Higgs, top color, and top condensate
models. However, they could also be weak isospin doublets including
top and a bottom partners ($T'$, $B'$,  $X_{5/3}$ and $Y_{-4/3}$ fermionic partners).  
The $Y_{-4/3}$ has a $T'$-like $W^- b$ final state, with distinction only possible through a challenging 
measurement of the $b$-quark jet charge.   Additional vector-like multiplets in 
higher representations are also possible, with the prediction of a wider 
range of $T'$-like exotica, with a collection of the possibilities 
outlined in~\cite{Cacciapaglia:2010vn}.  
Generically, models in which the SM fields propagate in an extra
 spatial dimension predict the existence of Kaluza-Klein towers of
 vector-like quarks.  The KK partner of the top quark, for example,
 will in general decay to primarily 3rd generation quarks and SM gauge
 bosons.   Additionally, ultraviolet completions of R-parity violating SUSY 
models that follow the philosophy of minimal flavor violation 
to protect against baryon number violating operators~\cite{Csaki:2011ge} contain such $T'$
 quarks~\cite{Krnjaic:2012aj}

Earlier optimized searches exist for special cases in which the $T'$
decays with 100\% branching ratio to the $W-b$ (as in the sequential
4th generation model)~\cite{Chatrchyan:2012vu,ATLAS:2012qe} or
$t-Z$~\cite{Chatrchyan:2012af} final states.  
For most of the well motivated constructions, three
final states $b-W$, $t-Z$, and $t-h$ may result from $T'$
decays. Note that other decays that involve the first two generation
quarks are in principle also possible, but are generally suppressed
in models that do not violate existing flavor constraints.  A recent
study which explores this possibility
 is~\cite{Cacciapaglia:2012dd}.


The benchmark scenario considered for the snowmass study 
takes into account the three decays allowed by different models, such as 
$T'\rightarrow t Z$, and $T' \rightarrow tH$. For this benchmark, 
Goldstone's theorem applies such that in  the heavy $T'$ limit, 
the branching ratios for the three processes asymptotically obey $\textrm{BF}(T'
\rightarrow b W) = 2\textrm{BF}(T' \rightarrow t Z) = 2 \textrm{BF}(T'
\rightarrow t h)$.  

Recent studies have sought to obtain more general limits such that
the three branching fractions of the $T'$ are
free parameters, albeit subject to the constraint that no other final
states are allowed, such that the model spans a ``triangle" of
branching fractions.  In fact, a large class of models follows a
specific trajectory within the triangle, with this trajectory
determined by quantum numbers of the $T'$.  


An analysis of a general set of top partner final states with 
optimization over various branching fractions in the triangular phase space
has been carried out by the CMS Collaboration\cite{CMS-PAS-B2G-12-015}.
Direct limits based on current data exclude vector-like quarks for
masses below 700-800 GeV, depending on their decay branching 
fractions~\cite{CMS-PAS-B2G-12-015,ATLAS-CONF-2013-060,CMS-PAS-B2G-12-012}. 
Earlier studies ~\cite{Rao:2012gf,Berger:2012ec,AguilarSaavedra:2009es},
based on ATLAS results have analyzed a more simplified  ``triangle" of
branching fractions, with only a few points considered.  
The second looks at the specific case of little Higgs models, 
where the $T'$ is taken to be a singlet.  
Given the existing mass limits it is not likely that the ILC
or TLEP could contribute significantly to the direct  study of the properties
of top partners. 

\begin{figure}[h!]
\subfigure{
\includegraphics[width=0.3\linewidth]{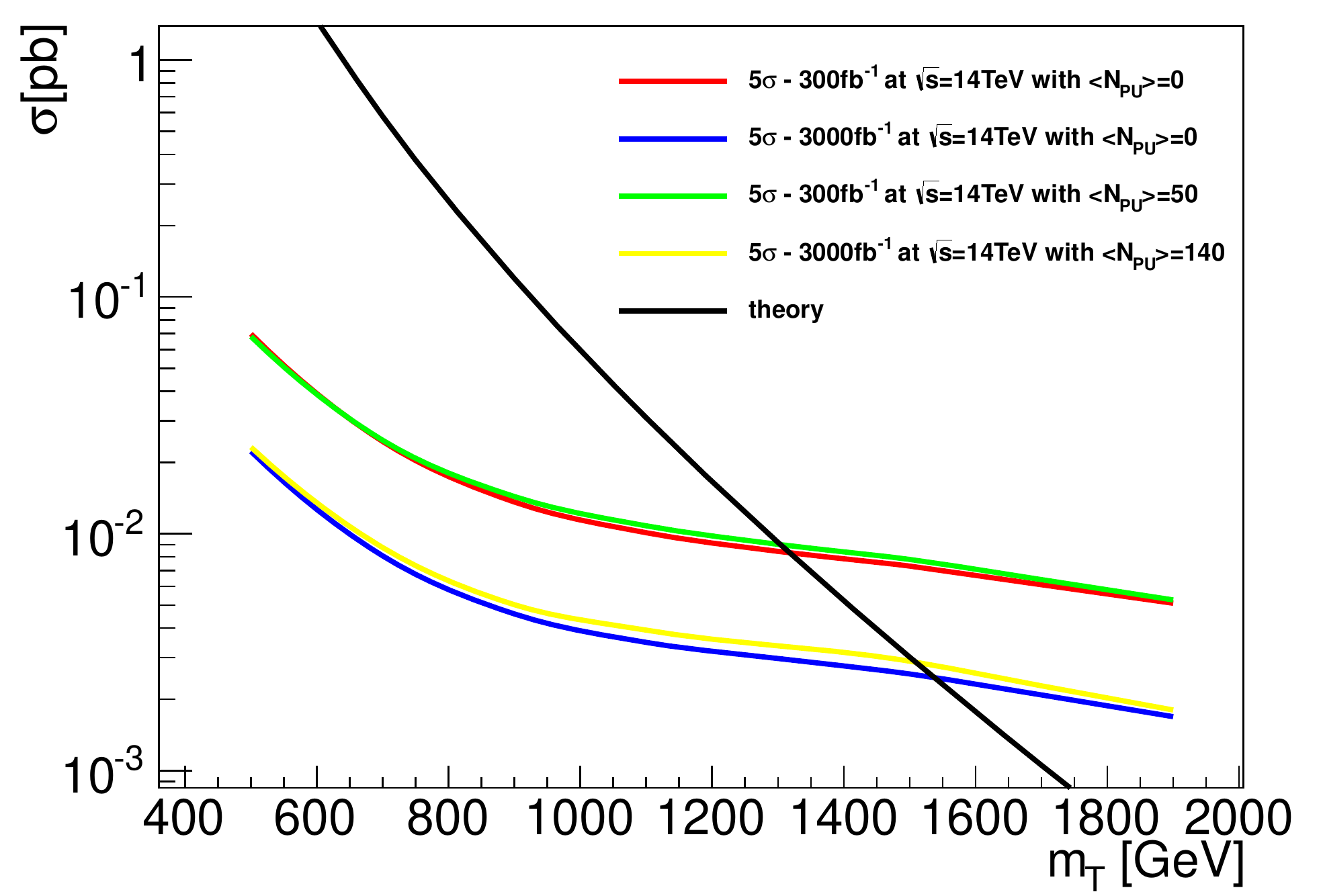} }
\subfigure{
\includegraphics[width=0.3\linewidth]{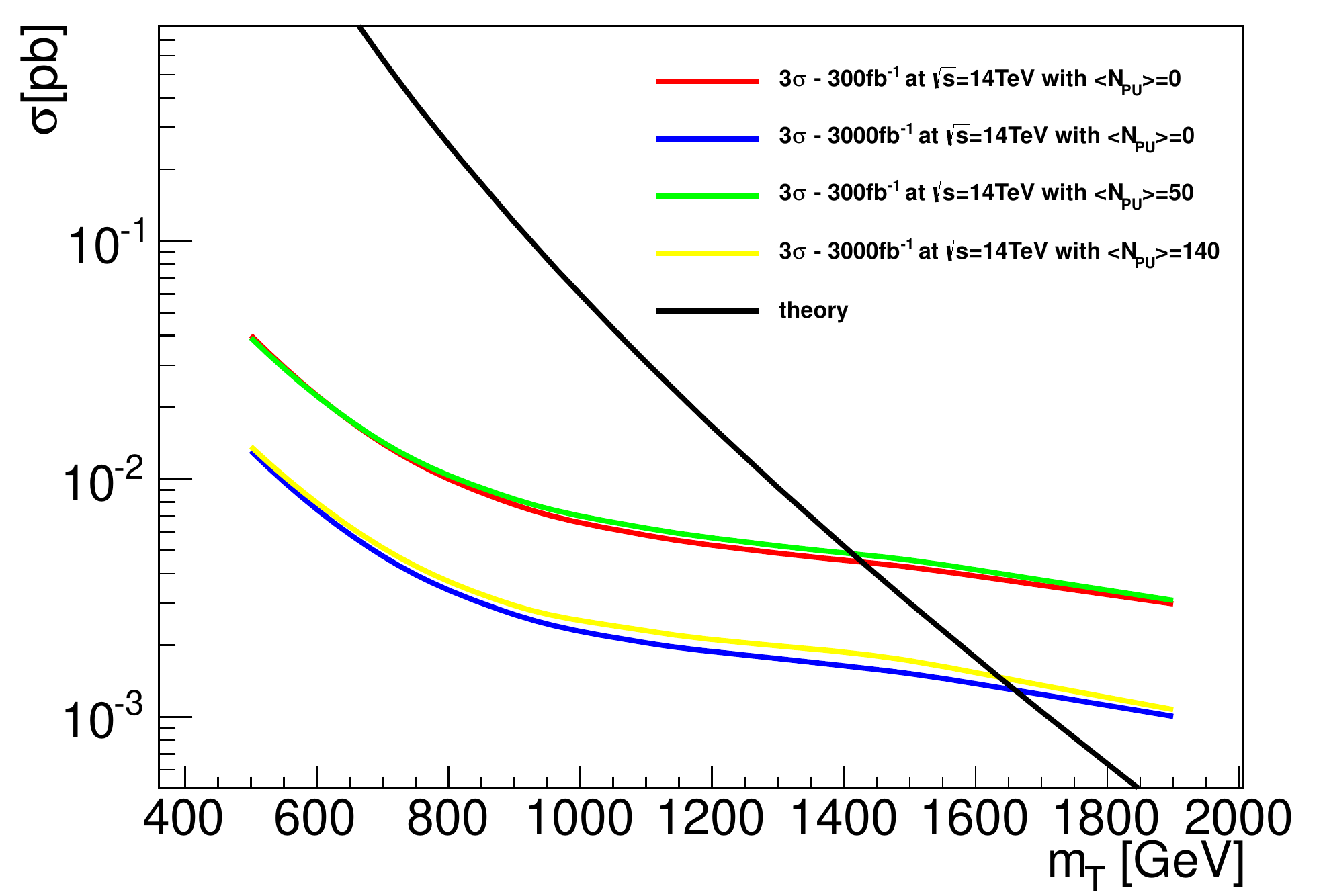}}
\subfigure{
\includegraphics[width=0.3\linewidth]{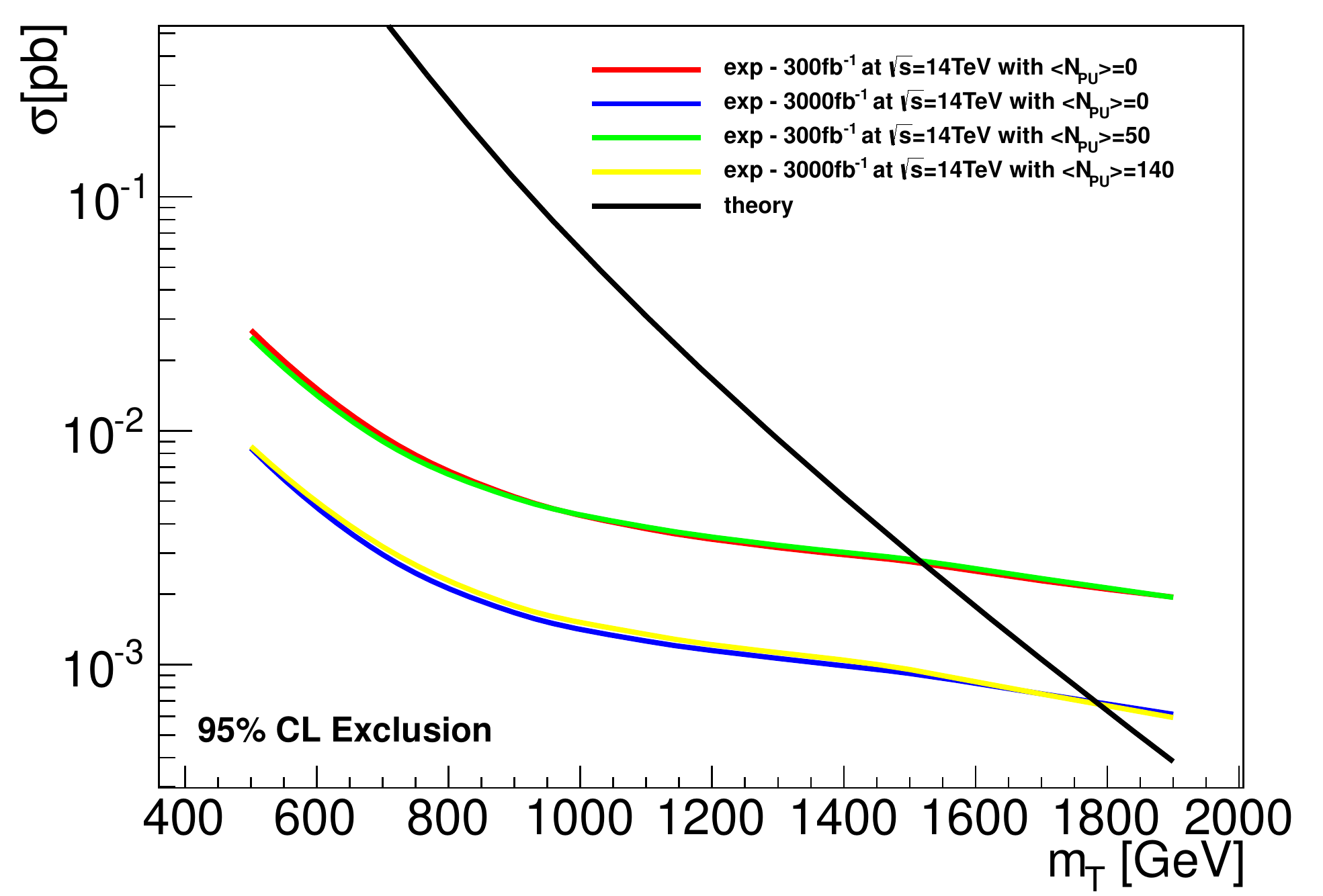}}
\caption[]{Discovery reach (left and middle panel) and exclusion (right panel) as a function of the mass of a
heavy vector-like quark at $\sqrt{s}=14$ TeV~\cite{DoubleTwhitepaper}. }
\label{fig:hQdoubleT}
\end{figure}

With LHC running at $\sqrt{s}=$ 14~TeV and a dataset 
corresponding to an integrated luminosity of 300 fb$^{-1}$, 
the reach for discovering heavy 
vector-like  quarks with charge 2/3 and exotic charge 5/3 
will be extended significantly. As demonstrated in Fig.~\ref{fig:hQdoubleT},
the   5$\sigma$ (3$\sigma$) reach for discovering heavy top-like quarks with mass around 1.3 (1.4)~TeV 
is achievable~\cite{DoubleTwhitepaper}. In the absence of such heavy quarks, 
we can probe masses up to 1.5~TeV~\cite{DoubleTwhitepaper,SingleTwhitepaper,T53whitepaper}. 
Similar conclusion is also reached in the whitepaper by CMS~\cite{CMS:2013xfa}.
At the HL-LHC with $\sqrt{s}=14$~TeV and 3000/fb vector-like quarks up to masses of around 
1.5 TeV can be observed or alternatively in the absence of such quarks we can 
exclude masses up to 1.8~TeV  (see Fig.\ref{fig:hQdoubleT}).
With HE-LHC, the reach increases to heavy top-like quark with masses up to 3.25 TeV.

For completeness, we note that there may be other exotic decays of
$T'$s to non-SM particles (or flavor violating decays) which may
reduce the sum of these three BF's below 1.   For example, in the
Littlest Higgs model with
T-parity~\cite{Cheng:2003ju,Cheng:2004yc,Hubisz:2004ft}, there is a
$T' \rightarrow T_-A_H \rightarrow t A_H A_H$, decay mode with the
$A_H$ playing the role of a ``neutralino."  This stop-like final state
reduces sensitivity in the $Wb$, $Zt$, and $Ht$ channels, but also
offers a complementary final state that is part of ongoing
searches~\cite{Aad:2012uu}.


If there is a vector-like $T$ quark with a mass of 1200~GeV an excess of events should appear at the LHC with 14 TeV $pp$ collisions after 300/fb have been collected. In events with a single electron or muon and several high-$p_T$ jets of which at least one shows substructure consistent with originating from a hadronic $W$- or $Z$-boson decay one may see an excess of 500 events over an expected background of about 2000 events. 

 
If such an excess is seen in a search for vector-like heavy quark one would
first want to determine the properties of the new particle, such as
production process (single or pair-production) and cross section,
mass, charge, decay modes and branching fractions. 
 The first order of business would be to
establish the nature of the new particle. Additional evidence for a new particle could come 
from events with two or more leptons. If the production cross
section is consistent with strong production the particle likely is
colored. One would identify whether the decay modes are consistent
with vector-like quarks. Vector-like quarks with charge 5/3 decay to
tW, those with charge 2/3 decay to $bW$, $tZ$, and $tH$,and those with
charge 1/3 decay to $tW$, $bZ$, and $bH$. 


Most interestingly, observation of
a vector-like quark would most likely indicate that there are other
heavy new particles. In little Higgs models there would be W and Higgs
boson partners, in compositeness models there would likely be other
vector-like quarks. A robust prediction of models with top-partners like composite or 
Little Higgs models is significant deviations of the Higgs couplings, in particular $hWW$ and $hZZ$, from the SM. 
The ILC-250 Higgs factory would measure these couplings with very high precision, providing 
a test of this interpretation of the excess. In addition, in most models, the top partner 
should be accompanied by additional new particles, some of which may be studied at the HL-LHC, HE-LHC, ILC at center-of mass energies of 500 GeV and 1 TeV. An example of this is the Littlest Higgs model with T-parity~\cite{Baer:2013vqa}, 
which shows that the achievable level of precision by ILC-500 
allows non-trivial tests of the model structure. 

Depending on the mass of the vector-like quark and the other new
particles, collisions at higher energy might be needed to produce the heavy
vector-like quarks in sufficient numbers to understand their properties. This
could be done at HE-LHC or VLHC pp colliders or at the CLIC e$^+$e$^-$
collider.

\label{sec:hquarks}

\subsection{Fermion Compositeness}
\label{sec:compositeness}

\def\missET {{\not\!\! E_T}}

High-energy particles are powerful probes of physics at small
scales. Experiments at escalating energy scales have historically
unveiled layers of substructure in particles previously considered as
fundamental, from Rutherfords probing of gold atoms which revealed the
presence of a central nucleus, to deep inelastic scattering of protons
which demonstrated the existence of quarks. In this section, we consider the extent to which the compositeness of quarks can be probed by
future collider facilities~\cite{Eichten:1983hw,Eichten:1984eu}.

Quarks as bound states of more fundamental particles
may explain current outstanding questions, such as the number of quark
generations, the charges of the quarks, or the symmetry between the
quark and lepton sectors~\cite{Harari:1979gi,Shupe:1979fv,Casalbuoni:1980pi}.

\begin{figure}
\centering
\includegraphics[width=0.5\linewidth]{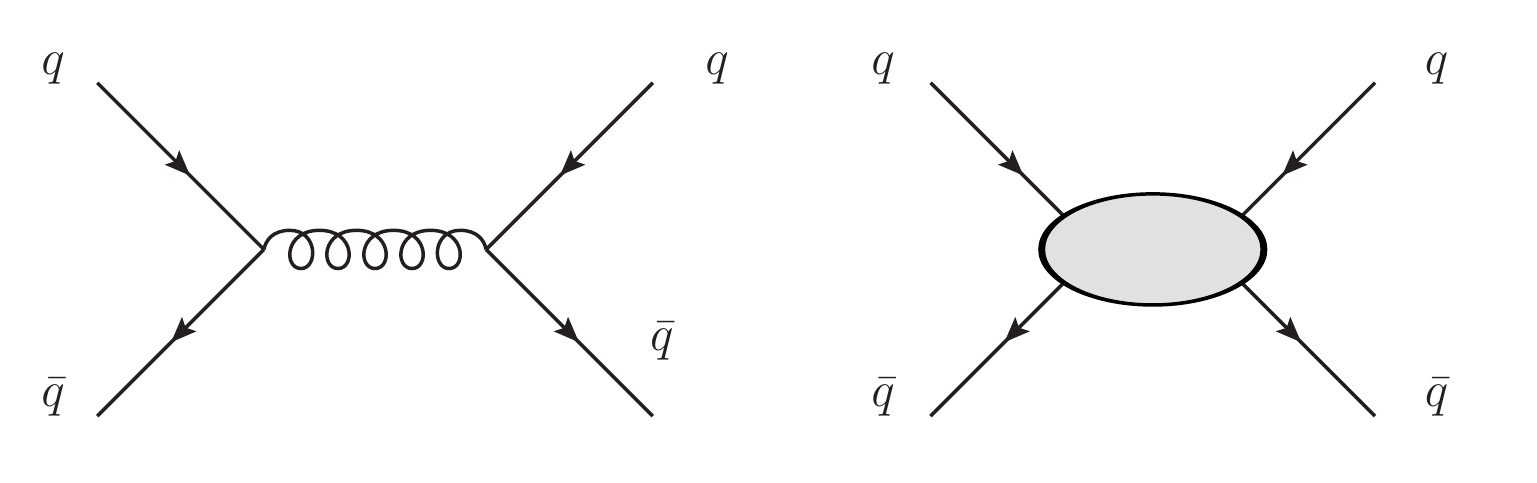}
\caption{Diagrams for QCD mediation of quark-quark interactions (left)
  and a four-fermion contact interaction describing an effective field
  theory for the mediation of a new interaction between quark
  constituents.}
\label{fig:diag}
\end{figure}

A typical approach to the study of quark compositeness~\cite{Chatrchyan:2012bf} is
to search for evidence of new interactions between quarks
at a large characteristic energy scale, $\Lambda$.  At interaction
energies below $\Lambda$, the details of the new interaction and
potential mediating particles can be integrated out to form a
four-fermion contact interaction model (see Fig~\ref{fig:diag}).  This is well-described by an
effective field theory approach~\cite{Beringer:1900zz}:


\begin{align} 
L_{qq} =  \frac{2\pi}{\Lambda^2}&[\eta_{LL}(\bar{q_L}\gamma^\mu q_L)(\bar{q_L}\gamma_\mu {q_L})   
+ \eta_{RR}(\bar{q_R}\gamma^\mu q_R)(\bar{q_R}\gamma_\mu q_R)    + 2\eta_{RL}(\bar{q_R}\gamma^\mu q_R)(\bar{q_L}\gamma_\mu q_L) ]
\end{align}

\noindent
where the quark fields have  chiral projections $L$ and $R$, and the
coefficients $\eta_{LL}$, $\eta_{RR}$, and $\eta_{RL}$ turn on and off
various interactions. In this study, we examine the cases of energy
scales $\Lambda^+_{LL}$, $\Lambda^+_{RR}$, and $\Lambda^+_{V-A}$ with
couplings $(\eta_{LL},\eta_{RR}, \eta_{RL}) = (1,0,0), (0,1,0)$ and
$(0,0,1)$, respectively, in order to demonstrate the center-of-mass
dependence of the sensitivity of possible future $pp$ facilities.

Evidence for contact interactions would appear as an enhancement of  dijet production 
 with large dijet invariant mass $m_{jj}$ and 
angle relative to the beam axis, $\theta^*$, in the center of mass frame.
Quantum chromodynamics (QCD) predominantly produces jets with small
$\theta^*$ peaked in the forward and backward directions. 

\begin{figure}
\centering
\includegraphics[width=0.4\linewidth]{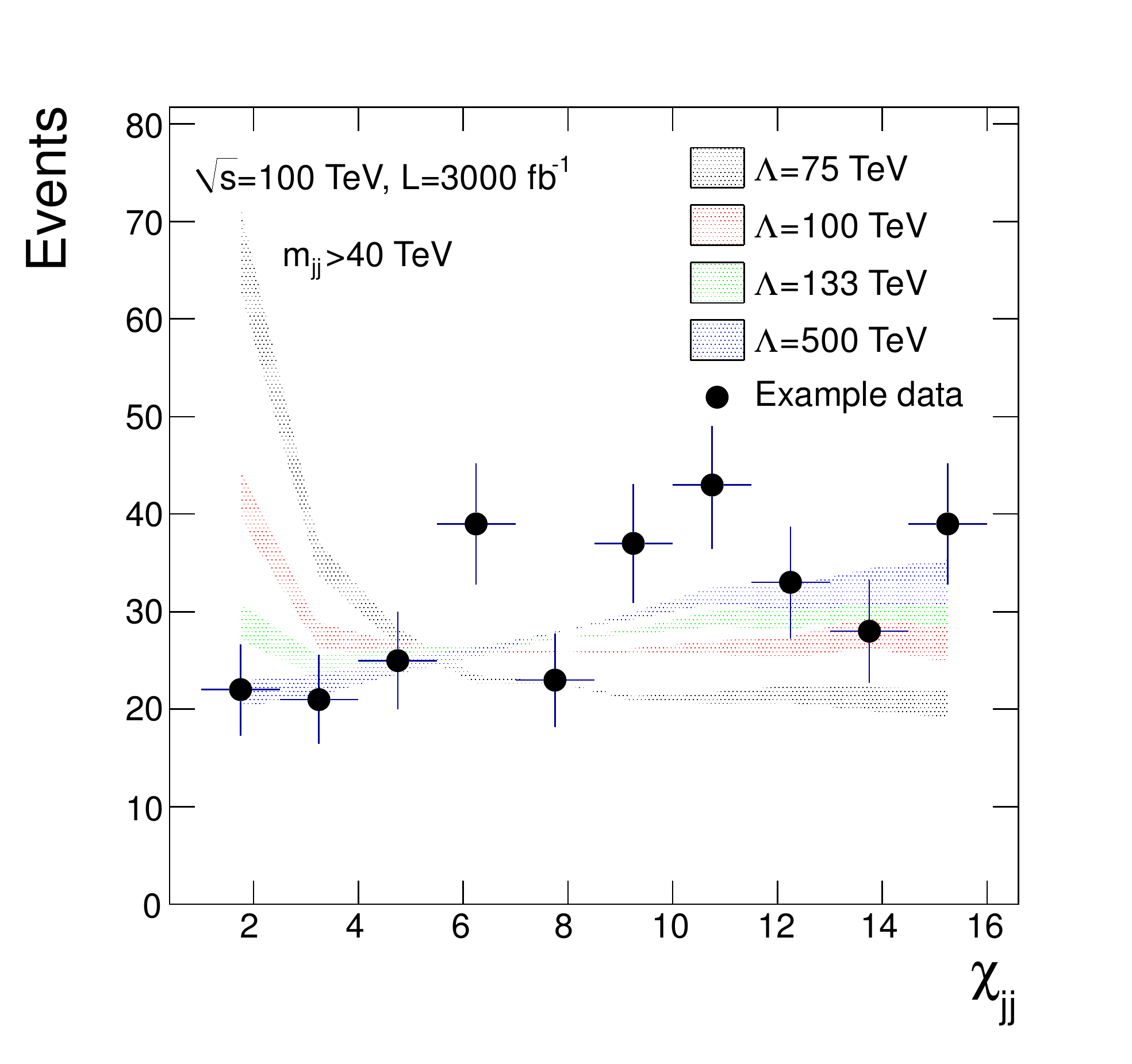}
\includegraphics[width=0.4\linewidth]{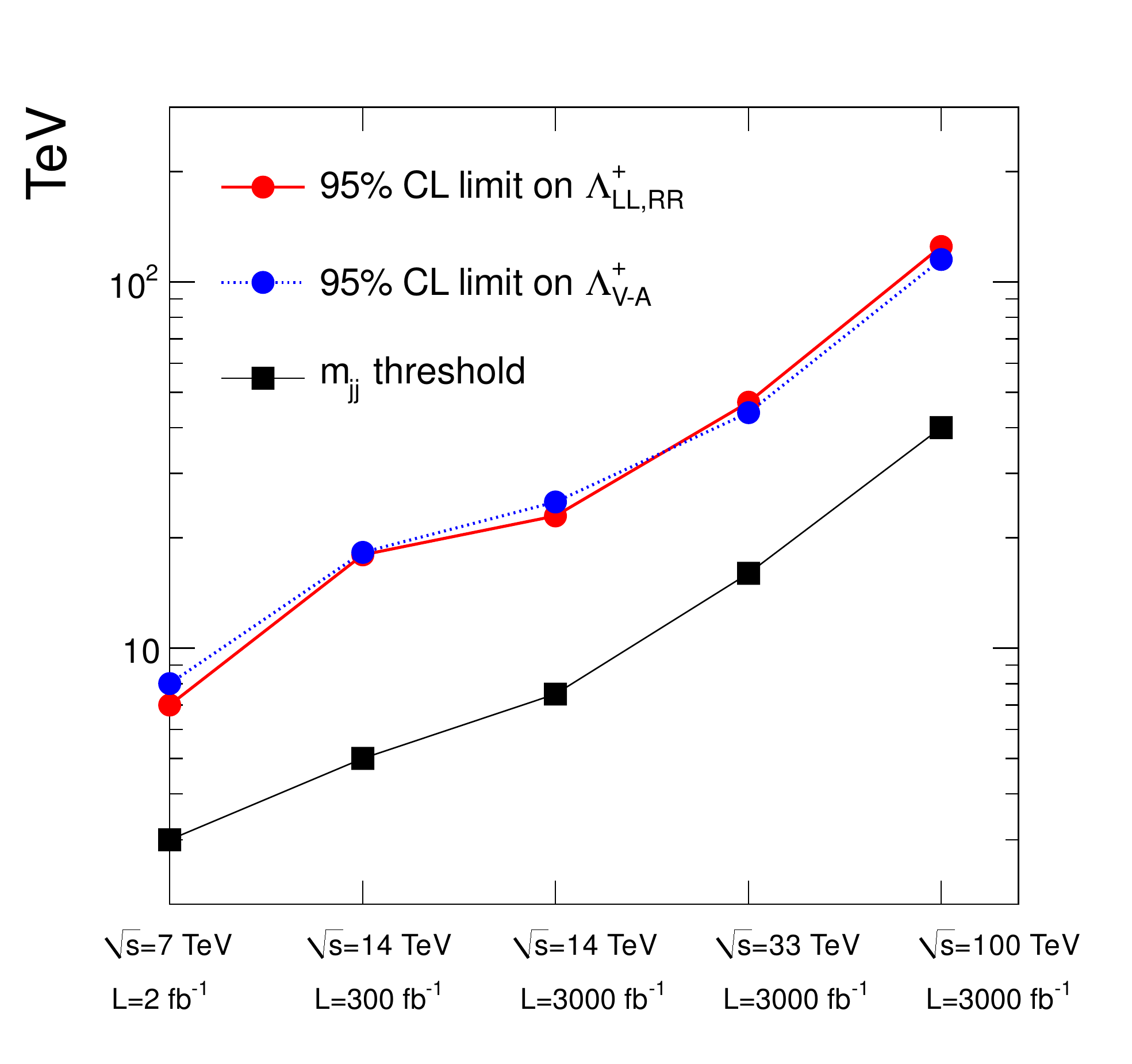}

\caption{ Left: distributions of $\chi_{jj}$ for QCD 
 and   contact interactions with a variety of choices of $\Lambda$ for the case of 
  $pp$ interactions with  with $\sqrt{s}=100$~TeV and
  $\mathcal{L}=3000$~fb$^{-1}$. Right: summary of $m_{jj}$ thresholds
  and sensitivity to the contact interaction scale $\Lambda$.}
\label{fig:quarkcom}
\end{figure}

While next-to-leading-order calculations of QCD~\cite{nlojet} and the contact
interactions expected from quark compositeness~\cite{nloci} are
available, they are computationally intensive and for the purpose of
this study, leading-order calculations are sufficient.  For events 
generated  at leading-order with {\sc
  madgraph}~\cite{Alwall:2011uj},  the showering and
hadronization is described by  {\sc pythia}~\cite{Sjostrand:2007gs} and the
detector response by {\sc delphes}~\cite{Ovyn:2009tx} for the
facilities described in Fig.~\ref{fig:quarkcom}.  

Based on Ref.~\cite{quarkcon} which follows the approach of Ref~\cite{Chatrchyan:2012bf}, 
the analysis variable is $\chi_{jj} = e^{|y_1 - y_2|}$ where $y_1$ and $y_2$ are the rapidities
of the two highest transverse momentum (leading) jets.  The
distribution for QCD interactions is slightly increasing with
$\chi_{jj}$, while contact interaction models predict angular
distributions that are strongly peaked at low values of $\chi_{jj}$. The distortion of the $\chi_{jj}$ shape is most distinct at large
$m_{jj}$. However, the cross section falls sharply with $m_{jj}$,
reducing the statistical power of the data.  These two effects are in
tension, and there is an optimum value of the minimum $m_{jj}$ threshold.

As seen in
Fig.~\ref{fig:quarkcom}, higher center-of-mass energies bring
significant increases in sensitivity to the mass scale, $\Lambda$,
such that a collider with $\sqrt{s}=100$ TeV would be expected to
probe scales above $\Lambda=125$ TeV.

If a deviation from QCD production is seen at the LHC with
$\sqrt{s}=14$ TeV, then a facility with higher energy will be 
needed  to directly produce the new heavy particle that mediates the interaction
of the quark constituents, dependening on the mass scale. This would
appear as a dijet resonance in $q\bar{q}\rightarrow q\bar{q}$ events.  Specifically, we can relate the
exclusion of the  compositeness scale $\Lambda$ to that of the mass of
a $Z'$ mediator as:

\[ \frac{g_{Z'}^2}{36 M_{Z'}^2} = \frac{2\pi}{\Lambda^2}. \]

For example, at $\sqrt{s}=14$ TeV with $\mathcal=3000$ fb$^{-1}$, an
exclusion of $\Lambda>18$ TeV would correspond to excluding a $Z'$
with $(m_{Z'}=1200$ GeV $, g_{Z'}=0.12)$. Figure \ref{fig:LHC_had}
shows the discovery sensitivity and current limits.





\subsection{Warped Extra Dimensions and Flavor}
\label{sec:rs_flavor}

\newcommand{\Mkk}{M_{\mathrm{KK}}}
\newcommand{\cws}{c_w^2 }
\newcommand{\sws}{s_w^2 }

Randall-Sundrum (RS) models with an IR localized Higgs provide an alternative 
solution to the gauge hierarchy problem which is closely related to the composite 
Higgs idea due to the AdS/CFT correspondence. An attractive feature of these 
models is that   
the hierarchical structure of the SM Yukawas can be 
explained by
order one shifts in the localization of the fermions along the extra dimension, 
if they propagate in the bulk. 
\\

Complete bulk RS models can therefore be probed in various ways, such as indirect 
tests, like measurements of modified 
Higgs couplings 
 and flavor violating observables, and direct  searches, 
including
searches for Kaluza-Klein (KK) partners of fermions, and KK resonances 
of
gauge bosons. The discovery of a light Higgs 
and the 
refined bounds from electroweak precision observables \cite{Baak:2011ze} 
universally constrain the mass scale in these models to $\Mkk> 5$ TeV for a SM 
bulk gauge group and   
 $\Mkk> 2$ TeV in the case of a custodial bulk $SU(2)_L\times SU(2)_R$ 
symmetry. 
This translates into 
 a mass bound for the first KK gluon of $m_{G^1} > 12.5\, (5)$ TeV, 
respectively. These resonances have a small production
cross section at colliders, because the coupling to initial state gluons is 
forbidden by parity and the 
coupling to light flavors is suppressed. The current LHC searches do therefore 
not yet 
constrain this mass scale directly \cite{Aad:2012raa}.\\

Whether a $33$ TeV or a $100$ TeV $pp$ machine will be able to see such a 
resonance in the $t\,\bar{t}$ final state
or in the dijet mass spectrum depends on 
the localization of the quarks, which control the branching fraction of the KK 
gluon as well as the flavor violating 
couplings of KK resonances to quarks. 
In contrast to supersymmetric models in which the flavor sector is in principle 
disconnected from collider observables, for RS models, 
measurements at the intensity frontier can set a roadmap for future direct 
searches.

\begin{figure}[t]
\centerline{
\begin{tabular}{cl} 
\includegraphics[width=0.46\textwidth]{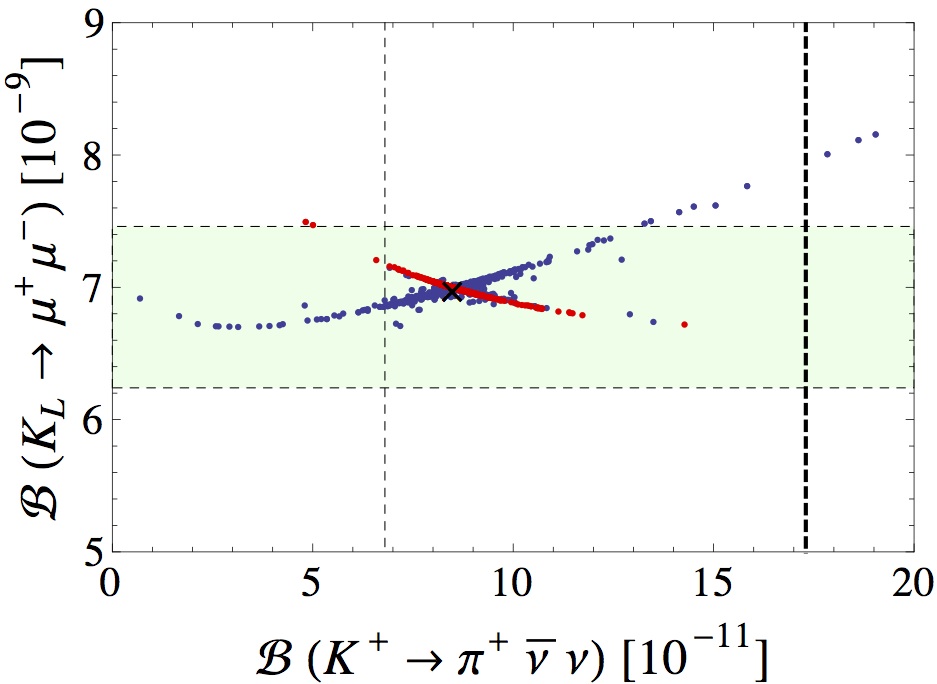} 
&\hspace{-.4cm}\includegraphics[width=.5\textwidth]{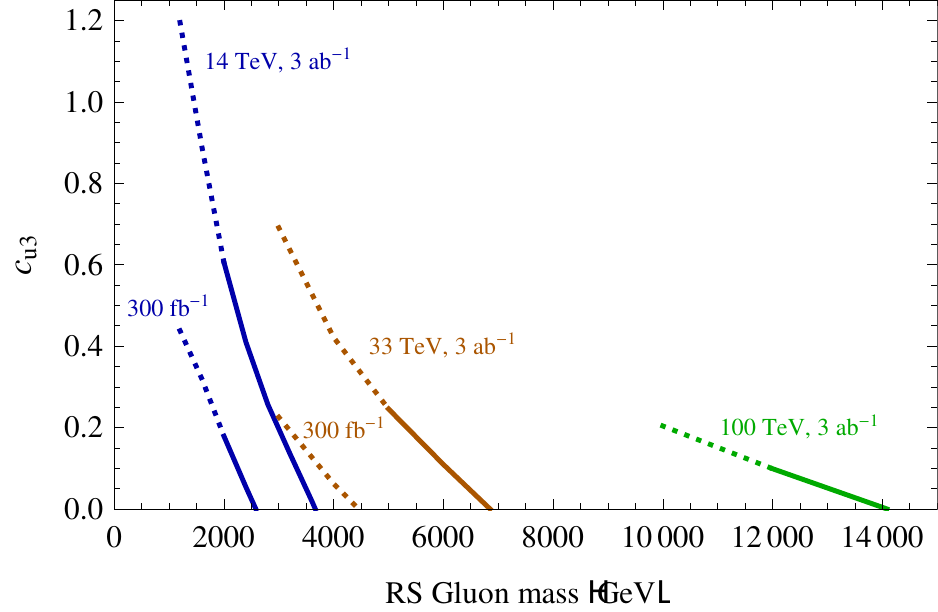}\\ 
\includegraphics[width=0.48\textwidth]{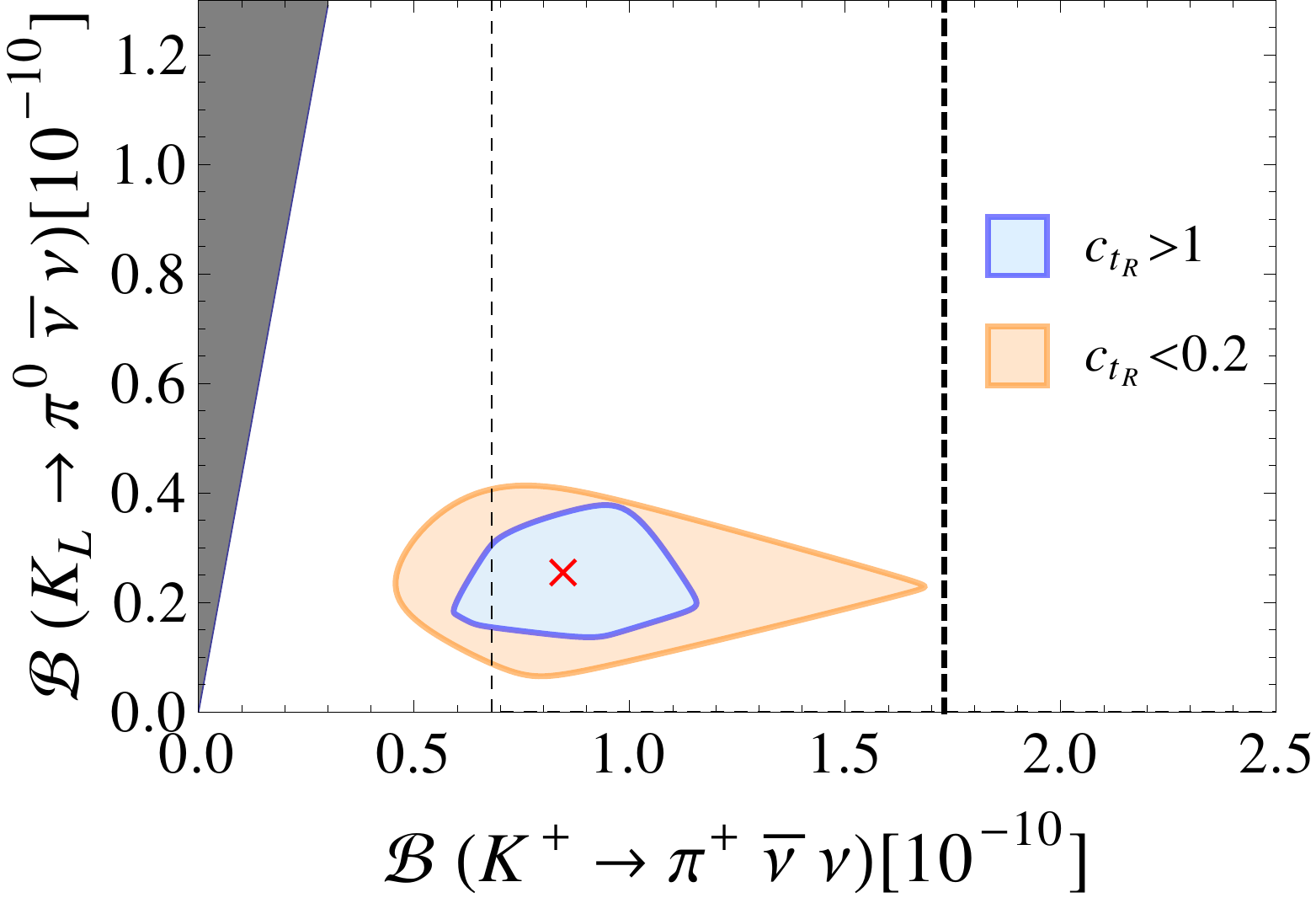} & 
\includegraphics[width=0.48\textwidth]{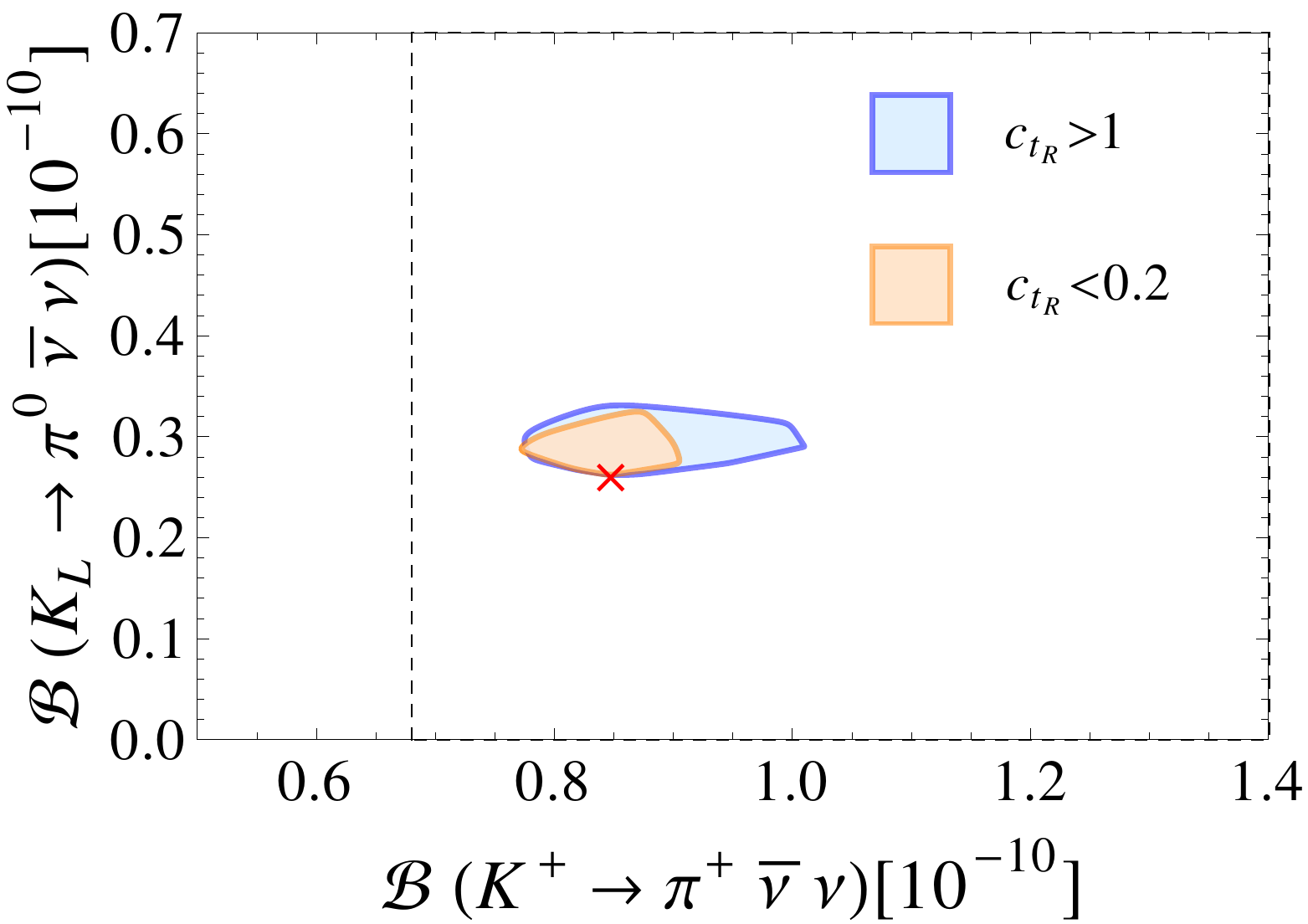} 
\end{tabular}}
\caption{The plots in the upper left panel shows opposite correlations in Kaon 
observables for the minimal (blue scatter points) and custodial (red scatter 
points) model. The upper right panel shows the reach of future collider
searches in dependence of the localization of the right-handed top quark 
$c_{u_3}$. The lower panel shows preferred parameter regions for models with
 small and large $c_{t_R} \equiv c_{u_3}$ in the minimal (left 
panel) and custodial (right panel) model. }
\label{fig:RS}
\end{figure}

Reference \cite{Agashe:2013kxa} shows, how the recent 
 discovery of the $B_s\to 
\mu^+\mu^-$ decay mode at LHCb constrains the parameter space 
of warped models and how
projected improvements in measurements of rare Kaon decays will further narrow 
down these parameters. In particular the 
expected precision measurements from ORKA \cite{Comfort:2011zz} and KEK 
\cite{KOTO} have the 
potential to not only 
determine these localization parameters but also identify the underlying 
electroweak bulk gauge group of 
the RS model. Figure \ref{fig:RS} shows the reach of dijet searches in 
dependence on the localization of the right-handed top quark ($c_{u_3}$),
the orthogonal correlation of $\mathcal{B} (K_L\to \mu^+\mu^-)$ and 
$\mathcal{B} (K^+\to \pi^+\nu\bar\nu)$ for the minimal and custodial model as 
well as the $c_{u_3}$ dependence of the 
size of effects in $\mathcal{B} 
(K^+\to \pi^+\nu\bar\nu)$ and $\mathcal{B} 
(K_L\to \pi^0\nu\bar\nu)$ in these two classes of models. 

Models with further structure in the gauge and flavor sectors can lead to lower 
KK mass scales, 
which could be in the reach of the $14$ TeV LHC \cite{Agashe:2013fda} or lead to 
interesting signatures like the flavor violating decay of the KK gluon into top 
and charm quarks, testable at the an upgraded $pp$ 
machine \cite{Drueke:2013wsa}.

%


%

\subsection{`Only' the Standard Model}
We now consider an `anti-discovery' scenario 
where LHC14 with 300~fb$^{-1}$ 
does not discover any additional particles or observe any anomalies.
Such a run will have significant acheivements: the LHC
will have not only discovered the
Higgs boson, but will have made impressive progress in the program
of precision Higgs measurements.
Projections for these are discussed in the Higgs working group report.
The scenario we are now considering also assumes that the improved 
measurements of Higgs couplings from LHC14 300~fb$^{-1}$
are consistent with their standard model values.
It also assumes that there is no discovery of physics beyond the standard
model from the intensity frontier program ({\it e.g.}~new flavor violation)
or the cosmic frontier program ({\it e.g.}~dark matter direct detection).
Any such discovery would be a sign of new physics that could be
at the TeV scale, giving additional motivation for continued exploration
of the energy frontier.
But if there is no discovery of new fundamental physics, 
is our motivation for exploring the TeV scale reduced?

As discussed throughout this report,
there are a number of big questions and big ideas that can be explored at
the TeV scale.
The big questions that have the strongest link to the TeV scale
are the origin of dark matter and the naturalness of the Higgs boson.
We discuss these questions in the context of the no-discovery
scenario below.

\subsubsection{Dark Matter}
Probably the best-motivated dark matter candidate is a weakly interacting massive
particle (WIMP). 
This requires only that the dark matter is a neutral stable particle that couples
weakly to the standard model, and that the dark matter particles are in thermal
equilibrium with the standard model particles in the early universe.
In this scenario, there is an upper limit on the WIMP mass
\begin{equation}
m_{\rm WIMP} \leq 2 {\rm \ TeV} \left( \frac{g_{\rm eff}^2}{0.3}\right),
\end{equation}
where $g_{\rm eff}$ is the coupling strength between dark matter and the SM particles,
which we have normalized to the weak coupling in the SM.
The dark matter mass can easily be at the TeV
scale, and LHC14 with 300~fb$^{-1}$ does not have sensitivity to direct production of 
non-colored states with this high mass.
Specific examples of this kind of dark matter are pure Higgsino or Wino dark matter
in SUSY with heavy superpartners.
Alternatively, the dark matter may be much lighter, but is not observed at
LHC14 with 300~fb$^{-1}$ because of `low-mass' loopholes such as suppressed
couplings or degenerate spectra leading to supressed missing energy.

%

The most model-independent collider search relies on the associated production 
of a pair of WIMPs together with hard radiation, {\it e.g.}~a jet, photon, {\it etc\/}.
This is discussed in \S\ref{sec:dmatter} above.
Assuming that the dark matter couples to colored particles, hadron colliders
can produce dark matter with a radiative jet, the monojet signature.
The LHC14 with 300~fb$^{-1}$ can probe dark matter masses in the range of
100~GeV to 1~TeV for thermal relic, depending on the nature of the interaction,
see Fig.~\ref{fig:d5d8}.
At the same time, a higher energy VLHC at 33/100 TeV can extend the reach 
of WIMPs into the TeV range and cover a significant part 
of the parameter region of the WIMP scenario.
This is illustrated in Figs.~\ref{fig:indirect}. 


If the dark matter is missed at LHC14 due to suppressed couplings to quarks
or low-mass loopholes, $e^+ e^-$ colliders retain the ability to 
discover dark matter up
to essentially $\sqrt{s}/2$, almost independently of the details of the
coupling.
Furthermore, they will be able to make precise measurements of the dark
matter mass and couplings to accessible states.
This is illustrated in Fig.~\ref{fig:ilc_mchi}. 



\subsubsection{Naturalness}
If nature is described by the standard model with an elementary Higgs boson
up to the Planck scale, then the observed Higgs boson mass is the sum of
different contributions that must cancel to an accuracy of
$\epsilon \sim (125~{\rm GeV} / M_{\rm Planck})^2 \sim 10^{-30}$.
This arises because the mass-squared parameter in the SM Lagrangian is
quadratically sensitive to large mass scales.
If this divergence is cut off by new physics at a scale $M_{\rm NP}$
the tuning is reduced to $\epsilon \sim (125~{\rm GeV} / M_{\rm NP})^2$.
This is the basic naturalness argument for new physics at the TeV
scale.
The normalization and quantitative interpretation of naturalness
estimates are not clear, but the quadratic scaling with $M_{\rm NP}$ is 
robust, and fine tuning can be used as a rough guide for where
to expect new physics.
This argument is independent of supersymmetry or any other scenario
for physics beyond the standard model.

In the standard model, the largest contribution to the Higgs mass
that must be cut off by new physics comes from the top loop.
Although this is a loop effect, the coefficient is large because
of the large top coupling and the QCD color factor.
This directly motivates searches for new physics in the top sector,
such as searches for stops in SUSY and fermionic top partners in
composite scenarios.
These are discussed in \S\ref{sec:lightstop} and \S\ref{sec:toppartners} of this report, respectively.
The summary is that LHC14 with 300~fb$^{-1}$ has sensitivity for these new
states to approximately the TeV scale.
Taken at face value, this implies roughly a tuning of $\epsilon \sim 1\%$.

Should this be taken as evidence that nature is unnatural?
A possibly useful historical analogy from cosmology is that 
in the early 1990s the quadrupole anisotropy of the cosmic microwave background 
appeared to be below expectations from cold dark matter cosmology.
This was arguably the `discovery mode' for this cosmological model, and 
the reason it was not found earlier is that it is coindidentally small,
with a probability from cosmic variance of roughly $1\%$.
The lesson may be that unfavorable accidents at the $1\%$ level do happen in
discovery modes for fundamental new physics.

We can therefore ask how well future experiments will probe naturalness.
A rough summary is that the 
HL-LHC increases the reach for new heavy particles by $10\%$ to $20\%$.
This does not make a dramatic impact on naturalness, although it should be
kept in mind that the new mass range that is being probed is in the
most interesting range in a wide range of well-motivated models,
as discussed above.
In addition, the HL-LHC can close many (but not all) low-mass loopholes
due to higher luminosity and improved systematics.

If we push to higher energies with a 100 TeV VLHC, we can probe colored SUSY
partners at the 10~TeV scale.
Based on the scaling of tuning, we expect this to probe tuning to the
level $\epsilon \sim 10^{-4}$.
This is a very strong motivation to expect the discovery of new physics.


On the other hand, in the scenario we are considering
it may be that the top partners or other new particles
have been missed at the LHC14 with 300~fb$^{-1}$ because 
of highly compressed spectrum or other low-mass loopholes.
At an $e^+ e^-$ collider, new particles can be searched for in
a nearly loophole-free way, typically for masses of up to $\sqrt{s}/2$.
In SUSY $e^+ e^-$ colliders can directly probe another source of tuning:
the Higgsino mass generically contributes directly to the Higgs mass,
and therefore SUSY models with heavy Higgsinos require tuning
at the level of $\epsilon \sim (125~{\rm GeV} / m_{\tilde{H}})^2$.
An $e^+ e^-$ collider can search for Higgsinos in a model-independent
way up to essentially $\sqrt{s}/2$.
At a 1~TeV $e^+ e^-$ collider, we can therefore probe tuning
at the level of $\epsilon \sim 1\%$ in SUSY~\cite{Baer:2013ava}.
Precision studies of the Higgsino sector may also allow 
indirect indications of the electroweak gaugino masses even
if the associated particles in the multi-TeV range \cite{Berggren:2013vfa}.

\subsubsection{Flavor, CP, and Precision Measurements}
Many models of new physics have potential contributions to flavor,
CP, and precision electroweak observables at a level that may point
to new physics at a scale of roughly 10~TeV.
For example, SUSY has additional sources of flavor mixing and
CP violation, and if they are not suppressed by special flavor
structure point to a scale of new physics above 10~TeV.
On the other hand, composite models generally give rise to
corrections to precision electroweak observables that also 
point to a scale above 10~TeV.
This scale is therefore the scale that will be probed by
precision electroweak and flavor studies, primarily at the 
intensity frontier.
Another promising indirect approach is increasing precision
on electroweak observables, discussed in detail the precision
electroweak portion of these proceedings.
An order-of-magnitude improvement on electroweak precision observables
is possible at lepton colliders, which may give important clues
about new physics in the multi-TeV range.
            
In conclusion, in the scenario we are considering, flavor, CP, precision
electroweak observables, and the energy frontier are arguably
exploring the 10~TeV energy scale.

\bibliography{newparticles}

\begin{thebibliography}{100}

\bibitem{KOTO}
{ KOTO collaboration}.
\newblock {\em \texttt{ http://koto.kek.jp}}.

\bibitem{Aad:2012uu}
Georges Aad et~al.
\newblock {Search for a heavy top-quark partner in final states with two
  leptons with the ATLAS detector at the LHC}.
\newblock {\em JHEP}, 1211:094, 2012.

\bibitem{ATLAS:2012hi}
Georges Aad et~al.
\newblock {Search for doubly-charged Higgs bosons in like-sign dilepton final
  states at $\sqrt{s}=7$ TeV with the ATLAS detector}.
\newblock {\em Eur.Phys.J.}, C72:2244, 2012.

\bibitem{ATLAS:2012ala}
Georges Aad et~al.
\newblock {Searches for heavy long-lived sleptons and $R$-hadrons with the
  ATLAS detector in $pp$ collisions at $\sqrt{s} = 7$~TeV}.
\newblock 2012.

\bibitem{atlaszz}
Georges Aad et~al.
\newblock {Measurement of $ZZ$ production in $pp$ collisions at $\sqrt{s}=7$
  TeV and limits on anomalous $ZZZ$ and $ZZ\gamma$ couplings with the ATLAS
  detector}.
\newblock {\em JHEP}, 1303:128, 2013.

\bibitem{atlasjet}
Georges Aad et~al.
\newblock {Search for dark matter candidates and large extra dimensions in
  events with a jet and missing transverse momentum with the ATLAS detector}.
\newblock {\em JHEP}, 1304:075, 2013.

\bibitem{atlasphoton}
Georges Aad et~al.
\newblock {Search for dark matter candidates and large extra dimensions in
  events with a photon and missing transverse momentum in $pp$ collision data
  at $\sqrt{s}=7$ TeV with the ATLAS detector}.
\newblock {\em Phys.Rev.Lett.}, 110:011802, 2013.

\bibitem{ATLAS:2012qe}
Georges Aad et~al.
\newblock {Search for pair production of heavy top-like quarks decaying to a
  high-pT $W$ boson and a $b$ quark in the lepton plus jets final state at
  $\sqrt{s}$=7 TeV with the ATLAS detector}.
\newblock {\em Phys.Lett.}, B718:1284--1302, 2013.

\bibitem{Aad:2012raa}
Georges Aad et~al.
\newblock {Search for resonances decaying into top-quark pairs using fully
  hadronic decays in $pp$ collisions with ATLAS at $\sqrt{s}=7$ TeV}.
\newblock {\em JHEP}, 1301:116, 2013.

\bibitem{clicwp}
H.~Abramowicz et~al.
\newblock {Physics at the CLIC e+e- Linear Collider -- Input to the Snowmass
  process 2013}.
\newblock 2013.

\bibitem{fermi}
M.~Ackermann et~al.
\newblock {Constraining Dark Matter Models from a Combined Analysis of Milky
  Way Satellites with the Fermi Large Area Telescope}.
\newblock {\em Phys.Rev.Lett.}, 107:241302, 2011.

\bibitem{Agashe:2013fda}
Kaustubh Agashe, Oleg Antipin, Mihailo Backović, Aaron Effron, Alex Emerman,
  et~al.
\newblock {Warped Extra Dimensional Benchmarks for Snowmass 2013}.
\newblock 2013.

\bibitem{Agashe:2013kxa}
Kaustubh Agashe, Martin Bauer, Florian Goertz, Seung~J. Lee, Luca Vecchi,
  et~al.
\newblock {Constraining RS Models by Future Flavor and Collider Measurements: A
  Snowmass Whitepaper}.
\newblock 2013.

\bibitem{Agashe:2006at}
Kaustubh Agashe, Roberto Contino, Leandro Da~Rold, and Alex Pomarol.
\newblock {A Custodial symmetry for Zb anti-b}.
\newblock {\em Phys.Lett.}, B641:62--66, 2006.

\bibitem{Agashe:2004rs}
Kaustubh Agashe, Roberto Contino, and Alex Pomarol.
\newblock {The Minimal composite Higgs model}.
\newblock {\em Nucl.Phys.}, B719:165--187, 2005.

\bibitem{Agrawal:2011ze}
Prateek Agrawal, Steve Blanchet, Zackaria Chacko, and Can Kilic.
\newblock {Flavored Dark Matter, and Its Implications for Direct Detection and
  Colliders}.
\newblock {\em Phys.Rev.}, D86:055002, 2012.

\bibitem{AguilarSaavedra:2009es}
J.A. Aguilar-Saavedra.
\newblock {Identifying top partners at LHC}.
\newblock {\em JHEP}, 0911:030, 2009.

\bibitem{Alwall:2011uj}
Johan Alwall, Michel Herquet, Fabio Maltoni, Olivier Mattelaer, and Tim
  Stelzer.
\newblock {MadGraph 5 : Going Beyond}.
\newblock {\em JHEP}, 1106:128, 2011.

\bibitem{zprime}
Haipeng An, Xiangdong Ji, and Lian-Tao Wang.
\newblock {Light Dark Matter and $Z'$ Dark Force at Colliders}.
\newblock {\em JHEP}, 1207:182, 2012.

\bibitem{Anandakrishnan:2013pja}
Archana Anandakrishnan, B.~Charles Bryant, Stuart Raby, and Akin Wingerter.
\newblock {Gluino bounds: Simplified Models vs a Particular SO(10) Model (A
  Snowmass white paper)}.
\newblock 2013.

\bibitem{SingleTwhitepaper}
Tim Andeen, Clare Bernard, Kevin Black, Taylor Childres, Lidia Dell'Asta,
  et~al.
\newblock {Sensitivity to the Single Production of Vector-Like Quarks at an
  Upgraded Large Hadron Collider}.
\newblock 2013.

\bibitem{SnowmassDetSim}
Jacob Anderson, Aram Avetisyan, Raymond Brock, Sergei Chekanov, Timothy Cohen,
  et~al.
\newblock {Snowmass Energy Frontier Simulations}.
\newblock 2013.

\bibitem{quarkcon}
Leonard Apanasevich, Suneet Upadhyay, Nikos Varelas, Daniel Whiteson, and Felix
  Yu.
\newblock {Sensitivity of potential future $pp$ colliders to quark
  compositeness}.
\newblock {\em arXiv:1307.7149}, 2013.

\bibitem{Artoni:2013zba}
Giacomo Artoni, Tongyan Lin, Bjoern Penning, Gabriella Sciolla, and Alessio
  Venturini.
\newblock {Prospects for collider searches for dark matter with heavy quarks}.
\newblock 2013.

\bibitem{SnowmassOSG}
A.~Avetisyan, S.~Bhattacharya, M.~Narain, S.~Padhi, J.~Hirschauer, et~al.
\newblock {Snowmass Energy Frontier Simulations using the Open Science Grid (A
  Snowmass 2013 whitepaper)}.
\newblock 2013.

\bibitem{T53whitepaper}
Aram Avetisyan and Tulika Bose.
\newblock {Search for top partners with charge 5e/3}.
\newblock 2013.

\bibitem{SnowmassBackgrounds}
Aram Avetisyan, John~M. Campbell, Timothy Cohen, Nitish Dhingra, James
  Hirschauer, et~al.
\newblock {Methods and Results for Standard Model Event Generation at
  $\sqrt{s}$ = 14 TeV, 33 TeV and 100 TeV Proton Colliders (A Snowmass
  Whitepaper)}.
\newblock 2013.

\bibitem{Baak:2011ze}
M.~Baak, M.~Goebel, J.~Haller, A.~Hoecker, D.~Ludwig, et~al.
\newblock {Updated Status of the Global Electroweak Fit and Constraints on New
  Physics}.
\newblock {\em Eur.Phys.J.}, C72:2003, 2012.

\bibitem{Baer:2013faa}
Howard Baer, Vernon Barger, Peisi Huang, Dan Mickelson, Azar Mustafayev, et~al.
\newblock {Leaving no stone unturned in the hunt for SUSY naturalness: A
  Snowmass whitepaper}.
\newblock 2013.

\bibitem{Baer:2013ava}
Howard Baer, Vernon Barger, Peisi Huang, Dan Mickelson, Azar Mustafayev, et~al.
\newblock {Naturalness, Supersymmetry and Light Higgsinos: A Snowmass
  Whitepaper}.
\newblock 2013.

\bibitem{Baer:2013yha}
Howard Baer, Vernon Barger, Peisi Huang, Dan Mickelson, Azar Mustafayev, et~al.
\newblock {Same sign diboson signature from supersymmetry models with light
  higgsinos at the LHC}.
\newblock 2013.

\bibitem{Baer:2013fva}
Howard Baer, Vernon Barger, Andre Lessa, and Xerxes Tata.
\newblock {SUSY discovery potential of LHC14 with 0.3-3~ab$^{-1}$ : A Snowmass
  whitepaper}.
\newblock 2013.

\bibitem{Baer:2013vqa}
Howard Baer, Mikael Berggren, Jenny List, Mihoko~M. Nojiri, Maxim Perelstein,
  et~al.
\newblock {Physics Case for the ILC Project: Perspective from Beyond the
  Standard Model}.
\newblock 2013.

\bibitem{Bai:2010hh}
Yang Bai, Patrick~J. Fox, and Roni Harnik.
\newblock {The Tevatron at the Frontier of Dark Matter Direct Detection}.
\newblock {\em JHEP}, 1012:048, 2010.

\bibitem{monow}
Yang Bai and Tim~M.P. Tait.
\newblock {Searches with Mono-Leptons}.
\newblock {\em Phys.Lett.}, B723:384--387, 2013.

\bibitem{Barbier:2004ez}
R.~Barbier, C.~Berat, M.~Besancon, M.~Chemtob, and A.~Deandrea.
\newblock {R-parity violating supersymmetry}.
\newblock {\em Phys.Rept.}, 420:1--202, 2005.

\bibitem{Barger:2013ofa}
Vernon Barger, Lisa~L. Everett, Heather~E. Logan, and Gabe Shaughnessy.
\newblock {Scrutinizing h(125) in Two Higgs Doublet Models at the LHC, ILC, and
  Muon Collider}.
\newblock 2013.

\bibitem{Bartels:2012ex}
Christoph Bartels, Mikael Berggren, and Jenny List.
\newblock {Characterising WIMPs at a future $e^+e^-$ Linear Collider}.
\newblock {\em Eur.Phys.J.}, C72:2213, 2012.

\bibitem{Beltran:2010ww}
Maria Beltran, Dan Hooper, Edward~W. Kolb, Zosia~A.C. Krusberg, and Tim~M.P.
  Tait.
\newblock {Maverick dark matter at colliders}.
\newblock {\em JHEP}, 1009:037, 2010.

\bibitem{Beltran:2008xg}
Maria Beltran, Dan Hooper, Edward~W. Kolb, and Zosia~C. Krusberg.
\newblock {Deducing the nature of dark matter from direct and indirect
  detection experiments in the absence of collider signatures of new physics}.
\newblock {\em Phys.Rev.}, D80:043509, 2009.

\bibitem{Berger:2012ec}
Joshua Berger, Jay Hubisz, and Maxim Perelstein.
\newblock {A Fermionic Top Partner: Naturalness and the LHC}.
\newblock {\em JHEP}, 1207:016, 2012.

\bibitem{Berggren:2013hda}
M.~Berggren, A.~Cakir, D.~KrŸcker, J.~List, A.~Lobanov, et~al.
\newblock {Non-Simplified SUSY: stau-Coannihilation at LHC and ILC}.
\newblock 2013.

\bibitem{Berggren:2013vna}
Mikael Berggren.
\newblock Simplified susy at the ilc.
\newblock 2013.

\bibitem{Berggren:2013vfa}
Mikael Berggren, Felix Brummer, Jenny List, Gudrid Moortgat-Pick, Tania Robens,
  et~al.
\newblock Tackling light higgsinos at the ilc.
\newblock 2013.

\bibitem{Berggren:2013bua}
Mikael Berggren, Tao Han, Jenny List, Sanjay Padhi, Shufang Su, et~al.
\newblock {Electroweakino Searches: A Comparative Study for LHC and ILC (A
  Snowmass White Paper)}.
\newblock 2013.

\bibitem{Beringer:1900zz}
J.~Beringer et~al.
\newblock {Review of Particle Physics (RPP)}.
\newblock {\em Phys.Rev.}, D86:010001, 2012.

\bibitem{Bharucha:2012ya}
Aoife Bharucha, Jan Kalinowski, Gudrid Moortgat-Pick, Krzysztof Rolbiecki, and
  Georg Weiglein.
\newblock One-loop effects on mssm parameter determination via chargino
  production at the lc.
\newblock {\em Eur.Phys.J.}, C73:2446, 2013.

\bibitem{DoubleTwhitepaper}
Saptaparna Bhattacharya, Jimin George, Ulrich Heintz, Ashish Kumar, Meenakshi
  Narain, et~al.
\newblock {Prospects for a Heavy Vector-Like Charge 2/3 Quark T search at the
  LHC with $\sqrt{s}$=14 TeV and 33 TeV. "A Snowmass 2013 Whitepaper"}.
\newblock 2013.

\bibitem{Blanke:2010cm}
Monika Blanke, David Curtin, and Maxim Perelstein.
\newblock {SUSY-Yukawa Sum Rule at the LHC}.
\newblock {\em Phys.Rev.}, D82:035020, 2010.

\bibitem{2HDMwp}
Eric Brownson, Nathaniel Craig, Ulrich Heintz, Gena Kukartsev, Meenakshi
  Narain, et~al.
\newblock {Heavy Higgs Scalars at Future Hadron Colliders (A Snowmass
  Whitepaper)}.
\newblock 2013.

\bibitem{Cacciapaglia:2010vn}
Giacomo Cacciapaglia, Aldo Deandrea, Daisuke Harada, and Yasuhiro Okada.
\newblock {Bounds and Decays of New Heavy Vector-like Top Partners}.
\newblock {\em JHEP}, 1011:159, 2010.

\bibitem{Cacciapaglia:2012dd}
Giacomo Cacciapaglia, Aldo Deandrea, Luca Panizzi, Stephane Perries, and Viola
  Sordini.
\newblock {Heavy Vector-like quark with charge 5/3 at the LHC}.
\newblock {\em JHEP}, 1303:004, 2013.

\bibitem{Cahill-Rowley:2013yla}
M.~Cahill-Rowley, J.L. Hewett, A.~Ismail, and T.G. Rizzo.
\newblock {pMSSM Studies at the 7, 8 and 14 TeV LHC}.
\newblock 2013.

\bibitem{Cahill-Rowley:2013gca}
Matthew~W. Cahill-Rowley, JoAnne~L. Hewett, Ahmed Ismail, Michael~E. Peskin,
  and Thomas~G. Rizzo.
\newblock {pMSSM Benchmark Models for Snowmass 2013}.
\newblock 2013.

\bibitem{Cao:2009uw}
Qing-Hong Cao, Chuan-Ren Chen, Chong~Sheng Li, and Hao Zhang.
\newblock {Effective Dark Matter Model: Relic density, CDMS II, Fermi LAT and
  LHC}.
\newblock {\em JHEP}, 1108:018, 2011.

\bibitem{Carena:2004xs}
Marcela~S. Carena, Alejandro Daleo, Bogdan~A. Dobrescu, and Timothy~M.P. Tait.
\newblock {$Z^\prime$ gauge bosons at the Tevatron}.
\newblock {\em Phys.Rev.}, D70:093009, 2004.

\bibitem{monoz}
Linda~M. Carpenter, Andrew Nelson, Chase Shimmin, Tim~M.P. Tait, and Daniel
  Whiteson.
\newblock {Collider searches for dark matter in events with a Z boson and
  missing energy}.
\newblock 2012.

\bibitem{Casalbuoni:1980pi}
R.~Casalbuoni and Raoul Gatto.
\newblock {SUBCOMPONENT MODELS OF QUARKS AND LEPTONS}.
\newblock {\em Phys.Lett.}, B93:47--52, 1980.

\bibitem{Chae:2012bq}
Yoonseok~John Chae and Maxim Perelstein.
\newblock {Dark Matter Search at a Linear Collider: Effective Operator
  Approach}.
\newblock {\em JHEP}, 1305:138, 2013.

\bibitem{Chang:2010en}
Spencer Chang, Neal Weiner, and Itay Yavin.
\newblock {Magnetic Inelastic Dark Matter}.
\newblock {\em Phys.Rev.}, D82:125011, 2010.

\bibitem{cmsjet}
Serguei Chatrchyan et~al.
\newblock {Search for dark matter and large extra dimensions in monojet events
  in $pp$ collisions at $\sqrt{s}=7$ TeV}.
\newblock {\em JHEP}, 1209:094, 2012.

\bibitem{cmsphoton}
Serguei Chatrchyan et~al.
\newblock {Search for Dark Matter and Large Extra Dimensions in pp Collisions
  Yielding a Photon and Missing Transverse Energy}.
\newblock {\em Phys.Rev.Lett.}, 108:261803, 2012.

\bibitem{Chatrchyan:2012vu}
Serguei Chatrchyan et~al.
\newblock {Search for pair produced fourth-generation up-type quarks in $pp$
  collisions at $\sqrt{s}=7$ TeV with a lepton in the final state}.
\newblock {\em Phys.Lett.}, B718:307--328, 2012.

\bibitem{Chatrchyan:2012bf}
Serguei Chatrchyan et~al.
\newblock {Search for quark compositeness in dijet angular distributions from
  $pp$ collisions at $\sqrt{s}=7$ TeV}.
\newblock {\em JHEP}, 1205:055, 2012.

\bibitem{Chatrchyan:2012dxa}
Serguei Chatrchyan et~al.
\newblock {Search for stopped long-lived particles produced in $pp$ collisions
  at $\sqrt{s}=7$ TeV}.
\newblock {\em JHEP}, 1208:026, 2012.

\bibitem{Chatrchyan:2012st}
Serguei Chatrchyan et~al.
\newblock {Search for third-generation leptoquarks and scalar bottom quarks in
  $pp$ collisions at $\sqrt{s}=7$ TeV}.
\newblock {\em JHEP}, 1212:055, 2012.

\bibitem{Chatrchyan:2012af}
Serguei Chatrchyan et~al.
\newblock {Search for heavy quarks decaying into a top quark and a $W$ or $Z$
  boson using lepton + jets events in $pp$ collisions at $\sqrt{s}$ = 7 TeV}.
\newblock {\em JHEP}, 1301:154, 2013.

\bibitem{Cheng:2003ju}
Hsin-Chia Cheng and Ian Low.
\newblock {TeV symmetry and the little hierarchy problem}.
\newblock {\em JHEP}, 0309:051, 2003.

\bibitem{Cheng:2004yc}
Hsin-Chia Cheng and Ian Low.
\newblock {Little hierarchy, little Higgses, and a little symmetry}.
\newblock {\em JHEP}, 0408:061, 2004.

\bibitem{Chizhov:2013hua}
M.V. Chizhov, V.A. Bednyakov, and J.A. Budagov.
\newblock {Hadron collider potential for excited bosons search: A Snowmass
  whitepaper}.
\newblock 2013.

\bibitem{Cohen:2013zla}
Timothy Cohen, Tobias Golling, Mike Hance, Anna Henrichs, Kiel Howe, et~al.
\newblock {A Comparison of Future Proton Colliders Using SUSY Simplified
  Models: A Snowmass Whitepaper}.
\newblock 2013.

\bibitem{Coleppa:2013dya}
Baradhwaj Coleppa, Felix Kling, and Shufang Su.
\newblock {Exotic Higgs Decay via AZ/HZ Channel: A Snowmass whitepaper}.
\newblock 2013.

\bibitem{CMS:2013xfa}
CMS Collaboration.
\newblock {Projected Performance of an Upgraded CMS Detector at the LHC and
  HL-LHC: Contribution to the Snowmass Process}.
\newblock 2013.

\bibitem{ATLAS-CONF-2013-011}
The~ATLAS Collaboration.
\newblock Search for invisible decays of a higgs boson produced in association
  with a z boson in atlas.
\newblock Technical Report ATLAS-CONF-2013-011, CERN, Geneva, Mar 2013.

\bibitem{ATLAS-CONF-2013-060}
The~ATLAS Collaboration.
\newblock Search for pair production of heavy top-like quarks decaying to a
  high-$p_{\rm t}$ $w$ boson and a $b$ quark in the lepton plus jets final
  state in $pp$ collisions at $\sqrt{s}=8$ tev with the atlas detector.
\newblock 2013.

\bibitem{CMS-PAS-B2G-12-015}
The~CMS Collaboration.
\newblock Inclusive search for a vector-like t quark by cms.
\newblock 2013.

\bibitem{CMS-PAS-HIG-13-018}
The~CMS Collaboration.
\newblock Search for invisible higgs produced in association with a z boson.
\newblock Technical Report CMS-PAS-HIG-13-018, CERN, Geneva, 2013.

\bibitem{CMS-PAS-B2G-12-012}
The~CMS Collaboration.
\newblock Search for t5/3 top partners in same-sign dilepton final state.
\newblock 2013.

\bibitem{Comfort:2011zz}
Joseph Comfort, Douglas Bryman, Luca Doria, Toshio Numao, Aleksey Sher, et~al.
\newblock {ORKA: Measurement of the $K^ \to \pi^+ \nu \bar{\nu}$ decay at
  Fermilab}.
\newblock 2011.

\bibitem{Contino:2003ve}
Roberto Contino, Yasunori Nomura, and Alex Pomarol.
\newblock {Higgs as a holographic pseudoGoldstone boson}.
\newblock {\em Nucl.Phys.}, B671:148--174, 2003.

\bibitem{Csaki:2011ge}
Csaba Csaki, Yuval Grossman, and Ben Heidenreich.
\newblock {MFV SUSY: A Natural Theory for R-Parity Violation}.
\newblock {\em Phys.Rev.}, D85:095009, 2012.

\bibitem{exotics}
David Curtin, Rouven Essig, Stefania Gori, Prerit Jaiswal, Andrey Katz, Tao
  Liu, Zhen Liu, David McKeen, Jessie Shelton, Matthew Strassler, Ze'ev
  Surujon, Brock Tweedie, and Yiming Zhong.
\newblock {Exotic Decays of the 125~GeV Higgs Boson}.
\newblock to appear.

\bibitem{Cvetic:1995zs}
Mirjam Cvetic and Stephen Godfrey.
\newblock {Discovery and identification of extra gauge bosons}.
\newblock 1995.

\bibitem{Cvetic:1997wu}
Mirjam Cvetic and P.~Langacker.
\newblock {Z-prime physics and supersymmetry}.
\newblock 1997.

\bibitem{Dawson:2012di}
S.~Dawson and E.~Furlan.
\newblock {A Higgs Conundrum with Vector Fermions}.
\newblock {\em Phys.Rev.}, D86:015021, 2012.

\bibitem{DeSimone:2012fs}
Andrea De~Simone, Oleksii Matsedonskyi, Riccardo Rattazzi, and Andrea Wulzer.
\newblock {A First Top Partner's Hunter Guide}.
\newblock {\em JHEP}, 1304:004, 2013.

\bibitem{delAguila:1989rq}
F.~del Aguila, L.~Ametller, Gordon~L. Kane, and J.~Vidal.
\newblock {VECTOR LIKE FERMION AND STANDARD HIGGS PRODUCTION AT HADRON
  COLLIDERS}.
\newblock {\em Nucl.Phys.}, B334:1, 1990.

\bibitem{DelAguila:1993rw}
F.~Del~Aguila and Mirjam Cvetic.
\newblock {Diagnostic power of future colliders for Z-prime couplings to quarks
  and leptons: e+ e- versus p p colliders}.
\newblock {\em Phys.Rev.}, D50:3158--3166, 1994.

\bibitem{DelAguila:1995fa}
F.~Del~Aguila, Mirjam Cvetic, and P.~Langacker.
\newblock {Reconstruction of the extended gauge structure from Z-prime
  observables at future colliders}.
\newblock {\em Phys.Rev.}, D52:37--43, 1995.

\bibitem{delAguila:1993ym}
F.~del Aguila, Mirjam Cvetic, and Paul Langacker.
\newblock {Determination of Z-prime gauge couplings to quarks and leptons at
  future hadron colliders}.
\newblock {\em Phys.Rev.}, D48:969--973, 1993.

\bibitem{Delannoy:2013ata}
Andres~G. Delannoy, Bhaskar Dutta, Alfredo Gurrola, Will Johns, Teruki Kamon,
  et~al.
\newblock {Probing Dark Matter at the LHC using Vector Boson Fusion Processes}.
\newblock 2013.

\bibitem{Delannoy:2013dla}
Andres~G. Delannoy, Bhaskar Dutta, Alfredo Gurrola, Will Johns, Teruki Kamon,
  et~al.
\newblock {Probing Supersymmetric Dark Matter and the Electroweak Sector using
  Vector Boson Fusion Processes: A Snowmass Whitepaper}.
\newblock 2013.

\bibitem{Diener:2009vq}
Ross Diener, Stephen Godfrey, and Travis~A.W. Martin.
\newblock {Discovery and Identification of Extra Neutral Gauge Bosons at the
  LHC}.
\newblock 2009.

\bibitem{Diener:2011jt}
Ross Diener, Stephen Godfrey, and Ismail Turan.
\newblock {Constraining Extra Neutral Gauge Bosons with Atomic Parity Violation
  Measurements}.
\newblock {\em Phys.Rev.}, D86:115017, 2012.

\bibitem{Dienes:2012yz}
Keith~R. Dienes, Shufang Su, and Brooks Thomas.
\newblock {Distinguishing Dynamical Dark Matter at the LHC}.
\newblock {\em Phys.Rev.}, D86:054008, 2012.

\bibitem{Dobrescu:2013cmh}
Bogdan~A. Dobrescu and Felix Yu.
\newblock {Coupling--mass mapping of di-jet peak searches}.
\newblock 2013.

\bibitem{CTA}
M.~Doro et~al.
\newblock {Dark Matter and Fundamental Physics with the Cherenkov Telescope
  Array}.
\newblock {\em Astropart.Phys.}, 43:189--214, 2013.

\bibitem{Dreiner:2012xm}
Herbert Dreiner, Moritz Huck, Michael Kramer, Daniel Schmeier, and Jamie
  Tattersall.
\newblock Illuminating dark matter at the ilc.
\newblock {\em Phys.Rev.}, D87:075015, 2013.

\bibitem{Drueke:2013wsa}
Elizabeth Drueke, Brad Schoenrock, Barbara~Alvarez Gonzalez, and Reinhard
  Schwienhorst.
\newblock {Searches for resonances in the tb and tc final states at the
  high-luminosity LHC}.
\newblock 2013.

\bibitem{RPVwhitepaper}
Daniel Duggan, Jared~A. Evans, James Hirschauer, Ketino Kaadze, David
  Kolchmeyer, et~al.
\newblock {Sensitivity of an Upgraded LHC to R-Parity Violating Signatures of
  the MSSM}.
\newblock 2013.

\bibitem{Dutta:2012xe}
Bhaskar Dutta, Alfredo Gurrola, Will Johns, Teruki Kamon, Paul Sheldon, et~al.
\newblock {Vector Boson Fusion Processes as a Probe of Supersymmetric
  Electroweak Sectors at the LHC}.
\newblock {\em Phys.Rev.}, D87:035029, 2013.

\bibitem{Eichten:1984eu}
E.~Eichten, I.~Hinchliffe, Kenneth~D. Lane, and C.~Quigg.
\newblock {Super Collider Physics}.
\newblock {\em Rev.Mod.Phys.}, 56:579--707, 1984.

\bibitem{Eichten:1983hw}
E.~Eichten, Kenneth~D. Lane, and Michael~E. Peskin.
\newblock {New Tests for Quark and Lepton Substructure}.
\newblock {\em Phys.Rev.Lett.}, 50:811--814, 1983.

\bibitem{Ellis:2006vu}
John~R. Ellis, Are~R. Raklev, and Ola~K. Oye.
\newblock {Gravitino dark matter scenarios with massive metastable charged
  sparticles at the LHC}.
\newblock {\em JHEP}, 0610:061, 2006.

\bibitem{Erler:2009jh}
Jens Erler, Paul Langacker, Shoaib Munir, and Eduardo Rojas.
\newblock {Improved Constraints on Z-prime Bosons from Electroweak Precision
  Data}.
\newblock {\em JHEP}, 0908:017, 2009.

\bibitem{Fairbairn:2006gg}
M.~Fairbairn, A.C. Kraan, D.A. Milstead, T.~Sjostrand, Peter~Z. Skands, et~al.
\newblock {Stable massive particles at colliders}.
\newblock {\em Phys.Rept.}, 438:1--63, 2007.

\bibitem{Fortin:2011hv}
Jean-Francois Fortin and Tim~M.P. Tait.
\newblock {Collider Constraints on Dipole-Interacting Dark Matter}.
\newblock {\em Phys.Rev.}, D85:063506, 2012.

\bibitem{Fox:2011pm}
Patrick~J. Fox, Roni Harnik, Joachim Kopp, and Yuhsin Tsai.
\newblock {Missing Energy Signatures of Dark Matter at the LHC}.
\newblock {\em Phys.Rev.}, D85:056011, 2012.

\bibitem{Fox:2002bu}
Patrick~J. Fox, Ann~E. Nelson, and Neal Weiner.
\newblock {Dirac gaugino masses and supersoft supersymmetry breaking}.
\newblock {\em JHEP}, 0208:035, 2002.

\bibitem{Freitas:2013xga}
A.~Freitas, K.~Hagiwara, S.~Heinemeyer, P.~Langacker, K.~Moenig, et~al.
\newblock {Exploring Quantum Physics at the ILC}.
\newblock 2013.

\bibitem{Friedland:2011za}
Alexander Friedland, Michael~L. Graesser, Ian~M. Shoemaker, and Luca Vecchi.
\newblock {Probing Nonstandard Standard Model Backgrounds with LHC Monojets}.
\newblock {\em Phys.Lett.}, B714:267--275, 2012.

\bibitem{nloci}
Jun Gao, Chong~Sheng Li, Jian Wang, Hua~Xing Zhu, and C.-P. Yuan.
\newblock {Next-to-leading QCD effect to the quark compositeness search at the
  LHC}.
\newblock {\em Phys.Rev.Lett.}, 106:142001, 2011.

\bibitem{Gershtein:2008bf}
Yuri Gershtein, Frank Petriello, Seth Quackenbush, and Kathryn~M. Zurek.
\newblock {Discovering hidden sectors with mono-photon $Z^\prime$o searches}.
\newblock {\em Phys.Rev.}, D78:095002, 2008.

\bibitem{Godfrey:2005pm}
Stephen Godfrey, Pat Kalyniak, and Alexander Tomkins.
\newblock {Distinguishing between models with extra gauge bosons at the ILC}.
\newblock 2005.

\bibitem{Goodman:2010ku}
Jessica Goodman, Masahiro Ibe, Arvind Rajaraman, William Shepherd, Tim~M.P.
  Tait, et~al.
\newblock {Constraints on Dark Matter from Colliders}.
\newblock {\em Phys.Rev.}, D82:116010, 2010.

\bibitem{Goodman:2010yf}
Jessica Goodman, Masahiro Ibe, Arvind Rajaraman, William Shepherd, Tim~M.P.
  Tait, et~al.
\newblock {Constraints on Light Majorana dark Matter from Colliders}.
\newblock {\em Phys.Lett.}, B695:185--188, 2011.

\bibitem{Graham:2011ah}
Peter~W. Graham, Kiel Howe, Surjeet Rajendran, and Daniel Stolarski.
\newblock {New Measurements with Stopped Particles at the LHC}.
\newblock {\em Phys.Rev.}, D86:034020, 2012.

\bibitem{Han:2013mra}
Tao Han, Paul Langacker, Zhen Liu, and Lian-Tao Wang.
\newblock {Diagnosis of a New Neutral Gauge Boson at the LHC and ILC for
  Snowmass 2013}.
\newblock 2013.

\bibitem{Harari:1979gi}
Haim Harari.
\newblock {A Schematic Model of Quarks and Leptons}.
\newblock {\em Phys.Lett.}, B86:83, 1979.

\bibitem{Hayden:2013sra}
Daniel Hayden, Raymond Brock, and Christopher Willis.
\newblock {Z Prime: A Story}.
\newblock 2013.

\bibitem{Hewett:1988xc}
JoAnne~L. Hewett and Thomas~G. Rizzo.
\newblock {Low-Energy Phenomenology of Superstring Inspired E(6) Models}.
\newblock {\em Phys.Rept.}, 183:193, 1989.

\bibitem{Hooper:2007qk}
Dan Hooper and Stefano Profumo.
\newblock {Dark matter and collider phenomenology of universal extra
  dimensions}.
\newblock {\em Phys.Rept.}, 453:29--115, 2007.

\bibitem{Hubisz:2004ft}
Jay Hubisz and Patrick Meade.
\newblock {Phenomenology of the littlest Higgs with T-parity}.
\newblock {\em Phys.Rev.}, D71:035016, 2005.

\bibitem{Jaeckel:2010ni}
Joerg Jaeckel and Andreas Ringwald.
\newblock {The Low-Energy Frontier of Particle Physics}.
\newblock {\em Ann.Rev.Nucl.Part.Sci.}, 60:405--437, 2010.

\bibitem{Kong:2013xta}
Kyoungchul Kong and Felix Yu.
\newblock {Discovery potential of Kaluza-Klein gluons at hadron colliders: A
  Snowmass whitepaper}.
\newblock 2013.

\bibitem{Kribs:2012gx}
Graham~D. Kribs and Adam Martin.
\newblock {Supersoft Supersymmetry is Super-Safe}.
\newblock {\em Phys.Rev.}, D85:115014, 2012.

\bibitem{kribsWP}
Graham~D. Kribs and Adam Martin.
\newblock {Dirac Gauginos in Supersymmetry -- Suppressed Jets + MET Signals: A
  Snowmass Whitepaper}.
\newblock 2013.

\bibitem{Krnjaic:2012aj}
Gordan Krnjaic and Daniel Stolarski.
\newblock {Gauging the Way to MFV}.
\newblock {\em JHEP}, JHEP04:064, 2013.

\bibitem{Langacker:2008yv}
Paul Langacker.
\newblock {The Physics of Heavy $Z^\prime$ Gauge Bosons}.
\newblock {\em Rev.Mod.Phys.}, 81:1199--1228, 2009.

\bibitem{Langacker:2009im}
Paul Langacker.
\newblock {The Physics of New U(1)-prime Gauge Bosons}.
\newblock {\em AIP Conf.Proc.}, 1200:55--63, 2010.

\bibitem{Leike:1998wr}
A.~Leike.
\newblock {The Phenomenology of extra neutral gauge bosons}.
\newblock {\em Phys.Rept.}, 317:143--250, 1999.

\bibitem{Lin:2013sca}
Tongyan Lin, Edward~W. Kolb, and Lian-Tao Wang.
\newblock {Probing dark matter couplings to top and bottom at the LHC}.
\newblock 2013.

\bibitem{cliccdr}
L.~Linssen et~al.
\newblock {Physics and Detectors at CLIC: CLIC Conceptual Design Report}.
\newblock 2012.

\bibitem{List:2013dga}
Jenny List and Benedikt Vormwald.
\newblock {Bilinear R Parity Violation at the ILC - Neutrino Physics at
  Colliders}.
\newblock 2013.

\bibitem{Liu:2013gea}
Tao Liu and C.T. Potter.
\newblock {Exotic Higgs Decay h to 2a at the International Linear Collider: a
  Snowmass White Paper}.
\newblock 2013.

\bibitem{nlojet}
Zoltan Nagy.
\newblock {Next-to-leading order calculation of three jet observables in hadron
  hadron collision}.
\newblock {\em Phys.Rev.}, D68:094002, 2003.

\bibitem{Nath:2010zj}
Pran Nath, Brent~D. Nelson, Hooman Davoudiasl, Bhaskar Dutta, Daniel Feldman,
  et~al.
\newblock {The Hunt for New Physics at the Large Hadron Collider}.
\newblock {\em Nucl.Phys.Proc.Suppl.}, 200-202:185--417, 2010.

\bibitem{Ovyn:2009tx}
S.~Ovyn, X.~Rouby, and V.~Lemaitre.
\newblock {DELPHES, a framework for fast simulation of a generic collider
  experiment}.
\newblock 2009.

\bibitem{Patt:2006fw}
Brian Patt and Frank Wilczek.
\newblock {Higgs-field portal into hidden sectors}.
\newblock 2006.

\bibitem{Petriello:2008pu}
Frank~J. Petriello, Seth Quackenbush, and Kathryn~M. Zurek.
\newblock {The Invisible $Z^\prime$ at the CERN LHC}.
\newblock {\em Phys.Rev.}, D77:115020, 2008.

\bibitem{Porod:2000hv}
W.~Porod, M.~Hirsch, J.~Romao, and J.W.F. Valle.
\newblock {Testing neutrino mixing at future collider experiments}.
\newblock {\em Phys.Rev.}, D63:115004, 2001.

\bibitem{Rajaraman:2011wf}
Arvind Rajaraman, William Shepherd, Tim~M.P. Tait, and Alexander~M. Wijangco.
\newblock {LHC Bounds on Interactions of Dark Matter}.
\newblock {\em Phys.Rev.}, D84:095013, 2011.

\bibitem{Rao:2012gf}
Kanishka Rao and Daniel Whiteson.
\newblock {Triangulating an exotic T quark}.
\newblock {\em Phys.Rev.}, D86:015008, 2012.

\bibitem{Schmaltz:2005ky}
Martin Schmaltz and David Tucker-Smith.
\newblock {Little Higgs review}.
\newblock {\em Ann.Rev.Nucl.Part.Sci.}, 55:229--270, 2005.

\bibitem{Shepherd:2009sa}
William Shepherd, Tim~M.P. Tait, and Gabrijela Zaharijas.
\newblock {Bound states of weakly interacting dark matter}.
\newblock {\em Phys.Rev.}, D79:055022, 2009.

\bibitem{Shupe:1979fv}
M.A. Shupe.
\newblock {A Composite Model of Leptons and Quarks}.
\newblock {\em Phys.Lett.}, B86:87--92, 1979.

\bibitem{squark-gluino-scan}
{Simplified Model Team}.
\newblock {\em Squark-Gluino Scan}.

\bibitem{Sjostrand:2007gs}
Torbjorn Sjostrand, Stephen Mrenna, and Peter~Z. Skands.
\newblock {A Brief Introduction to PYTHIA 8.1}.
\newblock {\em Comput.Phys.Commun.}, 178:852--867, 2008.

\bibitem{HiggsWG}
{Snowmass Higgs WG}.
\newblock 2013.

\bibitem{Suehara:2009bj}
Taikan Suehara and Jenny List.
\newblock {Chargino and Neutralino Separation with the ILD Experiment}.
\newblock 2009.

\bibitem{atlaswp}
{The ATLAS Collaboration}.
\newblock {\em LHC14 Whitepaper}.

\bibitem{ATLAS:sg}
{The ATLAS Collaboration}.
\newblock {\em ATLAS-CONF-2013-057}, 2013.

\bibitem{cmswp}
{The CMS Collaboration}.
\newblock {\em LHC14 Whitepaper}.

\bibitem{Walker:2013hka}
Devin G.~E. Walker.
\newblock {Unitarity Constraints on Higgs Portals}.
\newblock 2013.

\bibitem{Yu:2013wta}
Felix Yu.
\newblock {Di-jet resonances at future hadron colliders: A Snowmass
  whitepaper}.
\newblock 2013.

\bibitem{dmextrap}
Ning Zhou, David Berge, Tim M.~P. Tait, LianTao Wang, and Daniel Whiteson.
\newblock {Sensitivity of future collider facilities to WIMP pair production
  via effective operators and light mediators}.
\newblock {\em arXiv:1307.5327}, 2013.

\bibitem{dmcombo}
Ning Zhou, David Berge, and Daniel Whiteson.
\newblock {Mono-everything: combined limits on dark matter production at
  colliders from multiple final states}.
\newblock 2013.

\end{thebibliography}

\end{document}